UNIVERSITY OF OXFORD

DEPARTMENT OF COMPUTER SCIENCE

# Complexity of mixed equilibria in Boolean games

*Author:*
Egor Ianovski

*Supervisor:*
Dr. Luke Ong

# ABSTRACT

Boolean games are a succinct representation of strategic games wherein a player seeks to satisfy a formula of propositional logic by selecting a truth assignment to a set of propositional variables under his control. The difficulty arises because a player does not necessarily control every variable on which his formula depends, hence the satisfaction of his formula will depend on the assignments chosen by other players, and his own choice of assignment will affect the satisfaction of other players' formulae.

The framework has proven popular within the multiagent community and the literature is replete with papers either studying the properties of such games, or using them to model the interaction of self-interested agents. However, almost invariably, the work to date has been restricted to the case of pure strategies. Such a focus is highly restrictive as the notion of randomised play is fundamental to the theory of strategic games – even very simple games can fail to have pure-strategy equilibria, but every finite game has at least one equilibrium in mixed strategies.

To address this, the present work focuses on the complexity of algorithmic problems dealing with mixed strategies in Boolean games. The main result is that the problem of determining whether a two-player game has an equilibrium satisfying a given payoff constraint is NEXP-complete. Based on this result, we then demonstrate that a number of other decision problems, such as the uniqueness of an equilibrium or the satisfaction of a given formula in equilibrium, are either NEXP or coNEXP-complete. The proof techniques developed in the course of this are then used to show that the problem of deciding whether a given profile is in equilibrium is $\text{coNP}^{\#\text{P}}$-hard, and the problem of deciding whether a Boolean game has a rational-valued equilibrium is NEXP-hard, and whether a two-player Boolean game has an irrational-valued equilibrium is NEXP-complete. Finally, we show that determining whether the value of a two-player zero-sum game exceeds a given threshold is EXP-complete.

# CONTENTS

# LIST OF FIGURES

# LIST OF TABLES

# Part I

# GAMES, BOOLEAN AND OTHERWISE

# 1 INTRODUCTION

*It was done intentionally of course. The testers who tested this game went nuts. At first it was easier, but when the testers said "this is too difficult", I made it even more difficult.*

— Itagaki Tomonobu

The connection between games and logic is a natural one, and not merely because the game-theoretic concept of a "game" is broad enough to encompass just about every human endeavour; reason leads to argument, and an argument begs strategy to be won. It should come as no surprise that it was the highly litigious society of Athens that gave birth to the European logical tradition, nor that some of the most memorable arguments of Plato have a distinctly game-theoretic undertone.

With the increasing mathematisation of both disciplines in the nineteenth and twentieth century this connection too matured from the intuitive to the formal. Logics old and new find semantics in the interaction of strategic agents; epistemic properties of games are ready targets of logicians on the prowl; logic and games is the subject of various journals, conferences, lecture courses.

Among the fruits of this interaction is the framework of Boolean games (Harrenstein, van der Hoek, Meyer, *et al.* [2]), in which players assign truth values to propositional variables with the aim of satisfying a target formula. Boolean games function as succinct representations of strategic games, as the size of the truth function defined by a logical formula is potentially exponential in that of the formula. Such succinct representations are widely studied as the textbook normal form of strategic games scales too poorly to be used for anything but pedagogical purposes.

Logic is hard, however, and playing a game of itself is by no means trivial. Questions of complexity, thus, marched alongside Boolean games from the beginning, and at present much of the landscape is understood – in the case of pure strategies. The complexity of decision problems dealing with mixed strategies and mixed equilibria had been almost entirely neglected. This is the focus of the current work.

### 1.0.1 *Contribution*

We demonstrate that the the following problems are NEXP or coNEXP-complete for two-player Boolean games:

- ∃GuaranteeNash: deciding whether a Boolean game has an equilibrium satisfying a payoff constraint.

- ∀GuaranteeNash: deciding whether all equilibria of a Boolean game satisfy a payoff constraint.
- ∃NashSat: deciding whether a Boolean game has an equilibrium that satisfies a given formula.
- ∀NashSat: deciding whether all equilibria of a Boolean game satisfy a given formula.
- UniqueNash: deciding whether the equilibrium of a Boolean game is unique.
- IrrationalNash: deciding whether a Boolean game has an equilibrium where at least one strategy weight is irrational.

We also show that the following problem is NEXP-hard for Boolean games with at least three players:

- RationalNash: deciding whether a Boolean game has an equilibrium where all strategy weights are rational.

Finally, we show that IsNash, the problem of determining whether a given profile is an equilibrium, is $\text{coNP}^{\#\text{P}}$-hard and DValue, deciding whether the value of a two-player zero-sum Boolean game is above a given threshold, is EXP-complete.

## 1.1 THE STRUCTURE OF THE WORK

In Chapter 2 we introduce the framework we are working in and present some key results and definitions from the theory of strategic games that future arguments will take for granted. We then discuss how Boolean games and the results of this thesis relate to the wider literature.

The technical content begins in Chapter 3. This is our quarry, and the lemmata we prove here will serve as building blocks for the main results. These involve developing a system of shorthand to discuss numbers and arithmetic in propositional logic, as well as introducing two-player zero-sum gadget games of a fixed value that will allow us to simulate finer-grained preferences than what the win-lose nature of Boolean games allows.

We set the cornerstone in Chapter 4 by showing that the problem ∃GuaranteeNash is NEXP-complete for two-player Boolean games. This is the main technical result of the thesis, and it demonstrates how the vocabulary introduced in Chapter 3 can be used to simulate the computation of a Turing machine via a profile in mixed strategies. It is worth studying closely, as subsequent results will be variations on the same theme.

In Chapter 5 we exploit the result and techniques of the preceding chapter to broaden the edifice and demonstrate that UniqueNash, ∃NashSat, ∀NashSat ∀GuaranteeNash and IrrationalNash are either NEXP or coNEXP-complete. We also demonstrate that RationalNash is NEXP-hard for three-player Boolean games, and IsNash is $\mathrm{coNP}^{\#\mathrm{P}}$-hard if the number of players is unbounded.

The capstone, Chapter 6, shows that DValue is EXP-complete for two-player zero-sum Boolean games. This result demonstrates how the machinery developed thus far can be used to translate proofs from a wider class of succinct game representations to the case of Boolean games.

We top out in Chapter 7, presenting a summary of the results and a discussion of problems left unsolved.

No construction site is without its scrapheap, and the adventurous dumpster-diver can turn to Appendix A for proofs deemed unworthy to grace the pages of this thesis and Appendix B for lines of thought that did not lead to anything substantive.

Certain sections in Chapter 2 and Chapter 3 are marked as scholia.These are sections of reduced formality, intended to serve as justifications, explanations and asides. The results of these scholia either do not contribute to the technical content of the work, or are standard results that are assumed knowledge in the algorithmic game theory literature. Readers well acquainted with the field can either skip or skim the scholia with no detriment to the understanding of the main theorems.

# 2

# PRELIMINARIES

*No one can make aught thereof,*
*And wisdom must be wedded to his soul,*
*Who from these pieces can invent a game.*

— Shahnameh, C. 1725

## 2.1 FRAMEWORK

### 2.1.1 *Games and representations*

Boolean games can be, and have been, motivated in many different ways; perhaps the wide appeal of the concept is due precisely to the fact that scholars from the logic, multiagent, and game theory communities can all claim it as their own. The motivation we adopt in this work, then, is necessarily sectarian. We have little to say about what these games tell us about the algebra of logic, or their fecundity in modelling a multiagent environment. Instead, we place the stress firmly on their place in the wider algorithmic study of an object known as a finite strategic game.

The finite strategic game can be thought of as a game in the abstract, with all the fat stripped away: what is the smallest set of pieces needed to construct a game? Clearly, if a game is to be played, there need be players. Next, these players need to be endowed with some capacity of choice: some actions or strategies that they may choose to employ, else we could make do without players altogether. And finally, the game needs to have an outcome, which the players may interpret as their heart desires. It is these three elements, without any embellishments of time, state space, or information, that motivate our first definition.

**Definition 2.1.1.** Consider a finite set of $n$ players, $N = \{ 1, \dots, n \}$. For each $i \in N$, Player $i$ is equipped with a finite set of *pure strategies*, $S_i$. An $n$-tuple of pure strategies, one for each player, is called a *pure-strategy profile*. The space of pure-strategy profiles is denoted $\mathcal{S} = S_1 \times \cdots \times S_n$. Player $i$'s utility function maps pure-strategy profiles to the reals, $u_i : \mathcal{S} \to \mathbb{R}$. A *finite strategic game* is the triple $(N, \{ S_1, \dots, S_n \}, \{ u_1, \dots, u_n \})$.

*Our convention for pronouns will treat odd players as male and even players as female.*

There are two natural families of subclasses of such games. First, we could restrict the number of players; the class of games where $n$ is equal to $k$ gives us the *k-player games*. *Two-player games*, in particular, will be of interest to us in this thesis. On the other hand we could restrict the codomain of the utility functions, particularly since, given the biases of our field, we shall always have issues of computability in mind. It is then natural to consider the class of *rational-valued games*, in which every $u_i$ has its range restricted to $\mathbb{Q}$. In the more radical case where we restrict the range to $\{0, 1\}$, we term the result *win-lose games*.

A more complicated subclass, based on semantic rather than syntactic considerations, is that of *zero-sum games*. A game is said to be zero-sum just if there exists a $c$ such that for every $\boldsymbol{s} \in \mathcal{S}$, the following equation is satisfied:[1]

$$\sum_{i=1}^{n} u_i(\boldsymbol{s}) = c. \qquad \blacksquare$$

We would like to stress that what we have just defined is an abstract mathematical object and few people, if any at all, visualise games in the manner described. One would be tempted to sketch a state space, or a payoff matrix, or perhaps even concretely imagine a situation of actual beings interacting in a competitive setting. We study shadows on the wall to reconstruct the object from which they arise.

If we do not shy from pedantry (caveat lector: we won't), we may claim that even had one the desire to, to reason about a game in the abstract is simply not possible – after all, a set has no order, yet we write $S_1 = \{ s_1, \ldots, s_m \}$.

We wish to make this notion of representing an abstract mathematical object with a concrete data structure precise, due to the algorithmic nature of our enquiry; it is well and proper for a human being to conceive of a problem in whatever way convenient, provided they return to the definition once they `\begin` the proof environment, but a machine must be informed precisely how the sequence of symbols on its tape relates to the object in question. This brings us to the notion of a representation:

*If $\mathcal{R}$ is a representation of $\mathcal{G}$ and $f(R) = G$, we will overload terminology by referring to $R$ as the representation of $G$.*

**Definition 2.1.2.** Let $\mathcal{G}$ be a class of finite strategic games. A *representation* of $\mathcal{G}$ is a pair $(\mathcal{R}, f)$, where $\mathcal{R}$ is a class of finite data structures, and $f$ is a computable procedure that, given $R \in \mathcal{R}$, lists the players, pure strategies and graphs of the utility functions of some $G \in \mathcal{G}$.

A representation is said to be *adequate* with respect to $\mathcal{G}$ if for every $G \in \mathcal{G}$ there exists some $R \in \mathcal{R}$ such that $f$, run on $R$, lists the components of $G$. $\blacksquare$

It is clear that no representation can be adequate with respect to the class of all finite strategic games as the reals are non-denumerable.

1 Strictly speaking, we have defined a *constant-sum* game. In a zero-sum game $c$ ought to be zero. There is, however, no harm in such abuse of terminology, for reasons we will see in Section 2.1.1.2. The main benefit of this choice is a certain resonance with the wider literature; standard theorems state results for zero-sum games which in fact apply equally to constant-sum games, so this spares us the need for trivial corollaries; the conflation of zero/constant-sum nomenclature is by no means uncommon among authors, putting us in good company; and outside the technical literature of game theory the term "zero-sum" has come to signify a situation of complete competition, which of course applies equally to the constant-sum case. Moreover, a personal affection of the author for the euphonic character of the word "zero" may have swayed the decision.

**Player Two**

| **Player One** | $s^2_1$ | $\cdots$ | $s^2_m$ |
|---|---|---|---|
| $s^1_1$ | $u_1(s^1_1, s^2_1)$ $u_2(s^1_1, s^2_1)$ | $\cdots$ | $u_1(s^1_1, s^2_m)$ $u_2(s^1_1, s^2_m)$ |
| $\vdots$ | $\vdots$ | $\ddots$ | $\vdots$ |
| $s^1_k$ | $u_1(s^1_k, s^2_1)$ $u_2(s^1_k, s^2_1)$ | $\cdots$ | $u_1(s^1_k, s^2_m)$ $u_2(s^1_k, s^2_m)$ |

Figure 2.1: A two-player game in normal form.

Looking at a rational-valued game, however, a canonical choice seems evident: this is a finite object, so why not just write it down in its entirety? This leads us to the normal form:

**Definition 2.1.3.** The *normal form* is a representation of finite strategic games obtained by listing the components of a game, including the graphs of the utility functions, explicitly. In the case of two-player games this is frequently presented graphically in the form of a labelled table, as in Figure 2.1. ■

**Fact 2.1.4.** *The normal form is adequate with respect to rational-valued games.*

**Example 2.1.5.** *Prisoner's Dilemma* is a game where two prisoners have to decide whether to inform on their partner in crime or to remain silent ($S_1 = \{\text{Talk}^1, \text{Silent}^1\}$). If both prisoners are silent then the state will lack the evidence to prosecute, but the police will still be able to pin a minor offence on them, interning them for a short while ($u_1(\text{Silent}^1, \text{Silent}^2) = -1$). If one informs and one stays silent, the informer will be set free ($u_1(\text{Talk}^1, \text{Silent}^2) = 0$) and the other given the maximum sentence ($u_2(\text{Talk}^1, \text{Silent}^2) = -5$). If both talk, they will be given slightly shorter sentences in light of their cooperation ($u_1(\text{Talk}^1, \text{Talk}^2) = -4$).

*We use superscripts to record player ownership where necessary.*

*Battle of the Sexes* is a game where Player One and Two are trying to decide on a venue for a date. Player One prefers ballet and Player Two prefers boxing. In case of successful coordination, the player at their venue of choice gets a utility of 3 and the other player a utility of 2. The utility for standing outside the venue in the rain, waiting for someone who never comes, is 0.

*Matching Pennies* is a game where Player One and Two simultaneously display a coin either heads or tails up. If both coins have the same orientation, Player Two pockets a utility of 1 while Player One walks away with 0. If the orientations are different, Player One gets 1 and Player Two 0.

The normal forms of these games are displayed in the left column of Figure 2.2. ■

The tabular presentation of the normal form allows us to better visualise the strategic considerations the players are embroiled in. Every square represents a strategy profile, or an outcome of a game. Player One would naturally wish to find himself in a square where his utility is highest. However, he can only control the row of the table; it is up to Player Two to determine which column they play in. It is only when Player One finds himself in a position where he is in profile $\boldsymbol{s}$ and there is another $\boldsymbol{s'}$ in the same column for which $u_i(\boldsymbol{s'}) > u_i(\boldsymbol{s})$ that he can change his choice of strategy to improve his payoff – he has a *profitable deviation.*

The profiles where no profitable deviations for either player exist, then, are in some sense stable: we do not claim that a play of any game will naturally end up in such a profile, but if the players find themselves in such a position, and are aware of it, they will have no incentive to leave. This brings us to a fundamental solution concept in game theory, the *Nash equilibrium.*

*As Nash equilibria are the only kind of equilibria that will be of interest to us, we will refer to them simply as equilibria.*

**Definition 2.1.6.** A *pure-strategy Nash equilibrium* is a pure-strategy profile, $\boldsymbol{s}$, from which no player has a profitable deviation.

That is, if we use $\boldsymbol{s}_{-i}(s_j)$ to represent the strategy profile obtained by replacing Player $i$'s strategy in $\boldsymbol{s}$ with $s_j \in S_i$, then $\boldsymbol{s}$ is an equilibrium precisely when the following is satisfied:

$$\forall i \in N, \ \forall s_j \in S_i : \quad u_i(\boldsymbol{s}) \geq u_i(\boldsymbol{s}_{-i}(s_j)). \qquad ■$$

**Example 2.1.7.** Prisoner's Dilemma has one pure-strategy equilibrium: if Player Two is silent, Player One would rather talk as that would net him a utility of 0 as opposed to −1. If Player Two talks, then talking will get Player One a utility of −4, whilst being silent −5. Thus no matter what Player Two chooses, Player One would choose to talk – this is an example of what is called a *dominant strategy.* As the game is completely symmetric, the same argument applies to Player Two and in equilibrium both players dob each other in.

Battle of the Sexes has two pure-strategy equilibria: both players go to the boxing match or both players go to the ballet performance. In any profile where the players disagree on the venue, both have the incentive to change their strategy and increase their payoff from 0 to either 2 or 3. The players may disagree on the relative merits of the two equilibria, but they would certainly rather be at either of them than alone on Friday night.

Matching Pennies has no equilibria in pure strategies. Should Player One choose tails, Player Two will choose tails. Should Player Two choose tails, Player One will choose heads. There is no pure-strategy profile in which neither player can improve their payoff by deviating.

The equilibria, and the profitable deviations, are displayed in the right column of Figure 2.2. ■

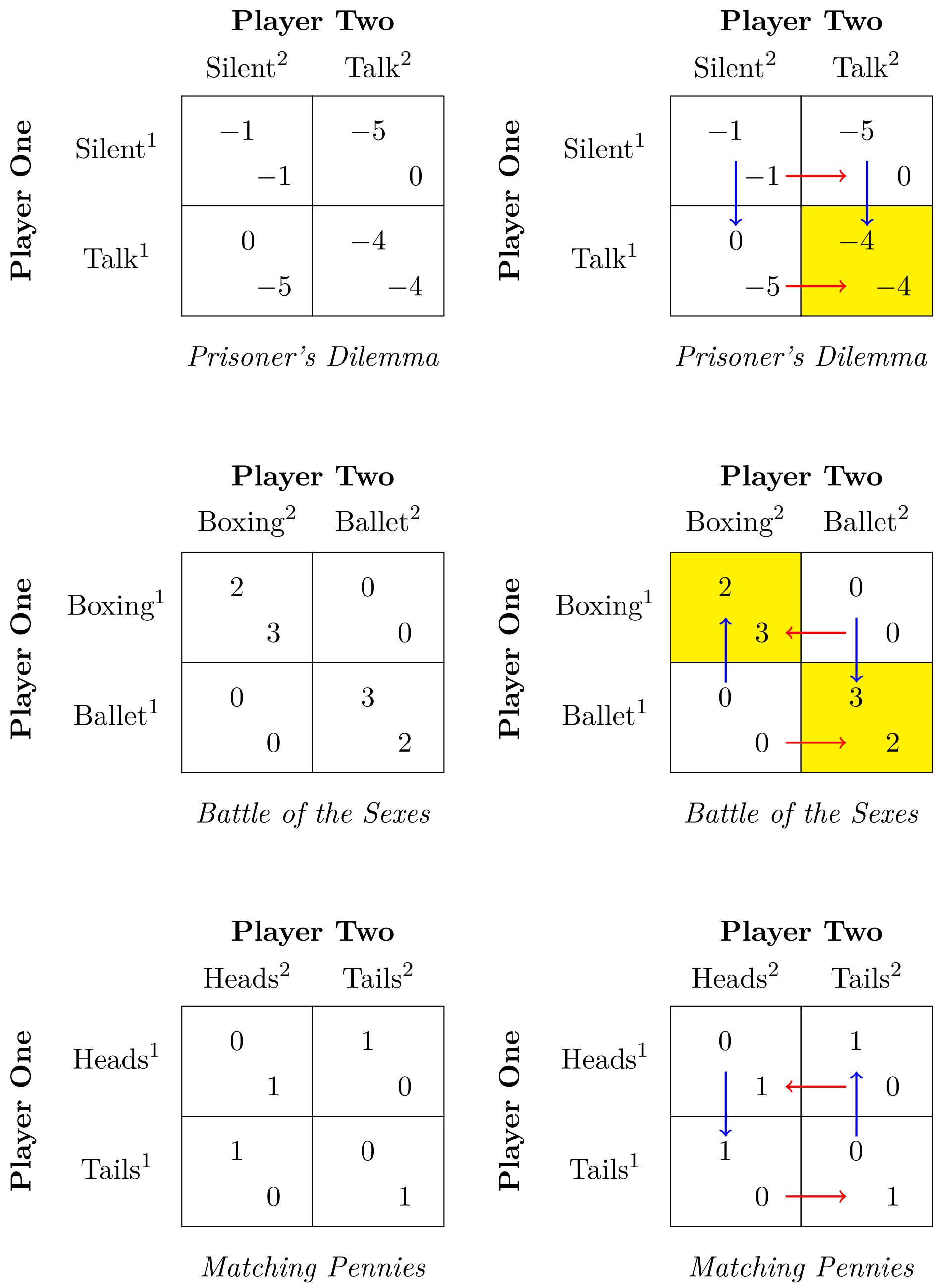

Figure 2.2: Three games in normal form, the players' deviations and resulting pure-strategy equilibria.

The case of Matching Pennies bears further consideration, as it demonstrates that even very simple games can fail to have a pure-strategy equilibrium; if we cannot account for this fact, then the limits of the mathematical treatment of games would appear to be grave indeed. Let us then consider where the problem lies. Player One does not want to play heads, as Player Two will respond with heads and win the game. Neither does he want to play tails, as Player Two can counter with tails as well.

Of course, this description of the situation is not faithful to our definition of a game – Player Two does not get a chance to observe Player One before responding, the coins are displayed simultaneously. In this sense, in a single match of Matching Pennies it does not really matter what strategy Player One chooses. Nevertheless, if Player One displays a certain predilection for heads, then over a sequence of ten matches (some nights are more boring than others) Player Two may discover this bias, and start responding with heads. Player One, of course, is not stupid. There may be no sure way to win the game, but there is a clear means to avoid losing: display heads or tails in an unpredictable pattern through the matches (James Bond fans may recall Bond's rock-paper-scissors match with Tanaka). This gives us a frequentist interpretation of randomised play, or what we call a mixed strategy:

**Definition 2.1.8.** A *mixed strategy* for Player $i$ is a probability distribution over $S_i$. An $n$-tuple of mixed strategies, one for each player, is called a *mixed-strategy profile.* If we let $\mathcal{P}(X)$ denote the space of probability distributions over $X$, then the space of mixed-strategy profiles is $\mathcal{P}(\mathcal{S})$ and the space of mixed strategies for Player $i$ is $\mathcal{P}(S_i)$.

We extend a utility function from the space of pure-strategy profiles to mixed-strategy profiles on the basis of expected utility. That is, the function $u_i : \mathcal{S} \to \mathbb{R}$ is extended to $U_i : \mathcal{P}(\mathcal{S}) \to \mathbb{R}$ in the following manner:

*We will conflate $u_i$ with $U_i$ where no confusion will arise*

$$U_i(\boldsymbol{\sigma}) = \sum_{\boldsymbol{s}\in\mathcal{S}} u_i(\boldsymbol{s})P(\boldsymbol{s} \mid \boldsymbol{\sigma}).$$

We can thus define a *mixed-strategy equilibrium* analogously to the pure strategy case: it is a strategy profile $\boldsymbol{\sigma}$ satisfying the following condition:

$$\forall i \in N,\ \forall \sigma_j \in \mathcal{P}(S_i) : \quad U_i(\boldsymbol{\sigma}) \geq U_i(\boldsymbol{\sigma_{-i}}(\sigma_j)). \qquad \blacksquare$$

**Example 2.1.9.** Matching Pennies has a unique mixed-strategy equilibrium where both players display heads with probability $1/2$. The utility Player One derives from this equilibrium is $1/2$: there is a $1/4$ chance of $(\text{Heads}^1, \text{Tails}^2)$ being realised, and a $1/4$ chance of $(\text{Tails}^1, \text{Heads}^2)$. This gives a $1/4 + 1/4$ probability of getting a utility of 1, and the other two profiles yield a utility of 0. If Player One were to deviate to playing $\text{Heads}^1$, then there would be a $1/2$ chance that Player Two would play $\text{Tails}^2$, giving Player One a $1/2$ probability of getting a utility of 1, and a $1/2$ chance of $\text{Heads}^2$, and hence a utility of 0. The same holds for a deviation to Tails$^1$, mutatis mutandis, so Player One has no incentive to deviate. As the game is symmetric, neither does Player Two.

Battle of the Sexes has an additional equilibrium to the two we have already seen in which Player One plays Boxing$^1$ with probability $^2/_5$ and Ballet$^1$ with probability $^3/_5$, and Player Two plays Ballet$^2$ with probability $^2/_5$, Boxing$^2$ with $^3/_5$. The payoff to Player One is $^2/_5 \cdot {^3/_5} \cdot 2 + {^3/_5} \cdot {^2/_5} \cdot 3 = {^6/_5}$. This is also the payoff he would get for playing Ballet$^1$ ($^2/_5 \cdot 3$) or Boxing$^1$ ($^3/_5 \cdot 2$).

Prisoner's Dilemma has no additional equilibria. This is a consequence of the presence of strictly dominant strategies – under any circumstance, Player One has a strict preference for squealing, and hence there can be no equilibrium in which he randomises over talking and staying silent. ■

The reader will note two things. First, unlike with pure-strategy equilibria, we made no attempt to motivate how players reach these states. We have already mentioned that equilibria are best thought of as states in which the game stays, rather than states which the game reaches. This will be useful to keep in mind when we construct the complex equilibria needed to prove our results – it does not matter how players arrive at such profiles, only that once they do, they will not deviate.

Second, in the example above we have only considered the case of players deviating to pure strategies, whereas the definition of equilibrium requires there to be no profitable deviation in the entire continuous space $\mathcal{P}(S_i)$. There is no generality lost in doing so, as we shall quickly verify.

**Fact 2.1.10.** *A strategy profile is in equilibrium if and only if no player can strictly increase his payoff by deviating to a pure strategy. This also implies that a necessary condition for equilibrium is that a player need be indifferent between deviating to any pure strategy in the support of his mixed strategy.*

*Proof.* Finding a profitable deviation for Player $i$ is the problem of maximising the function $U_i(\boldsymbol{\sigma}_{-i}(\sigma_j))$ with respect to the variable $\sigma_j$, over the simplex $\mathcal{P}(S_i)$. Following Definition 2.1.8:

$$U_i(\boldsymbol{\sigma}_{-i}(\sigma_j)) = \sum_{s \in \mathcal{S}} \boldsymbol{\sigma}_{-i}(\sigma_j)(\boldsymbol{s}) u_i(\boldsymbol{s}).$$

This is a linear function, and will therefore find its extrema at the boundary points of the domain – the pure strategies. Q.E.D.

Mixed-strategy equilibria are the subject of the most famous theorem of game theory.

**Theorem 2.1.11** (Nash [4]). *Every finite strategic game has an equilibrium in mixed strategies.*

It is, however, worth noting that game theory predates the concept of equilibrium. Indeed, "what is a stable outcome of the game?" is a much less natural question to ask than "what is the best way to play?".[2]

There is, of course, no reason to suppose that in an arbitrary game an "optimal" way to play should exist. A game may consist of many players pursuing different goals, and were Player One to have an optimal choice of strategy, that would imply that the game is in some sense trivial: the other players do not matter at all – Player One's strategy does not depend on theirs.

The case of two-player zero-sum games, however, is special in that the notion of optimal play *is* well defined. In such a case it makes sense to ask how much utility this optimal strategy would secure for Player One, which brings us to the next theorem:

**Theorem 2.1.12** (von Neumann [5])**.** *Player One obtains the same utility in every equilibrium of a finite two-player zero-sum game. This is called the* value *of the game, and Player One obtains it by playing a* maxmin strategy*. That is, the optimum strategy for Player One is the following:*

$$\operatorname*{arg\,max}_{\sigma^1 \in \mathcal{P}(S_i)} \min_{s^2 \in S_2} U_1(\sigma^1, s^2).$$

It should be noted that the theorems of Nash and von Neumann apply to games rather than representations – a finite strategic game will have an equilibrium whether it is presented in normal form or otherwise. So far the normal form is the only representation we have dealt with, as it is perhaps the most natural way to view a finite strategic game,[3] but this generality is of importance to us as it will save us the trouble of restating variants of Nash's theorem for every representation we shall come across (and we will examine a fair number in Section 2.2). There are two other representations we wish to introduce before the end of this section, one for its significance for the algorithmic study of games, and the other being the star of the show itself.

We have described the normal form of a two-player game as a payoff table. Perhaps a more common term in the literature is a payoff matrix. We avoided this usage as there does in fact exist a game representation that is based on matrices in the strict mathematical sense of the word.

**Definition 2.1.13.** A *bimatrix game* is a representation of a two-player game consisting of two matrices, $A$ and $B$. The matrices are $m \times k$,

2 The motivating question of von Neumann [5] was: "Wie muß einer dieser Spieler, $S_m$, spielen, um dabei ein möglichst günstiges Resultat zu erzielen?".

3 That said, it is worth noting that in both his 1928 paper and 1947 book, von Neumann begins by defining a much more complicated game form, incorporating alternating turns and chance moves, before demonstrating how such a game reduces to a much simplified "Normalform" – namely, rather than letting players make a move at each stage of the game, simply have them choose a single response function, a "Spielmethode", upfront. That is why we call the members of $S_i$ strategies rather than moves or actions.

where $|S_1| = k$ and $|S_2| = m$. The $(i, j)$ entry of $A$, $a_{i,j}$, is $u_1(s^1_i, s^2_j)$, and the $(i, j)$-entry of $B$, $b_{i,j}$, is $u_2(s^1_i, s^2_j)$.

If the game is zero-sum, then only one matrix is required. This is sometimes called a *matrix game.* ■

**Fact 2.1.14.** *Bimatrix games are adequate with respect to rational-valued two-player games.*

One may well ask what the point of this definition is – all we did was split the table of the normal form in two. However, this representation provides us with a very convenient way of evaluating utilities. Given mixed strategies $\sigma^1 \in \mathcal{P}(S_1)$ and $\sigma^2 \in \mathcal{P}(S_2)$, interpreted as vectors of probabilities, the utility Player One obtains from $\boldsymbol{\sigma} = (\sigma^1, \sigma^2)$ is $(\sigma^1)^{\mathrm{T}} A \sigma^2$, whereas Player Two gets $(\sigma^1)^{\mathrm{T}} B \sigma^2$.

Equilibrium, then, is equivalent to two conditions:

$$\begin{aligned} \forall \tau \in \mathcal{P}(S_1) : \quad & (\sigma^1)^{\mathrm{T}} A \sigma^2 \geq \tau^{\mathrm{T}} A \sigma^2, \\ \forall \tau \in \mathcal{P}(S_2) : \quad & (\sigma^1)^{\mathrm{T}} B \sigma^2 \geq (\sigma^1)^{\mathrm{T}} B \tau. \end{aligned}$$

It is precisely this characterisation of equilibria as solutions to linear systems that enabled much fruitful research into the algorithmic properties of two-player games. This is relevant for us in that it tells us that two-player games have rational-valued equilibria, as we shall discuss in Section 3.3.

Our main focus, however, is on a game representation that looks very different from the normal form, in which the preferences of the players are expressed via propositional logic.

**Definition 2.1.15.** A Boolean game is a representation of a win-lose game consisting of $n$ disjoint sets of propositional variables, $\Phi_1, \ldots, \Phi_n$, and $n$ formulae of propositional logic, $\gamma_1, \ldots, \gamma_n$, defined over $\Phi = \biguplus_{i \in N} \Phi_i$.

We refer to $\gamma_i$ as Player $i$'s *goal formula* and, as the name suggests, the goal of Player $i$ is to satisfy $\gamma_i$. To do this Player $i$ is given control over the variables in $\Phi_i$. The game is played by having all players, independently and simultaneously, choose a truth assignment to the variables under their control, and then evaluating their goal formula on the resulting truth assignment.

That is to say, a pure strategy for Player $i$ is a truth assignment $\nu_i : \Phi_i \to \{\mathit{true}, \mathit{false}\}$. The set of pure strategies available to him is thus $2^{\Phi_i}$. The truth assignments to $\Phi_1, \ldots, \Phi_n$ are combined into a joint assignment, $\boldsymbol{\nu}$, to $\Phi$ in the natural way, and Player $i$'s utility function tracks the satisfaction of $\gamma_i$:

$$u_i(\boldsymbol{\nu}) = \begin{cases} 1 & : \boldsymbol{\nu} \models \gamma_i, \\ 0 & : \boldsymbol{\nu} \not\models \gamma_i. \end{cases}$$ ■

**Fact 2.1.16.** *Boolean games are adequate with respect to the class of win-lose games where the cardinality of every player's strategy set is a power of two.*

*Proof.* A direct consequence of the expressive completeness of propositional logic: every truth function on a finite domain can be represented by a propositional formula. Q.E.D.

**Example 2.1.17.** Of the three example games we have considered, only Matching Pennies satisfies the conditions of Fact 2.1.16, and indeed this game has a Boolean representation:

$$\begin{aligned} \Phi_1 &= \{\, p \,\}, \\ \Phi_2 &= \{\, q \,\}, \\ \gamma_1 &= \neg(p \leftrightarrow q), \\ \gamma_2 &= p \leftrightarrow q. \end{aligned}$$

Battle of the Sexes relies on a tripartite distinction between outcomes, so we cannot faithfully reproduce it in Boolean form. We could, however, represent a simplified "pure coordination" variant, where the players are indifferent to the venue insofar as they are at the same place:

$$\begin{aligned} \Phi_1 &= \{\, p \,\}, \\ \Phi_2 &= \{\, q \,\}, \\ \gamma_1, \gamma_2 &= p \leftrightarrow q. \end{aligned}$$

Such a game will preserve the equilibrium structure of Battle of the Sexes, but will lose the fact that players are not indifferent between them.

Prisoner's Dilemma does not have a Boolean form, and there is no reasonable way to supply it with one. The tripartite distinction between outcomes is vital. Player One *has* to strictly prefer $(\text{Talk}^1, \text{Silent}^2)$ to $(\text{Silent}^1, \text{Silent}^2)$, or else he would see no reason to talk if paired with a cooperative codetainee, Player Two *has* to strictly prefer $(\text{Talk}^1, \text{Talk}^2)$ to $(\text{Talk}^1, \text{Silent}^2)$, else she would be content to suffer in silence. ■

Despite several attempts to wed the two disciplines, it is fair to say that logic and probability do not coexist peacefully in most people's minds. As such it is natural to ask whether mixed strategies are as vital to the theory of Boolean games as strategic games in general. The short answer is yes; Boolean games are an adequate representation of a sufficiently large class of strategic games, and hence inherit all their properties, such as the necessity of mixed strategy play for the existence of equilibria, values of zero-sum games, and so on. In the above example, both constructed games have the expected equilibria in mixed strategies.

If such an answer is unsatisfactory, it helps to remember that at their core, mixed strategies are a mechanism for coping with uncertainty. What is sometimes regarded as the first formal theorem in game theory (Zermelo [7]) demonstrates that without chance and uncertainty, a game is determined by pure strategy play. So let us consider uncertainty.

**Example 2.1.18.** Consider a family of two-player zero-sum Boolean games $G(\varphi)$ parametrised by a formula $\varphi$. To play $G(\varphi)$ Player Two chooses a truth assignment to $\varphi$ $\nu_2$, and Player One states whether $\nu_2 \vDash \varphi$. That is to say, the game is as follows:

$$\begin{aligned}\Phi_1 &= \{ p \},\\ \Phi_2 &= \text{ all the variables of } \varphi,\\ \gamma_1 &= (p \to \varphi) \wedge (\neg p \to \neg\varphi),\\ \gamma_2 &= \neg\gamma_1.\end{aligned}$$

Consider first the case without uncertainty: $\varphi$ is a tautology or a contradiction. In this case Player One can confidently proceed with a pure strategy: setting $p$ to *true* or *false* as the case may be. He can do so safely because he knows for a fact that no matter what truth assignment Player Two chooses, it will not affect the truth value of $\varphi$.

Now suppose that $\varphi$ is neither. Suddenly, committing to either *true* or *false* becomes a terrible idea, as Player Two could counter by picking a contrary assignment. The pure equilibria disappear: if Player One states that $\varphi$ is *true* or *false*, Player Two can deviate by making it *false* or *true*, and any assignment by Player Two to $\varphi$ cannot help but either satisfy the formula or fail to, hence Player One can deviate by setting $p$ to the appropriate value.

Player One simply does not know what the value of $\varphi$ will be, and he has no reason to prefer one to the other. He may as well set $p$ to *true* with probability $1/2$.[4] As it turns out, this is in fact his maxmin strategy. ■

A key feature of this representation, in contrast to the normal form, is that Boolean games are *succinct*: there exist strategic games whose normal form is exponentially larger than their smallest[5] Boolean representation. It is, however, difficult to say more than this. The size of the normal form, the canonical representation of a strategic game, is $O(n \max_i |S_i|^n \cdot \max_{j,s} |u_j(\boldsymbol{s})|)$. The size of a Boolean game is determined by the variable sets and formula lengths, which is $O(n \max_i |\Phi_i| + n \max_j |\gamma_j|)$. How does one compare this? The dependency on the number of players is no longer exponential, to be sure, but without knowing the magnitude of the other factors this does not tell us anything about the total size. We could, of course, put very crude bounds on these; in the best case, every formula is of constant size and we only need $\log_2 |S_i|$ variables to represent $|S_i|$ strategies. Our game is of size:

*Of course for win-lose games, $|u_j(\boldsymbol{s})| = 1$.*

$$O(n \log_2 |S_i| + nc) = O(n \log_2 |S_i|).$$

Clearly, an improvement on the $O(n|S_i|^n)$ of the normal form.

4 A less handwavy justification could invoke the principle of maximum entropy.

5 The Boolean representation is inherently non-unique, as every formula is logically equivalent to infinitely many others.

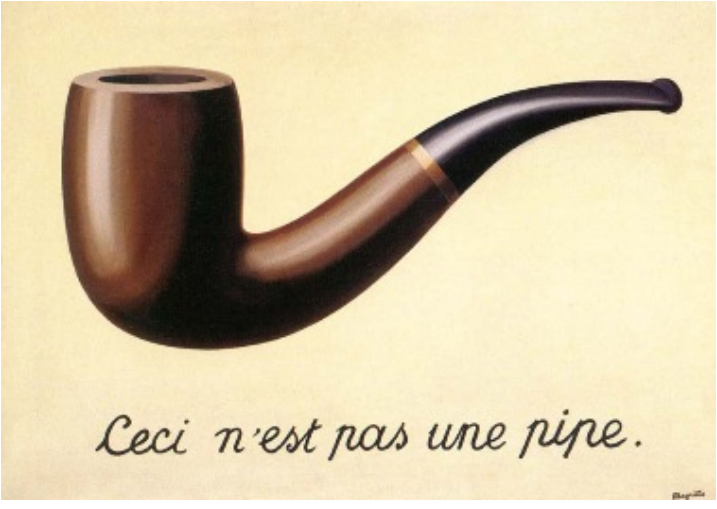

$$\begin{aligned}
\Phi_1 &= \{\, p_1, p_2 \,\}, \\
\Phi_2 &= \{\, q_1, q_2, q_3 \,\}, \\
\gamma_1 &= (p_1 \wedge q_1) \vee (q_2 \wedge q_3), \\
\gamma_2 &= \neg(q_1 \wedge q_2 \wedge q_3) \wedge (p_1 \vee p_2).
\end{aligned}$$

Figure 2.3: An image of a pipe, and a representation of a game.

For the other extreme, faced with a game with no logical structure whatsoever we might have to resort to the expedient of writing out the payoff matrix explicitly and giving each player a gigantic DNF formula picking out the non-zero entries. In this case the size our game would be:

$$O(n \log_2 |S_i| + n|S_i|^n \log_2 \mathcal{S}) = O(n|S_i|^n \log_2 \mathcal{S}).$$

Which would take us a logarithmic factor above the explicit representation.

This wide range need not surprise us; there is no universal compression function so any representation adequate with respect to a sufficiently comprehensive class of games cannot make them all smaller. The question is, which case is typical? Do there exist interesting games with concise Boolean representations, or may we just as well use the normal form each time $\gamma_i \neq \top$? This question could be seen as motivation for the results presented in this thesis: were Boolean games incapable of providing concise representations of complicated games, then we would expect their algorithmic complexity to be a lot lower – after all, we could always expand the game into its normal form and run the standard algorithm on that. The fact that we demonstrate an exponential jump in complexity for problems pertaining to the Boolean representation would suggest that a great many algorithmically rich games can indeed be concisely represented as a Boolean game.

#### 2.1.1.1 *Scholium: Ceci n'est pas un jeu*

> *The famous pipe. How people reproached me for it! And yet, could you stuff my pipe? No, it's just a representation, is it not? So if I had written on my picture "This is a pipe", I'd have been lying!*
>
> — René Magritte

The framework we have adapted is quite similar to that of Schoenebeck and Vadhan [9] and Àlvarez, Gabarró, and Serna [10], and not unlike that of Papadimitriou [11]. However, it does differ substantially from the way game are usually presented in the literature of both economics and computer science. Most authors do not distinguish between a game and its representation. What we have, after Osborne and Rubinstein [12], termed a strategic game, is commonly known as a normal-form game, and the two terms are treated as synonyms. The various representations we introduce in this thesis are generally defined simply as games in their own right – although rarely do authors verify that these new games inherit the classical results of game theory, upon which much of their analysis is based.

After comparing Definition 2.1.1 and Definition 2.1.3, however, the reader may well be tempted to take their side: the two state the exact same thing except we call one a "game" and the other a "representation". One could make the argument that the distinction we are trying to enforce is no more than petty sophistry. We hope to dissuade the reader of any such notion.

Even though various representations of games tend to be identified with the games themselves, the fact that some intuitive affinity between them is lurking beneath the veil is ineluctable. However, given the choice of definitions this intuition is ultimately problematic. Economics students, after being introduced to extensive games with imperfect information, are often asked to demonstrate that these are the same as normal-form games. This task is complicated by the fact that the claim, and others like it, are patently false – a table of values is obviously not the same thing as a labelled tree, there is no meaningful way to evaluate equality between a matrix and a Boolean circuit, and Nash's theorem certainly has nothing to say about the behaviour of a Turing machine.

Of course, everyone knows this. What is really going on is that an unspoken "it is intuitively obvious how to translate game X into game Y that preserves property Z" is inserted before any such statement, and indeed it is. But if we were to formalise this intuition we would end up with a plethora of notions of similarity between every pair of games in the literature, and we feel that the approach adopted in this thesis is cleaner: it is the responsibility of the representation itself to tell us how it relates to the concept of a game, the other representations need not know of its existence. Then, if we want to compare two representations, we can apply the computable procedure to obtain the game itself. This reduces the problem of comparing arbitrary algorithmic data structures to comparing finite sets and functions.

The distinction between a game and its representation is especially convenient given the algorithmic nature of this enquiry. Consider the problem of deciding whether a game has a pure-strategy equilibrium. Clearly, the answer to this question should not depend on the game representation used any more than the odour of a rose upon its name. Equally clearly, the complexity of the problem could differ substantially. Our convention allows us to use locutions like "∃PNE is in P for games in normal form, but $\Sigma_2^p$-complete for Boolean games", rather than defining ∃PNENorm and ∃PNEBool as separate problems.

More importantly, it allows us to compare the complexity of representations with respect to the entire class of such algorithmic problems, as we demonstrate below:

**Definition 2.1.19.** A game representation $(\mathcal{R}, f)$ is *polynomial-time embeddable* into $(\mathcal{T}, g)$ just if there exists a polynomial-time $h : \mathcal{R} \to \mathcal{T}$ such that for every $R \in \mathcal{R}$, $f(R) = g(h(R))$. ■

We have used "=" in the definition above. This brings us to the next question: what does it mean for two games to be the same? Standard notions in the literature include strict equality – are the strategy sets and utility functions identical? Strong isomorphism – can we rename the strategies and players to obtain equality? And the Harsanyi-Selten isomorphism – does there exist an isomorphism modulo an affine transformation of utilities?

The notion we will use lies somewhere in between equality and isomorphism: we allow the renaming of strategies, but not of players.[6] The symbol we use for this notion is =, and when $f(R) = g(T)$ we say that the representations $R$ and $T$ *represent the same game.*

*This will not cause confusion as we will never have need to distinguish = from strict equality.*

There is a natural class of algorithmic problems that do not distinguish between two representations that represent the same game.

**Definition 2.1.20.** Let $\mathcal{A}$ be an algorithmic problem about strategic games. $\mathcal{A}$ is said to be *representation-invariant* if the answer to the problem does not depend on the game representation used.

More concretely, given representations $(\mathcal{R}, f)$ and $(\mathcal{T}, g)$, if $\mathcal{A}$ is a decision problem then for all $R \in \mathcal{R}, T \in \mathcal{T}$, if $f(R) = g(T)$ then $R \in \mathcal{A}$ if and only if $T \in \mathcal{A}$. If $\mathcal{A}$ is a function problem then in the same conditions $\mathcal{A}(R) = \mathcal{A}(T)$. ■

**Fact 2.1.21.** *If $(\mathcal{R}, f)$ is polynomial-time embeddable into $(\mathcal{T}, g)$, then for any representation-invariant algorithmic problem $\mathcal{A}$, the $(\mathcal{R}, f)$ inherits the upper bounds on the complexity of $\mathcal{A}$ for $(\mathcal{T}, g)$, and $(\mathcal{T}, g)$ inherits the lower bounds on the complexity of $\mathcal{A}$ for $(\mathcal{R}, f)$, in terms of $\leq_m^{\mathrm{P}}$-degrees.*

*Proof.* For the upper bound on $(\mathcal{R}, f)$, note that $R \in \mathcal{A}$ if and only if $h(R) \in \mathcal{A}$. For the lower bound on $(\mathcal{T}, g)$, if $\mathcal{A}$ belongs to a $\leq_m^{\mathrm{P}}$-degree $\mathcal{D}$, and one can determine whether some $R \in \mathcal{R}$ belongs to $\mathcal{A}$ by applying $h$ to it and running the $(\mathcal{T}, g)$ algorithm on $h(R)$, then the $(\mathcal{T}, g)$ case cannot be easier than $\mathcal{D}$, otherwise the $(\mathcal{R}, f)$ case would belong to a lower degree. Q.E.D.

Note that if there exists a polynomial-time embedding both ways, i.e. that $(\mathcal{R}, f)$ and $(\mathcal{T}, g)$ are *biembeddable*, then from the vantage point of

6 The names of strategies are immaterial to us, so we choose not to distinguish between them. The numbering of the players, whilst unimportant to the mathematical side of the enquiry, we have opted to retain as they can highlight certain links between logical and game theoretic notions, such as negation and role switching.

mainstream algorithmic game theory there is little sense to distinguish between the two representations, as every representation-invariant complexity result will be the same. Thus when the word "equivalent" is used in the literature, it often means "polynomial-time biembeddable" within our framework.

#### 2.1.1.2 *Scholium: Cardinal utility*

In Definition 2.1.1, by defining $u_i$ as mapping to $\mathbb{R}$ we have made the assumption that a player is able to associate a real number with every outcome of the game, thereby expressing how satisfied he is with that outcome. In other words we have made the assumption of cardinal utility, and this choice begs an apology.

The history of this concept straddles the breadth of the history of economics, from the philosophical arguments of classical political economy, through various mathematical treatments in neoclassical economics, to the exciting new insights shed on the matter by the burgeoning field of neuroeconomics; we shall make no attempt to cover it here. To the interested reader we recommend the first chapter of von Neumann and Morgenstern [6] as a starting point, although we should note that the authors of that œuvre themselves plead that an adequate treatment of the subject is not the purpose of their book.

However, we would be doing the reader a disservice were we not to expose them to the von Neumann-Morgenstern utility theorem, especially as a familiarity with its consequences will aid the understanding of certain proofs in the sequel.

It is clear that cardinal preferences are a stronger assumption than ordinal: given a utility function $u_i$ for Player $i$, we can construct a preference relation for the player, $\succeq_i$, defined over all possible outcomes of the game (the space of mixed-strategy profiles, $\mathcal{P}(\mathcal{S})$), simply by setting $\boldsymbol{\sigma} \succeq_i \boldsymbol{\tau}$ for all those $\boldsymbol{\sigma}$, $\boldsymbol{\tau}$ for which $U_i(\boldsymbol{\sigma}) \geq U_i(\boldsymbol{\tau})$. The point of the theorem is that under certain assumptions, it is possible to do the reverse.

*The strict and symmetric counterparts, $\succ_i$ and $\sim_i$, are defined in the natural way.*

Suppose, then, that we start with a player equipped only with a preference relation, $\succeq_i \subseteq \mathcal{P}(\mathcal{S}) \times \mathcal{P}(\mathcal{S})$. Fortune favours the bold, so we assume our player is decisive enough to assert $\boldsymbol{\sigma} \succeq_i \boldsymbol{\tau}$ or $\boldsymbol{\tau} \succeq_i \boldsymbol{\sigma}$ for any pair of profiles: his preferences are complete. The very term of ordinal preferences suggests an order, and order implies transitivity, so we require that if the player asserts $\boldsymbol{\sigma} \succeq_i \boldsymbol{\tau}$ and $\boldsymbol{\tau} \succeq_i \boldsymbol{\upsilon}$, he also affirms that $\boldsymbol{\sigma} \succeq_i \boldsymbol{\upsilon}$. Continuity is not typically a virtue to which computer scientists attach a lot of value, but in deference to our economics colleagues we may yield the issue, and suppose that it is possible to find some mix of a good outcome and a bad outcome that is equally appealing to an intermediate outcome: given $\boldsymbol{\sigma} \succeq_i \boldsymbol{\tau} \succeq_i \boldsymbol{\upsilon}$, we can find a $p \in [0, 1]$ for which $p\boldsymbol{\sigma} + (1-p)\boldsymbol{\upsilon} \sim_i \boldsymbol{\tau}$.[7] Finally, as mixed-strategy profiles are in essence lotteries, and will ultimately resolve into one and only one outcome, it seems reasonable to assume that the presence of outcomes that are not realised in the lottery should not affect the lottery's value: the outcomes are independent. Thus if $\boldsymbol{\sigma}$ is preferred to $\boldsymbol{\tau}$, then for any $\boldsymbol{\upsilon}$, and any $p \in [0,1]$, the player asserts $p\boldsymbol{\sigma} + (1-p)\boldsymbol{\upsilon} \succeq_i p\boldsymbol{\tau} + (1-p)\boldsymbol{\upsilon}$.

We seek not to convince the reader that these assumptions are reasonable, but rather to make clear what the assumptions are. Completeness may well be unreasonable when the players are not rational entities, but psychologically driven creatures: I am not indifferent between pork and chicken, but I spend a long time staring at the shelves in Tesco trying to make up my mind. Transitivity could fail where the players do not represent unitary beings but societies of agents utilising a voting mechanism; enter Condorcet. Continuity, among other things, implies a certain attitude to risk: most people would prefer £1,000,000 to £999,999 to a kick in the teeth, but few would be willing to trade £999,999 for a $p$ chance of getting £1,000,000, no matter how close to 1 $p$ may be. Independence, as the name suggests, is about independence. Suppose you would prefer sunbathing in the Bahamas to skiing in the Alps. Your travel agent offers you a lottery: you inform him of your preference, drive to the airport and with probability $p$ your plane will take you to your destination of choice, and with $(1-p)$ to an Antarctic adventure. If you stick to your guns, you run the risk of finding yourself armed with snorkel and swimming togs, surrounded by penguins.

Nevertheless, in a circumstance where the four assumptions are appropriate we can rely on the following theorem:

**Theorem 2.1.22** (von Neumann and Morgenstern [6], appendix)**.** *Consider a relation $\succeq$ defined over $\mathcal{P}(\mathcal{S})$, satisfying four conditions:*

1. *For all $\boldsymbol{\sigma}, \boldsymbol{\tau} \in \mathcal{P}(\mathcal{S})$, $\boldsymbol{\sigma} \succeq \boldsymbol{\tau}$ or $\boldsymbol{\tau} \succeq \boldsymbol{\sigma}$ (completeness).*
2. *For all $\boldsymbol{\sigma}, \boldsymbol{\tau}, \boldsymbol{\upsilon} \in \mathcal{P}(\mathcal{S})$, $\boldsymbol{\sigma} \succeq \boldsymbol{\tau}$ and $\boldsymbol{\tau} \succeq \boldsymbol{\upsilon}$ imply $\boldsymbol{\sigma} \succeq \boldsymbol{\upsilon}$ (transitivity).*
3. *For all $\boldsymbol{\sigma}, \boldsymbol{\tau}, \boldsymbol{\upsilon} \in \mathcal{P}(\mathcal{S})$, if $\boldsymbol{\sigma} \succeq \boldsymbol{\tau} \succeq \boldsymbol{\upsilon}$, then there exists a $p \in [0,1]$ for which $p\boldsymbol{\sigma} + (1-p)\boldsymbol{\upsilon} \sim \boldsymbol{\tau}$ (continuity).*
4. *For all $\boldsymbol{\sigma}, \boldsymbol{\tau}, \boldsymbol{\upsilon} \in \mathcal{P}(\mathcal{S})$, if $\boldsymbol{\sigma} \succeq \boldsymbol{\tau}$, then for any $p \in [0,1]$, $p\boldsymbol{\sigma} + (1-p)\boldsymbol{\upsilon} \succeq p\boldsymbol{\tau} + (1-p)\boldsymbol{\upsilon}$ (independence).*

*There exists a $u : \mathcal{S} \to \mathbb{R}$ that induces $\succeq$. Moreover, $u$ is unique up to an affine transformation.*

Given this uniqueness result it is tempting to declare that games that differ only by an affine shift of the utility functions are fundamentally

7 Addition and scaling of mixed-strategy profiles is understood to be componentwise; clearly, a convex combination of probability distributions is in itself a probability distribution.

"the same", but that would be an unpolitical declaration to make in light of our long argument on the importance of being precise about the meaning of "sameness" in the previous section. We can however state the following, which is sufficient for our purposes:

**Fact 2.1.23.** *Consider games* $G = (N, \{S_i\}_{i\in N}, \{u_i\}_{i\in N})$ *and* $G' = (N', \{S'_i\}_{i\in N'}, \{u'_i\}_{i\in N'})$ *for which there exist* $a_i, b_i \in \mathbb{R}$, $a_i \neq 0$, *satisfying the following:*

$$N = N',$$
$$\forall i.S_i = S'_i,$$
$$\forall i.u_i = a_i u'_i + b_i.$$

*Then* $\boldsymbol{\sigma}$ *is an equilibrium of* $G$ *if and only if it is also an equilibrium of* $G'$.

*Proof.* Observe that given the definition of $\succeq_i$, $\boldsymbol{\sigma}$ is an equilibrium of $G$ if and only if:

*This is, in fact, how equilibrium is commonly defined.*

$$\forall i \in N.\forall \sigma_j \in \mathcal{P}(S_i).\boldsymbol{\sigma} \succeq_i \boldsymbol{\sigma}_{-i}(\sigma_j).$$

By Theorem 2.1.22, we know that $u_i$ and $u'_i$ give rise to the same $\succeq_i$. Q.E.D.

### 2.1.2 *Extensive games*

If the first chapter of a textbook introduces a game in normal form, it is a safe bet that the next will deal with the extensive form. The second most famous game representation is not directly related to the technical content of this thesis, and as such this section can be safely skipped without hindering understanding. However, a familiarity with the concept is vital to navigate the literature surrounding Boolean games. As we shall see in Section 2.2.2, the original formulation of a Boolean game was actually as an extensive game, and not in the manner of Definition 2.1.15; moreover the extensive game is often studied with other succinct representations, as we see in Section 2.2.1.1.

Even more so than before, we would like to stress that the presentation that follows is by no means standard – the overwhelming majority of authors do not distinguish between an extensive game and the extensive form representation.[8] We insist on the distinction because while the extensive *game* is quite separate from a strategic game – it is an alternative model of strategic behaviour incorporating notions of time and information – the extensive *form*, a representation, can be used to represent both extensive and strategic games. This will allow us to

8 The only exception that comes to mind is Àlvarez, Gabarró, and Serna [13]. Our definition of the extensive game is modelled on theirs; what we define as the extensive form, or game tree, is perhaps a more common definition of an extensive game in the literature.

avoid statements of the form "extensive game $A$ is equivalent to strategic game $B$" in favour of "the extensive form $T$ represents the strategic game $B$ as well as the extensive game $A$".

**Definition 2.1.24.** A *finite extensive game* consists of the following components:

- A set $N$ of $n$ players, each equipped with a finite, disjoint set of *actions*, $A_i$. We denote the combined set of actions $\mathcal{A} = \biguplus_{i \in N} A_i$.
- A finite set $H$ of *histories*, $H \subseteq \mathcal{A}^*$, that is closed under prefixes. $Z \subseteq H$ is the set of *terminal* histories, i.e. $h \in Z$ only if $ha \notin H$ for any $a \in \mathcal{A}$.
- An *ownership* function mapping non-terminal histories to players, $\Omega : (H \setminus Z) \to N$. Whenever $\Omega(h) = i$ we require that $ha \in H$ only if $a \in A_i$. I.e., only Player $i$ may make a move from $h$.
- A partition of $H \backslash Z$ into some number of information sets denoted $\mathcal{I}_1, \ldots, \mathcal{I}_m$. Every history in an information set is owned by the same player, i.e. there exists an $i$ such that $h \in \mathcal{I}_j$ only if $\Omega(h) = i$, and the same set of actions is available at every history in a given information set: for all $h, h' \in \mathcal{I}_j$, $ha \in H$ if and only if $h'a \in H$. If every $\mathcal{I}_j$ is a singleton we say the game is one of *perfect information.*
- Finally, every player has a utility function mapping terminal histories to the reals, $u_i : Z \to \mathbb{R}$.

A strategy for Player $i$ is a function mapping histories owned by Player $i$ to Player $i$'s actions, $s : \{ h \mid h \in H \text{ and } \Omega(h) = i \} \to A_i$, satisfying two conditions:

1. $s(h) = a$ only if $ha \in H$. That is, extending a history by Player $i$'s choice yields another history.
2. For $h, h' \in \mathcal{I}_j$, $s(h) = s(h')$. If two histories are in the same information set, then Player $i$ must choose the same action at either of them.

The reader will note that a pure strategy profile will uniquely identify some $h \in Z$. ■

**Definition 2.1.25.** The *extensive form*, or *game tree*, of an extensive game is the tree induced by $H$. That is, the smallest tree $T$ containing a branch for every $h \in H$. Every node of $T$ is thus an action, and we can label every internal node $a$ with $\Omega(a)$. Every leaf branch $h$ is labelled with $u_i(h)$ for every player $i$. Finally, we add *indistinguishability* arcs between strategies in the same information set, labelled by the player who cannot distinguish between those strategies. ■

We make the following observation:

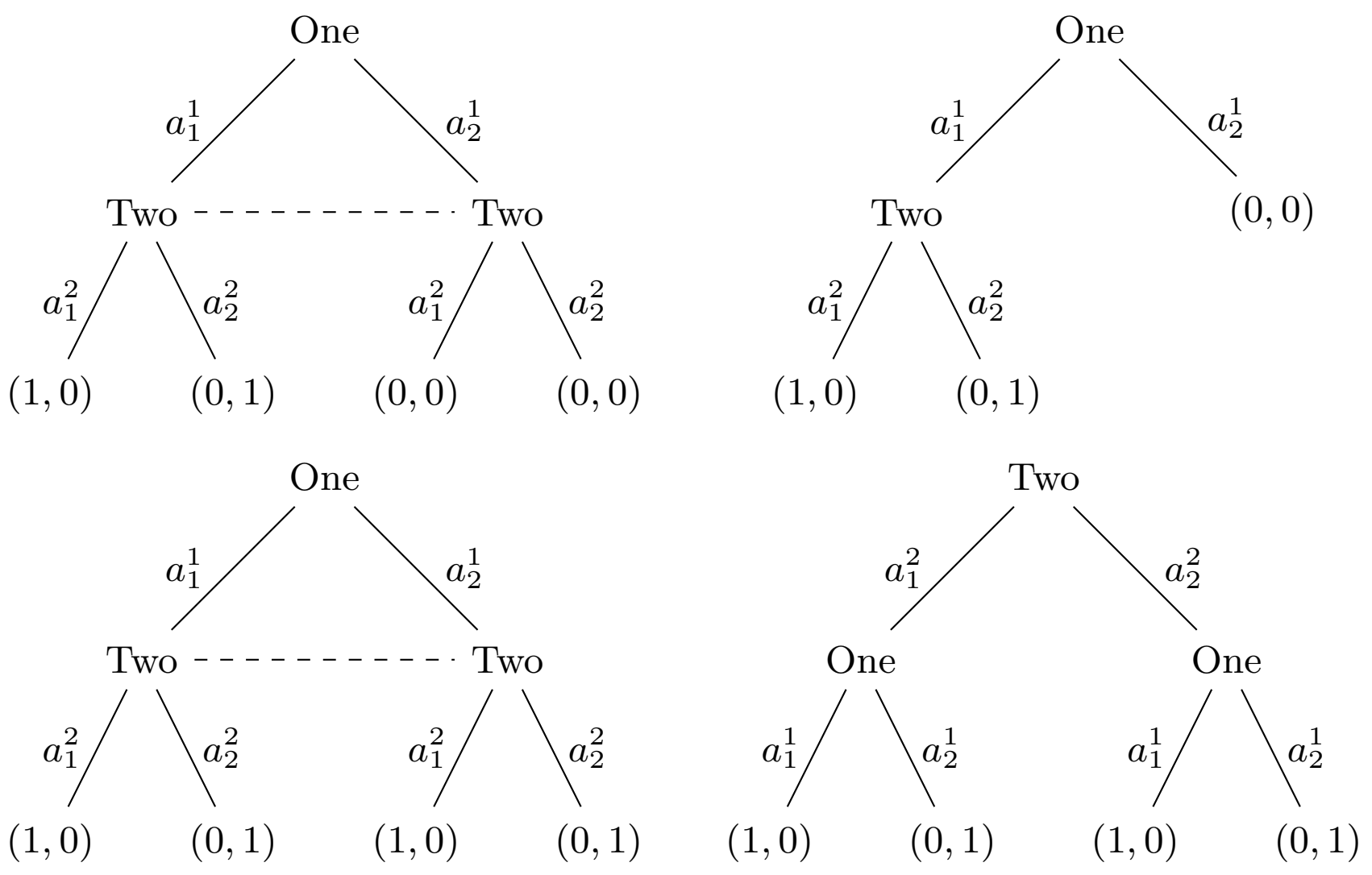

Figure 2.4: The top two games induce the same strategic game, the bottom two do not.

**Fact 2.1.26.** *Every player in a finite extensive game has a finite number of strategies.*

This of course means that should we choose to throw away all the elements specific to an extensive game, such as information sets and order of play, what we are left with is a strategic game.

**Definition 2.1.27.** Let $G = (N, H, \Omega, \{ u_1, \dots, u_n \})$ be an extensive game with strategy sets $S_1, \dots, S_n$ we say that $G$ *induces* a strategic game $G' = (N, \{ S_1, \dots, S_n \}, \{ u'_1, \dots, u'_n \})$ just if $u'_i(\boldsymbol{s}) = u_i(h)$, where $h$ is the outcome of $G$ identified by $\boldsymbol{s}$. ■

**Fact 2.1.28.** *Every strategic game is induced by some extensive game.*

It might be tempting to thus identify an extensive game with the strategic game it induces, but such a course of action has complications – different extensive games may induce the same strategic game, and morally similar extensive games may end up looking rather different if their strategic counterparts are considered; consider Figure 2.4. These examples may seem trivial, but the fact is that the question of what it means for extensive games to be equivalent is a lot more complicated than for strategic games, and this has been often asked in the literature (Thompson, van Benthem, Goranko [14]–[16]).

This issue aside, what Definition 2.1.27 means is that the extensive form, apart from being a representation of an extensive game $G$, can also serve as a representation of a strategic game – the one induced by $G$. Such a representation has the advantage of being succinct – a game tree with $k$ nodes and branching factor two induces a strategic game with $\mathcal{S}$ of size $O(2^k)$ – and as such within the literature the extensive form is at times treated purely as a concise means of representing a game.

However, there is reason to suspect just how effective the extensive form is in that rôle as it is all too easy to construct game trees inducing a vast amount of strategies, the vast majority of which are redundant. Consider a game that begins with Player One choosing to play in the left or right subtree, each containing $m$ binary choices for Player One, and Player One chooses to play left. A strategy for Player One is a choice at *every* node, so even though the game will never progress into the right subtree, he still has to specify a decision at all $m$ points. As a result, every "actual" strategy in the left subtree ends up being represented as $2^m$ strategies in the induced strategic game.

We shall add to these doubts regarding the succinctness of the extensive form as a representation of strategic game in the literature review, particularly in Section 2.2.2 where we demonstrate that there exist games which admit a Boolean representation that is exponentially more succinct than the game tree, but that any game that admits a Boolean representation admits one that is at most polynomial in the size of the game tree.

### 2.1.3 *Algorithmic problems*

In Figure 2.5 we present the problems studied in Chapter 4 and Chapter 5. Other standard problems in algorithmic game theory are given in Figure 2.6. With the exception of DValue, these are not treated in this work.

RationalNash and IrrationalNash invoke a concept we have yet to define, so we give it here:

**Definition 2.1.29.** An equilibrium $\boldsymbol{\sigma}$ is *rational* if every strategy weight used is rational. It is *irrational* if at least one strategy weight is not. ■

Computability issues arise for these problems whenever the number of players is three or greater, as there exist rational-valued three-player games where no rational equilibria exist. This, of course, does not imply that they are uncomputable, but any algorithm dealing with them would need to deal with or represent irrational numbers in some manner, and that is an issue many authors try to avoid. For this reason a weaker notion of equilibrium is prominent in the literature:

**Definition 2.1.30.** Given an $\epsilon > 0$, an *$\epsilon$-equilibrium* is a strategy profile $\boldsymbol{\sigma}$ satisfying the following condition:

$$\forall i \in N,\ \forall \sigma_j \in \mathcal{P}(S_i) : \quad U_i(\boldsymbol{\sigma}) + \epsilon \geq U_i(\boldsymbol{\sigma}_{-i}(\sigma_j)).$$

That is, a player can get at most $\epsilon$ utility by deviating. ■

It is important to note that an $\epsilon$-equilibrium is *not* an approximation of an equilibrium – it is true that every point sufficiently close to an

| ∃GuaranteeNash | |
|---|---|
| Input: | A representation of a game $G$ and a vector of payoffs $\boldsymbol{v}$. |
| Output: | YES if there exists an equilibrium $\boldsymbol{\sigma}$ of $G$ such that $u_i(\boldsymbol{\sigma}) \geq \boldsymbol{v}[i]$ for all $i$, NO otherwise. |

| ∀GuaranteeNash | |
|---|---|
| Input: | A representation of a game $G$ and a vector of payoffs $\boldsymbol{v}$. |
| Output: | YES if for all equilibria $\boldsymbol{\sigma}$ of $G$, $u_i(\boldsymbol{\sigma}) \geq \boldsymbol{v}[i]$ for all $i$, NO otherwise. |

| UniqueNash | |
|---|---|
| Input: | A representation of a game $G$. |
| Output: | YES if $G$ has a unique equilibrium, NO otherwise. |

| ∃NashSat† | |
|---|---|
| Input: | A Boolean game $G$ and a propositional formula $\varphi$. |
| Output: | YES if there exists an equilibrium $\boldsymbol{\sigma}$ of $G$ in which $\varphi$ holds with probability 1, NO otherwise. |

| ∀NashSat† | |
|---|---|
| Input: | A Boolean game $G$ and a propositional formula $\varphi$. |
| Output: | YES if in all equilibria $\boldsymbol{\sigma}$ of $G$, $\varphi$ holds with probability 1, NO otherwise. |

| IsNash | |
|---|---|
| Input: | A representation of a game $G$ and (the non-zero entries of) a strategy profile $\boldsymbol{\sigma}$. |
| Output: | YES if $\boldsymbol{\sigma}$ is an equilibrium of $G$, NO otherwise. |

| RationalNash | |
|---|---|
| Input: | A representation of a game $G$. |
| Output: | YES if $G$ has a rational equilibrium, NO otherwise. |

| IrrationalNash | |
|---|---|
| Input: | A representation of a game $G$. |
| Output: | YES if $G$ has an irrational equilibrium, NO otherwise. |

Figure 2.5: Algorithmic problems studied. Those not marked with a dagger are representation-invariant.

| ∃PNE | |
|---|---|
| Input: | A representation of a game $G$. |
| Output: | YES if $G$ has a pure-strategy equilibrium, NO otherwise. |

| MaxPayoff | |
|---|---|
| Input: | A representation of a game $G$ and some $v \in \mathbb{Q}$. |
| Output: | YES if there exists an equilibrium $\boldsymbol{\sigma}$ of $G$ such that $u_i(\boldsymbol{\sigma}) \geq v$ for all $i$, NO otherwise. |

| SubsetInNash | |
|---|---|
| Input: | A representation of a game $G$ and some $T \subseteq \bigcup S_i$. |
| Output: | YES if there exists an equilibrium $\boldsymbol{\sigma}$ of $G$ such that all strategies in $T$ are included in the support of $\boldsymbol{\sigma}$. |

| NashInSubset | |
|---|---|
| Input: | A representation of a game $G$ and some $T \subseteq \bigcup S_i$. |
| Output: | YES if there exists an equilibrium $\boldsymbol{\sigma}$ of $G$ such that all strategies in the support of $\boldsymbol{\sigma}$ are included in $T$. |

| DValue | |
|---|---|
| Input: | A representation of a two-player, zero-sum game $G$ and some $v \in \mathbb{Q}$. |
| Output: | YES if the value of $G$ is at least $v$, NO otherwise. |

| Value | |
|---|---|
| Input: | A representation of a two-player, zero-sum game $G$. |
| Output: | The value of $G$. |

| Evaluation | |
|---|---|
| Input: | A representation of a game $G$ and (the non-zero entries of) a strategy profile $\boldsymbol{\sigma}$.. |
| Output: | $u_i(\boldsymbol{\sigma})$ for ever player $i$. |

| FindNash | |
|---|---|
| Input: | A representation of a game $G$. |
| Output: | Some equilibrium of $G$. |

Figure 2.6: Other standard problems about games. Aside from DValue, these are not addressed in this thesis.

equilibrium is an $\epsilon$-equilibrium, but an $\epsilon$-equilibrium could be arbitrarily distant from any actual equilibrium in the space of strategy profiles – see Corollary 11 of Etessami and Yannakakis [17].

It is common practice to study algorithmic problems either in the case of two-players and exact equilibria, or multiple players and $\epsilon$-equilibria, as that facilitates tight complexity bounds. RationalNash is an exception to this, as the two-player case trivialises away, and an $\epsilon$ version of the problem is rather meaningless. IrrationalNash, too, does not admit a relaxation to $\epsilon$-equilibria, though the problem remains non-trivial in the case of two-players.[9]

Some of the decision problems in Figure 2.5 closely resemble the ones in Figure 2.6. It is easy to show that the problems we study are reducible from the standard problems in the literature.

**Proposition 2.1.31.** *The following reductions hold for Boolean games:*

- MaxPayoff $\leq_m^{\mathrm{P}}$ $\exists$GuaranteeNash.
- NashInSubset $\leq_m^{\mathrm{P}}$ $\exists$NashSat.
- SubsetInNash $\leq_m^{\mathrm{P}}$ *the complement of* $\forall$NashSat.

*Proof.* Clearly $(G, u) \in$ MaxPayoff if and only if $(G, [u, u])$ is a positive instance of $\exists$GuaranteeNash.

Given $(G, \{s_i : i \in T\})$, where each $s_i$ is a pure strategy (and hence a truth assignment), let $\varphi_i$ be the formula that is true if and only if the truth assignment is $s_i$. I.e., $\varphi_i$ is a conjunction of positive literals for those variables made true by $s_i$ and negative literals for those variables made false. Now observe that:

$$(G, \{s_i : i \in T\}) \in \textsc{NashInSubset} \iff (G, \bigvee_{i \in T} \varphi_i) \in \exists\textsc{NashSat},$$

as if there exists a Nash equilibrium $\boldsymbol{\sigma}$ the support of which is contained in $\{s_i : i \in T\}$, no matter which $s_i$ is realised the corresponding $\varphi_i$ makes the disjunction true.

Likewise:

$$(G, \{s_i : i \in T\}) \in \textsc{SubsetInNash} \iff (G, \bigvee_{i \in T} \neg\varphi_i) \notin \forall\textsc{NashSat},$$

because if there exists an equilibrium realising each $s_i$ with non-zero probability, it cannot be the case that in every equilibrium at least one $\varphi_i$ is false with probability 1. Q.E.D.

The reason why we study $\exists$GuaranteeNash rather than MaxPayoff is to make our results comparable to the work of Schoenebeck and

9 Bilò and Mavronicolas [18] claim that every two-player game has only rational equilibria, and hence do not study the two-player case of IrrationalNash. This is only true of non-degenerate games, however, and determining whether a game is non-degenerate is no easy task (Campbell and Knight [19]).

Vadhan [20]. ∃NashSat and ∀NashSat are preferred to the subset problems as they are more natural in the Boolean games setting – it is hard to envisage why we would care whether certain arbitrary truth assignments are present in an equilibrium, but with ∃NashSat the social planner could specify a policy as a propositional formula and then verify whether it is possible to implement.

## 2.2 THE WORK OF OTHERS

### 2.2.1 *Algorithmic game theory*

The realisation that games are hard is by no means novel; after all, Nash concluded his 1951 paper by stating:

> The complexity of the mathematical work needed for a complete investigation increases rather rapidly, however, with increasing complexity of the game; so that analysis of a game much more complex than the example given here might only be feasible using approximate computational methods.

Initially, the algorithmic study of strategic games was based on the relationship between Nash equilibria and solutions to linear systems in the setting of bimatrix games. It was in this framework that the most famous algorithm for computing the equilibria of a (two-player) game was given (Lemke and Howson [21]), and the complementarity pivot theory developed (Cottle and Dantzig [22]). While our interest in the present work is not so much practical algorithms as complexity bounds, the relationship to linear programming yields a result vitally relevant to us – the existence of rational-valued equilibria in rational-valued two-player games. We shall expound on this in greater detail in Section 3.3. Furthermore, bireducibility between linear programming and Value established that the problem is FP-complete (Dantzig [23]).[10]

Gilboa and Zemel [25] were the first to study decision problems about games. The authors demonstrate that MaxPayoff, SecondNash, NashInSubset, SubsetInNash, MaxSupport and MinSupport are all NP-complete in the setting of two-player games in normal form. These results were strengthened by Conitzer and Sandholm [26], who showed that approximation variants of these problems remain hard, as well as the #P-hardness of counting equilibria. This is the origin of the established wisdom in algorithmic game theory – 'non-trivial questions about Nash equilibria are NP-hard. To simplify presentation, we denote such a "non-trivial question" as the pseudo-problem ?Nash, although it should be noted that there does not exist a unifying result à

10 To be precise, at the time of writing Dantzig only showed that Value is equally hard to linear programming. The fact that linear programming could be done in polynomial time was not known until Khachiyan [24] la Rice's Theorem regarding precisely which problems ?NASH encompasses. NP-hardness of RATIONALNASH and IRRATIONALNASH, which as we have argued are a bit different from ?NASH, was established by Bilò and Mavronicolas [18].

| DECISION | COMPLEXITY | FUNCTION | COMPLEXITY |
|---|---|---|---|
| ∃PNE | $AC^0$ | VALUE | FP-c |
| ?NASH | NP-c | FINDNASH | PPAD-c |
| RATIONALNASH† | NP-h | APPROXNASH† | SQRTSUM-h |
| IRRATIONALNASH† | NP-h | #NASH | #P-h |

Table 2.1: Complexity of algorithmic problems about games in normal form. Results are for the two-player case unless marked by dagger.

The complexity of FINDNASH remained an open problem for a long time before the $\epsilon$ version of the problem was finally shown to be PPAD-complete by Daskalakis, Goldberg, and Papadimitriou [27]. The original proof was for four players, and in fact it was already known that whatever the complexity for the general case, four-player games are no easier (Goldberg and Papadimitriou [28]); the three-player case was proved shortly after (Daskalakis and Papadimitriou, Chen and Deng [29], [30]), and the PPAD-completeness of (exact) FINDNASH for two-player games was established by Chen and Deng [31]. The hardness of the problem persists even if most of the entries in the bimatrix representation of the game are zero (Chen, Deng, and Teng [32]). The problem of approximating an actual equilibrium, as opposed to computing an $\epsilon$-equilibrium, was shown to be hard for SQRTSUM (a class believed to lie between #P and FPSPACE) by Etessami and Yannakakis [17].

One may wonder to what extent these results are pertinent to Boolean games; these are, after all, a very restricted class of games – the payoffs are either 1 or 0. It would be reasonable to assume that some of the difficulty of reasoning about games concerns graded preferences, and it is tempting to ask whether the win-lose case is therefore easier. This does not appear to be the case. Even before the difficulty of FINDNASH was finalised, it was known to be no harder in the general case than for win-lose games (Abbott, Kane, and Valiant [33]), and Chen and Teng [34] extended this result to an approximation scheme. Decision problems were studied by Codenotti and Štefankovič [35] and Bilò and Mavronicolas [36], which affirm the general pattern of things: nontrivial problems are NP-hard. There do, however, remain problems known to be hard for rational-valued games that have not yet been shown to be hard in the win-lose case, notably IRRATIONALNASH.

#### 2.2.1.1 *Succinct representations*

It was not lost on the algorithmic game theory community that the normal form has scaling problems, and a wide variety of succinct representations of games exist in the literature. Loosely speaking these can be split into two general categories: those which attain succinctness by exploiting game-theoretic properties, such as player dependency, symmetry of moves, information structure; and those that attain succinctness by exploiting computational properties – ultimately, whether or not a certain string of 1s and 0s can be written in a more concise form than it is. Boolean games fall into the second camp: some data has a concise logical representation, while an algorithmically random sequence of bits does not; this has little to nothing to do with the actual properties of the game. In this section we will focus mainly on the first category, as we shall deal with Boolean and related game representation shortly.

The most famous of these, perhaps, is the extensive form. In Section 2.1.2 we have discussed how extensive games could be treated either as games in their own right, or simply as compact representations of strategic games. The second stance is taken by Koller, Megiddo, and von Stengel [37], who demonstrate that the complexity of VALUE for games in extensive form is in FP (an alternative proof of completeness can be found in the appendix of Fortnow, Kabanets, Impagliazzo, *et al.* [38]). Note that this is no harder than for games in normal form; we shall see that this is a recurring pattern, suggesting that the representation complexity of games (at least in terms of worst case, pathological behaviour) lies not in their intuitive, game-theoretic properties, but rather in the purely algorithmic structure of the players' preferences.

Graphical games were first introduced by Kearns, Littman, and Singh [39] and achieve succinctness by focusing on the dependencies between players – Player $i$ is said to depend on Player $j$ just if there exists a strategy profile $\boldsymbol{\sigma}$ and a $s_j \in S_j$ such that $u_i(\boldsymbol{\sigma}) \neq u_i(\boldsymbol{\sigma}_{-j}(s_j))$. A graphical game is represented by the dependency graph, which has a vertex for every player and an edge between $i$ and $j$ just if Player $i$ depends on Player $j$. Every vertex is then labelled with a payoff table consisting only of the strategies of adjacent vertices. The size of such a representation is $O(n \max_i |S_i|^d \cdot \max_{j,\boldsymbol{s}} |u_j(\boldsymbol{s})|)$, where $d$ is the maximum degree in the dependency graph. This means that should the dependency graph have bounded degree, the representation size of the graphical game would be polynomial in the number of players.

Here, too, increased succinctness does not seem to lead to increased complexity. The main focus of Kearns, Littman, and Singh, in fact, is on the existence of efficient algorithms for graphical games with simple dependency graphs rather than on the succinctness achieved by such representations. Later, Goldberg and Papadimitriou [28] present a reduction from FINDNASH for graphical games with a degree bound of $d$ to FINDNASH for normal-form games with $d^2$ players.

Papadimitriou [11] presents a generalised scheme for reasoning about succinct representations of games. He defines a *succinct game* to be the triple $(I, T, U)$. $I$ is a set of acceptable inputs to $T$ and $U$, both of which are polynomial-time functions. $T$ on input $w$ returns $(|N|, |S_1|, \ldots, |S_n|)$, i.e. the number of players and strategies in the game. $U$ on input $(w, i, \boldsymbol{s})$ returns $u_i(\boldsymbol{s})$ in the game identified by $w$.

This notion is very similar to Definition 2.1.2. As far as the formalisation is concerned, the only real difference is that $T$ and $U$ are required to be polynomial time, whereas $f$ can be any computable function. In terms of intended purpose the difference is more significant. The succinct games of Papadimitriou [11] are intended to be used as an interface – $T$ and $U$ are supposed to tell us all we need to know to play the game, and how these algorithms are implemented is unimportant. Definition 2.1.2 is more abstract in nature – it tells us that in principle we could recover an explicit representation of the game, but it is up to the representation itself to provide us convenient ways to evaluate the outcome of the game.

*Boolean games are not of polynomial type*

The main focus of the paper is on games of *polynomial type*, that is games for which $|N|$ and $|S_i|$ are polynomial in $|w|$. This encompasses most of the typical succinct representations (symmetric, anonymous, polymatrix, etc). It is shown that for all games of polynomial type the complexity of computing a correlated equilibrium is in FP – the same as for games in normal form. Daskalakis, Fabrikant, and Papadimitriou [40] add to this by showing a reduction to FindNash in a two-player, normal-form game for all games of polynomial type where expected utilities can be calculated by an arithmetic circuit of a certain kind.

While there certainly are problems and representations in these frameworks which have a higher computational complexity than the normal form – Evaluation for games of polynomial type is #P-hard – in general, these representations do not appear to be sufficiently succinct to reflect in higher complexity results. As we shall shortly see, this stands in stark contrast to Boolean games and related representations.

### 2.2.2 *Boolean games*

The concept of a Boolean game was first introduced by Harrenstein, van der Hoek, Meyer, *et al.* [2] (but see also Harrenstein's 2004 thesis [41]). At first glance, these entities do not appear to have much in common with Definition 2.1.15, so we feel we ought to demonstrate how the original definition evolved into the one commonly known today before the reader shuts the cover of this thesis in disgust.

**Definition 2.2.1.** An *old-style Boolean game* is defined inductively from two disjoint sets, $A_1$ and $A_2$, termed Player One's and Player Two's *actions*:

- **1** and **0** are games.

- If $G_1$ and $G_2$ are games, and $\alpha \in A_1$ or $A_2$, then $\alpha(G_1, G_2)$ is a game.

*A given $\alpha$ can appear any number of times in the definition.*

**1** is the game where Player One wins, and **0** where he loses. The game $\alpha(G_1, G_2)$, for $\alpha \in A_i$, allows Player $i$ to choose whether to continue play in $G_1$ or $G_2$. We interpret performing action $\alpha$ as choosing $G_1$, and refraining from $\alpha$ as $G_2$. As a result the set of pure strategies of Player One is $2^{A_1}$, and of Player Two is $2^{A_2}$. Interpreting a win as a utility of 1 and a loss as a utility of 0, the utility function can be recovered by traversing the induced game tree. ∎

This ought to remind the reader of the extensive form – an old-style Boolean game is just a game tree with payoffs in $\{0, 1\}$ and a branching factor of two.[11] This leads to the following observation:

**Fact 2.2.2.** *Old-style Boolean games can be viewed as a representation of extensive games (and, via Definition 2.1.27, of strategic games), by means of interpreting a sentence of the form $\alpha(G_1, G_2)$ as a game tree with children $G_1$ and $G_2$, and the root owned by whichever player owns $\alpha$. This representation is adequate with respect to the class of two-player win-lose games where the cardinality of the strategy sets is a power of two.*

*Moreover, if an extensive game of this form has a game tree with $m$ nodes then it has an old-style Boolean game representation the induced tree of which has at most $m \log m$ nodes.*

*Proof.* Recall that a strategy in a game tree is a choice of action at every node. It follows that if $|S_i| = 2^k$, then it must be the case that the number of actions available to Player $i$ at every node is a power of two. This means that whenever Player $i$ has $2^j$ actions available at an information set we could just as well replace them with a sequence of $j$ binary decisions, i.e. actions in the sense of old-style Boolean games. As the maximum number of actions Player $i$ could have available to him is bounded by $m$, the number of nodes, we would not need to introduce more than $m \log m$ new decisions. Q.E.D.

The old-style Boolean game, however, has a connection with logic that was noted by Harrenstein, van der Hoek, Meyer, *et al.* [2] in Section 7 of their work, where the authors demonstrate the relationship between winning a game and satisfying a propositional formula. This sounds a lot closer to home, and indeed we demonstrate how this allows us to embed an old-style Boolean game into the sort of Boolean game used in this thesis.

11 Although do note that the notion of an action in an old-style Boolean game is not the same thing as an action in an extensive game – a player that has only two choices available to him would have two actions in an extensive game but only one in a Boolean game, because the player has the power to not perform the action in the latter.

**Fact 2.2.3** (Harrenstein, van der Hoek, Meyer, *et al.* [2])**.** *Old-style Boolean games are linear-time embeddable into Boolean games.*

*Proof.* The key is to inductively associate a formula $\mathit{Form}(G)$ with every game, $G$. This formula will be the goal formula of Player One. The set of Player One's variables, $\Phi_1$, is set to $A_1$ and $\Phi_2$ to $A_2$.

For the base case, let $\mathit{Form}(\mathbf{1}) = \top$ and $\mathit{Form}(\mathbf{0}) = \bot$.[12] For $\alpha(G_1, G_2)$, $\mathit{Form}(\alpha(G_1, G_2)) = \mathit{Form}(\alpha \wedge \mathit{Form}(G_1)) \vee \mathit{Form}(\neg\alpha \wedge \mathit{Form}(G_1))$.

The desired Boolean game is:

$$\begin{aligned}\Phi_1 =& A_1,\\ \Phi_2 =& A_2,\\ \gamma_1 =& \mathit{Form}(G),\\ \gamma_2 =& \neg\gamma_1.\end{aligned}$$

Q.E.D.

We do note, however, that whilst the authors certainly understand the connection between Boolean games and propositional formulae, they stop short of identifying the one with the other – the game is *related* to the formula, but it is never *identified* with it.[13] The step of dropping the game tree and keeping only the formula was taken only by Bonzon, Lagasquie-Schiex, Lang, *et al.* [42], from whence Definition 2.1.15 derives. After the redefinition by Bonzon, Lagasquie-Schiex, Lang, *et al.*, old-style Boolean games disappear from the literature.

#### 2.2.2.1 *Algorithmic properties*

The original paper on Boolean games dealt with their algebraic properties as well as the notion of relativised validity – for Player One to have a winning strategy means that there is an assignment to $\Phi_1$ relative to which $\gamma_1$ is a tautology, which is clearly a property of logical as well as game-theoretic interest. Neither of these topics are the focus of this thesis, though we do briefly note a connection between the algebra of Harrenstein, van der Hoek, Meyer, *et al.* and a family of auxiliary games we use in our proofs in Section 3.2.2.

Of more direct interest to us is the first algorithmic study of Boolean games – that of Dunne and van der Hoek [43]. As algorithms need clearly defined input, this work raised the question of how a Boolean game is to be represented – the original definition, after all, is inductive. In order to study the algorithmic properties of Boolean games, therefore, Dunne and van der Hoek made a similar distinction between

12 For Harrenstein, van der Hoek, Meyer, *et al.*, any propositional tautology or contradiction would suffice. We, on the other hand, require that it be a propositional constant, otherwise old-style **1** would have one strategy for Player One, while the Boolean game image two. These two strategies are in some sense immaterial, $p \vee \neg p$ is true regardless of what Player One does with $p$, but our notion of game equivalence requires the number of strategies to be the same.

13 If this sounds like a philosophical distinction, then that is because Harrenstein is a philosopher.

a game and a representation to that of ours, except the underlying object for which they sought representations was not the finite strategic game but the old-style Boolean game itself.

Dunne and van der Hoek suggest two ways to represent such a game. Firstly, the "extensive form" is the canonical one suggested by the statement of the definition, i.e. a sequence of symbols of such as $\alpha(\beta(\mathbf{1},\mathbf{0}),\theta(\mathbf{0},\mathbf{0}))$ is interpreted as the implied game tree. Second, the "concise form" is a sentence in the algebraic language defined by Harrenstein, van der Hoek, Meyer, *et al.* [2] which uniquely identifies some such game. This concise form is morally equivalent to the representation we use; while on the surface the concise form is built up using symbols like $+$ and $\cdot$, whereas ours is in the language of propositional logic, the isomorphism between the algebra of old-style Boolean games and the Lindenbaum algebra of propositional logic proved by Harrenstein, van der Hoek, Meyer, *et al.* [2] means that we can replace such a sentence with an equivalent propositional formula in linear time. As a result, complexity results for concise form old-style Boolean games apply equally to the framework adopted here. On the other hand, we can construct simple examples of games whose (smallest) extensive form is exponentially larger than the (smallest) concise form.

**Fact 2.2.4.** *Let $G$ be a zero-sum Boolean game with $\Phi_1 = \{\, p_1, \dots, p_k \,\}$, $\Phi_2 = \{\, q_1, \dots, q_k \,\}$, and $\gamma_1 = \bigwedge(p_i \leftrightarrow q_i)$. The concise form of this game is of order $|\gamma_1|$, that is $O(k)$, but the smallest game tree has at least $2^k$ nodes.*

*Proof.* We argue that the tree must have a leaf for every satisfying assignment of $\gamma_1$, of which there are $2^k$.

Observe that in the game described Player One has a *winning response* to every strategy of Player Two. That is, if we fix any pure strategy $\nu^2$ for Player Two, there exists a $\nu^1$ for Player One such that Player One wins in the profile $(\nu^1, \nu^2)$. Moreover, such a $\nu^1$ is unique with respect to a given $\nu^2$, and $\nu^2$ is unique with respect to $\nu^1$. In other words, if we fix $\nu^2$ then there exists exactly one $\nu'$ for which $\nu', \nu^2 \vDash \gamma_1$, and likewise if we fix $\nu^1$ then there exists exactly one $\nu''$ for which $\nu^1, \nu'' \vDash \gamma_1$.

Now consider a game tree with less than $2^k$ leaves and two profiles, $(\nu_1^1, \nu_1^2)$ and $(\nu_2^1, \nu_2^2)$, such that Player One wins in both. There are $2^k$ such profiles, so we can assume without loss of generality that the plays induced by $(\nu_1^1, \nu_1^2)$ and $(\nu_2^1, \nu_2^2)$ terminate in the same leaf node, $l$. Recall that a play is a path from the root to a leaf, and such a path is unique in a tree, ergo $(\nu_1^1, \nu_1^2)$ and $(\nu_2^1, \nu_2^2)$ must induce the same play.

This gives us our contradiction – a strategy in a game tree is a choice at every node, but decisions made at nodes that do not lie along the play path cannot affect the outcome of the play. This means $(\nu_1^1, \nu_2^2)$ and $(\nu_2^1, \nu_1^2)$ must also terminate at $l$, as they only differ in choices made on nodes not lying on the path. This is impossible, because Player One ought to lose at these profiles. Q.E.D.

The core results of Dunne and van der Hoek concern four problems: WS, WE, EQUIV and GE. WS concerns the existence of a winning strategy for a given player, which we define below:

**Definition 2.2.5.** A *winning strategy* for Player $i$ in a win-lose game $G$ is a strategy, $s_j$, such that $u_i(\boldsymbol{\sigma}_{-i}(s_j)) = 1$ for all $\boldsymbol{\sigma}$. I.e., $s_j$ guarantees Player $i$ a payoff of 1 regardless of the play of the other players. ■

**Fact 2.2.6.** *For a two-player zero-sum win-lose game, the following three propositions are equivalent:*

1. *Some player has a winning strategy.*
2. *The value of the game is 1 or 0.*
3. *The game has a pure-strategy equilibrium.*

The notion of a winning strategy is defined for all win-lose games, but is most natural for the two-player zero-sum case; in more complicated games winning strategies are rare and not particularly interesting (if a player can win irrespective of the choices of the other $n-1$ players, what purpose do those players serve?). Note also that there is no generality lost in assuming the winning strategy is pure: if a mixed strategy is winning, so is every strategy in its support.

WE takes as input a Boolean game without a partition of actions[14], and asks whether there is *some* partition of variables that would provide Player $i$ with a winning strategy. EQUIV and GE deal with two notions of game equivalence.

For concise form representations: WS is $\Sigma_2^p$-complete, WE is NP-complete, EQUIV is coNP-complete and GE is co$D_2^p$-complete[15]. For extensive form representations, all four problems are in P – this ought not to surprise us, as we have already established that the concise form is more succinct.

As we have already mentioned, Bonzon, Lagasquie-Schiex, Lang, *et al.* [42] introduce the newer definition of Boolean games, generalising the concept to variable-sum games with any number of players. In addition, they show that ∃PNE, and deciding whether a strategy is dominated, are both $\Sigma_2^p$-complete. The authors also observe that if the goal formulae of the players, $\gamma_i$, are restricted to easier fragments of propositional logic then the complexity of ∃PNE can decrease. This line of inquiry was later taken up by Dunne and Wooldridge [44]. We summarise their results in Table 2.2.

To address the difficulty of ∃PNE, Dunne and Wooldridge [44] consider an alternative solution concept: a $k$-bounded equilibrium, from which no player has incentive to deviate without switching the value

14 Such objects are called "Boolean forms" in the early literature on Boolean games.
15 A rather arcane complexity class. A language in $D_2^p$ is the intersection of $L_1 \in \Sigma_2^p$ and $L_2 \in \Pi_2^p$. Not to be confused with $\Delta_2^p$, which is contained within $\Sigma_2^p \cap \Pi_2^p$, i.e. $L \in \Delta_2^p$ only if $L$ is in *both* $\Sigma_2^p$ and $\Pi_2^p$.

| Constraints on goal | Complexity |
|---|---|
| None | $\Sigma_2^p$-complete |
| DNF | NP-complete |
| Satisfiability in P | NP$^\dagger$ |
| Horn-renameable DNF | P |
| Affine form | P |
| 2CNF | P |
| Monotone CNF | P |
| $k$-clause CNF | P$^\dagger$ |

Table 2.2: Complexity of ∃PNE under various restrictions on goal formulae. Results marked by dagger are due to Dunne and Wooldridge [44], unmarked to Bonzon, Lagasquie-Schiex, Lang, *et al.* [42].

assigned to at most $k$ variables. Such a solution concept clearly only makes sense in the setting of Boolean games, and represents the idea that players themselves may be operating under computational constraints, and thus only capable of considering a small number of local deviations. For a constant $k$, such an equilibrium can be found in polynomial time, and in the case where every player's goal is in CNF with at most $k$ clauses, then every pure equilibrium is $k$-bounded; hence the result in Table 2.2.

An alternative approach to tackling the complexity of Boolean games, presented by Bonzon, Lagasquie-Schiex, and Lang [45], is based on the dependency structure of the game. Player dependency is used in the same sense as earlier, Player $i$ is dependent on Player $j$ if the utility Player $i$ receives from a profile depends on the strategy chosen by Player $j$, and there is reason to expect that a game with a sparse network of dependencies ought to be tractable – we would have little trouble predicting the outcome of two hundred rock-paper-scissors matches taking place in tandem, even though the normal form of such a game is enormous. We have seen that in algorithmic game theory proper this is the motivation behind graphical games; the work of Bonzon, Lagasquie-Schiex, and Lang is hence the Boolean game equivalent.

In the framework of Boolean games dependency between players has a logical characterisation: Player $i$ is *not* dependent on Player $j$ if and only if $\gamma_i$ is logically equivalent to a $\gamma_i'$ that does not contain any variables from $\Phi_j$. Granted, this is hardly a trivial problem to solve, but seeing how the normal-form alternative would be to enumerate each and every single possible strategy profile, this could be seen as an improvement. Bonzon, Lagasquie-Schiex, and Lang note two other possible notions of dependency; the first, peculiar to Boolean games, is to simply consider which variables appear in $\gamma_i$ and assume that every one of those is a dependency. Clearly, this is sufficient but not necessary for dependency proper, and in the worst case this could be arbitrarily bad – one could construct a Boolean game where every formula contains every variable but depends on none of them. However, this notion has the advantage of being trivial to compute, and in the case where the formulae can be assumed to be close to minimal this is a very cheap way to construct a dependency graph. The second, stronger, notion is based on the observation that dependency is not the same thing as need. Under the definition given Player $i$ with $\gamma_i = p \vee q$ and $\Phi_i = \{ p \}$ is dependent on the player controlling $q$, but clearly, Player $i$ does not *need* any other player – he can satisfy $\gamma_i$ unilaterally. This stronger notion, of need rather than dependency, has not, to date, been studied in the literature.

Given a dependency graph, Bonzon, Lagasquie-Schiex, and Lang [45] demonstrate that if the irreflexive fragment of the graph is acyclic, then the game is guaranteed a pure-strategy equilibrium; moreover, by splitting the graph into cliques the number of operations necessary to compute such an equilibrium can be reduced.

#### 2.2.2.2 *Applications*

Boolean games have proven to be popular in the multiagent community, perhaps because whilst being concise they nevertheless retain an intuition of play – it is a lot easier to get an intuitive grasp of what is going in inside a Boolean game than a Turing machine that spits out utilities. However, as the basic Boolean model is perhaps too simple to model many phenomena of interests, research has also focused around various variations on the theme.

This body of work can be roughly divided into three areas: the study of the cooperation and coalition formation in Boolean games, incentive engineering for Boolean games, and the use of Boolean games as models of concrete phenomena of interest. We will address the three in turn.

The first paper to study cooperation in a Boolean games setting is Dunne, van der Hoek, Kraus, *et al.* [46]. The authors study the ability of players to form coalitions to better achieve their goals, the primary solution concept studied being the core:

**Definition 2.2.7.** We say that a strategy profile $\boldsymbol{\nu}$ is *blocked* by a coalition $C \subseteq N$ through a profile $\boldsymbol{\nu'}$ if the players in $C$ strictly prefer $\boldsymbol{\nu'}$ to $\boldsymbol{\nu}$, and can deviate from $\boldsymbol{\nu}$ to $\boldsymbol{\nu'}$ unilaterally. That is:

1. $\boldsymbol{\nu'} \succ_i \boldsymbol{\nu}$ for all $i \in C$.
2. For all $p \notin \bigcup_{i \in C} \Phi_i$, $\boldsymbol{\nu'} \models p$ iff $\boldsymbol{\nu} \models p$.

The set of strategy profiles not blocked by any coalition is called the *core*. ■

The authors are also the first to introduce a *cost function* into a Boolean game, which proved to be a fertile innovation – virtually every paper that followed included costs in one way or another. A cost function is some $c : \Phi \to \mathbb{R}_{\geq 0}$, with $c(p)$ indicating the cost Player $i$ incurs for setting $p$ to *true*. Player $i$ wants to satisfy $\gamma_i$, yet at the same time wishes to minimise the costs he incurs in doing so, which allows to represent more fine-grained distinctions between profiles than the purely dichotomous preferences of Boolean games proper. However, a notion of dichotomy is preserved in that *every* profile that satisfies $\gamma_i$ is preserved to those that do not, regardless of cost.[16]

The complexity of deciding whether a given truth assignment belongs to the core is shown to be coNP-complete, deciding whether the core is non-empty $\Sigma_2^p$-complete and checking whether a given truth assignment characterises the core $\Pi_2^p$-complete. Further work on cooperative Boolean games was done by Bonzon, Lagasquie-Schiex, and Lang [47] in their study of effectivity functions, and Sauro, Torre, and Villata [48] extended the study of dependency in Boolean games to the cooperative setting.

Ben-Naim and Lorini [49] study the power of players in this setting – that is, which agents are essential to a coalition achieving its goal, and which are essentially freeloaders.

A variation of cooperative Boolean games was considered by Popovici and Dobre [50], where every player receives a certain amount of currency upon satisfying his formula. In this sense this framework is closer to the cooperative games of classical game theory, which are characterised by a function $f : 2^N \to \mathbb{R}$ which determines the payoff every coalition can achieve by itself. However, the similarities end here. Popovici and Dobre do not have much to say about the standard question of cooperative game theory proper: how should coalition $C$ distribute $f(C)$ among its members to prevent deviation? Instead, their notion of the core is based on a coalition trying to maximise its total value.

An alternative model of coalition formation is a hedonic game (Drèze and Greenberg [51]). In a hedonic game each player has his own preference order on possible coalitions, i.e. subsets of $N$, and seeks to find himself in the best coalition possible. Solution concepts are partitions of $N$ into coalitions satisfying various notions of stability. As every player needs to rank all $2^N$ coalitions, succicnt preference representation plays a key role in the study of hedonic games; thus Aziz, Harrenstein, Lang, *et al.* [52] introduce the Boolean hedonic game, where every player is equipped with a propositional formula telling him which coalitions are acceptable and which are not.

On the incentive engineering front, the framework typically consists of a Boolean game (with costs) and a social planner wishing to imple-

16 It is not necessarily clear how to extend this notion to mixed profiles, but every paper discussed in this section deals with pure profiles only.

ment some social goal, which is generally a special formula $\Gamma$ that the planner wishes to see in some or every equilibrium. In the setting of taxation schemes (Endriss, Kraus, Lang, *et al.*, Wooldridge, Wooldridge, Endriss, Kraus, *et al.* [53]–[55]) the planner does this by choosing a *taxation policy*, $\tau : \Phi \to \mathbb{R}^+$, which adds to the players' costs. Deciding whether it is possible to implement $\Gamma$ is shown to be $\Sigma_2^p$-complete. The problem of actually computing a taxation scheme is studied experimentally by Levit, Grinshpoun, Meisels, *et al.* [56].

Turrini [57] considers transfers rather than taxation schemes – the difference being that these are not decided on centrally, but rather every player individually decides how much of his utility he would like to share with the other players. Deciding whether there exists a transfer scheme guaranteeing a pure-strategy equilibrium is also $\Sigma_2^p$-complete.

In light of these two directions, Harrenstein, Turrini, and Wooldridge [58] introduce the notion of a *hard equilibrium*, an equilibrium that cannot be eliminated by transfers or taxes, as opposed to a *soft equilibrium* that can. The authors provide a purely logical characterisation of the notion.

Kraus and Wooldridge [59] consider a Boolean game where the variable sets $\Phi_i$ do not totally partition $\Phi$ – certain variables remain unallocated. It is the social planner's task to distribute those variables in a manner that ensures $\Gamma$ is satisfied in some or every equilibrium. Both problems are $\Sigma_2^p$-complete, and an optimisation version lies somewhere within $\text{FP}^{\Sigma_p^2}$.

The framework of Grant, Kraus, Wooldridge, *et al.* [60] also involves unallocated variables, but these are *environment* variables, $\Phi_E$, which are not handed out to players but instead are preset to some value. This value is not known by the agents; every agent only has a belief function, $b_i : \Phi_E \to \mathbb{B}$, and plays the game according to those. The planner may attempt to influence players' behaviour by informing them, privately or publicly, of the true state of some or all of $\Phi_E$. In doing so the planner hopes to satisfy $\Gamma$, and the complexity results are similar: deciding whether the desired announcement exists is $\Sigma_2^p$-complete, computing a minimal one is within $\text{FP}^{\Sigma_p^2}$.

An alternative take on cost is presented by Harrenstein, Turrini, and Wooldridge [61] in the form of electric Boolean games, where costs do not affect preferences, however the total costs incurred by a player must be less than his *endowment*. The metaphor used is that players are constrained by a finite battery charge – they do not care how much energy they expend to meet their goal, but once they run out they can take no further actions. The planner may redistribute endowments with the hope of inducing the players achieve a certain equilibrium. This problem lies somewhere within $\Delta_2^p$.

The use of Boolean games to actually model scenarios in the multiagent setting has been sparse, but varied. Lukasiewicz and Ragone [62] consider a variation of Boolean games, where the players have goals in description logic, to model agents in the semantic web, Galafassi and Bazzan [63] model a Boolean network as a Boolean game to implement an evolutionary algorithm for the network and Levit, Grinshpoun, and Meisels [64] consider an iterative[17] version of Boolean games with costs and environmental variables to model the charging of autonomous electric vehicles.

The work of Bradfield, Gutierrez, and Wooldridge [66] does not easily fit into any of the three categories we have identified. The authors impose an event structure on a Boolean game, i.e. a directed acyclic graph on $\Phi$ that determines the information a player can take into account before assigning a truth value to a variable. As such, a strategy for Player $i$ is not a choice of a *true* or *false* value for each $p \in \Phi_i$, but a choice of a function $f_p : 2^{\Phi_p} \to \{ \mathit{true}, \mathit{false} \}$, where $\Phi_p$ is the set of predecessors of $p$ in the graph. Note that this framework differs from an extensive form in two ways: first, the graph is not necessarily a tree; and second, a player's choice can only depend on the immediate predecessors, rather than the entire history to date.

### 2.2.3 *Related games and extensions*

#### 2.2.3.1 *Boolean games with richer preferences*

An often voiced concern about Boolean games is that, being only able to represent win-lose games, they inevitably trivialise strategic interaction. Now, while we have seen that win-lose games are certainly far from trivial computationally, it is nevertheless true that their breadth of application is somewhat limited – after all, even Prisoner's Dilemma has no Boolean representation. The cost function introduced above is a popular solution, but there have been numerous attempts to introduce graded preferences at a more fundamental level.

The first such attempt was that of Bonzon, Lagasquie-Schiex, and Lang [67]. They consider two extensions. The first is a Boolean game with prioritised goals, or a PG-Boolean game. A PG-Boolean game equips each player, $i$, with $k$ sets of propositional formulae: $\Gamma_1^i, \ldots, \Gamma_k^i$. The formulae in $\Gamma_j^i$ are perceived to have a higher priority than the formulae in $\Gamma_{j'}^i$, for $j < j'$, and Player $i$ would rather satisfy the formulae with higher priority. PG-Boolean games can be seen as a generalisation of distributed evaluation games (Harrenstein [41], Section 8.4) to more than one priority level. The second approach equips each player with a CP-net over the variable set $\Phi$, yielding a CP-Boolean game. Players choose a truth assignment to the variables under their control, and then consult their CP-net to decide how they feel about the outcome. Bonzon, Lagasquie-Schiex, and Lang observe that a certain acyclicity condition on the CP-nets involved can guarantee the exis-

*There are several ways to formalise this as a preference order, which we will not dwell on.*

*Without a way to compare mixed profiles, in these games a mixed equilibrium is ill-defined.*

17 Not to be confused with the iterat*ed* games of Gutierrez, Harrenstein, and Wooldridge [65]

tence of a unique (pure-strategy) equilibrium. Following this, Bonzon, Devred, and Lagasquie-Schiex [68] present algorithms that transform an argumentation framework into a CP-Boolean game.

The reader will note that these two extensions involve purely ordinal, and not necessarily complete, preferences. As such, even though PG- and CP-Boolean games contain finite strategy sets, they are not representations of finite strategic games in the sense of Definition 2.1.1.

Mavronicolas, Monien, and Wagner [69] present a cardinal, rather than ordinal, extension in the form of weighted Boolean formula games. Players assign truth values to a set of variables as with Boolean games, but in lieu of a goal formula each player possesses a set, $\Gamma_i$, of formula/integer pairs. The utility Player $i$ derives from $\boldsymbol{\nu}$ is:

*$(\boldsymbol{\nu} \models \gamma)$ is 1 if $\boldsymbol{\nu} \models \gamma$ and 0 otherwise.*

$$\sum_{(\gamma,\alpha)\in\Gamma_i} \alpha(\boldsymbol{\nu} \models \gamma),$$

Mavronicolas, Monien, and Wagner consider the complexities of pure-strategy and payoff-dominant equilibria, and identify a subclass of weighted Boolean formula games that bears a relation to the congestion games of Rosenthal [70].

Bilò [71] independently developed a framework virtually identical to the weighted Boolean formula game; the only difference in a satisfiability game of Bilò is that a player's strategy set is $S_i \subseteq 2^{\Phi_i}$, i.e. some subset of possible assignments, while in weighted Boolean formula games $S_i = 2^{\Phi_i}$. Bilò shows that in a restricted fragment (the goal formulae are all CNF or all DNF and contain at most two literals, every player controls only one variable or is only allowed to set one variable to *true*) it is possible to compute a mixed equilibrium in polynomial time. We note that this is the only result on mixed strategies in a setting closely related to Boolean games outside of the present work.

#### 2.2.3.2 *Games with other logics*

Boolean games can be seen to belong to a wider family of game representations in which players possess goals in some logical language $\mathcal{L}$, and strategy profiles induce models of $\mathcal{L}$ in some natural way. An alternative way to represent richer player preferences, therefore, could be to choose a richer logic.

Iterated Boolean games (Gutierrez, Harrenstein, and Wooldridge [65]) involve players possessing preferences in LTL. The game is played by requiring each player to submit a response function. As in the case of Boolean games, Player $i$ controls the variables in $\Phi_i$; however, he does not choose a single truth assignment, but a truth assignment for every possible time period, and the choice of assignment for period $k$ can depend on everything that happened in periods $j < k$. By playing these response functions against each other, an $\omega$-word over $2^{\Phi}$ is generated, which is a model of LTL. It should be noted, that while it is natural to interpret an iterated Boolean game as playing a Boolean game over

*Linear temporal logic, introduced by Pnueli [72].*

and over again, it is certainly *not* equivalent to a repeated game in the standard sense of economics.

An important thing to note about iterated Boolean games is that, unlike almost every other game representation we have considered, the games these represent are *not finite*. Player $i$'s set of strategies contains every function $f : (2^{\Phi})^* \to 2^{\Phi_i}$, which is not even denumerable. The authors restrict themselves to functions representable by a finite-state transducer, and a consequence of Pnueli and Rosner [73] is that if a player cannot achieve his goal with such a strategy then he cannot achieve it at all, but that nevertheless leaves an infinite number, albeit of finitely representable, strategies to choose from.

An iterated version of electric Boolean games is also studied by Harrenstein, Turrini, and Wooldridge [61].

*As the logic is finitely-valued, this is indeed a finite strategic game.*

The Łukasiewicz games of Marchioni and Wooldridge [74] are based on a finitely-valued Łukasiewicz logic: the players assign a value in $\{ 1/k, \ldots, k/k \}$ to every variable under their control, and reap utility equal to the degree to which their goal formula is satisfied. This of course allows players to possess far more intricate preferences than in the case of Boolean games, as every continuous piecewise linear function $f : [0,1]^k \to [0,1]$ with integral coefficients can be expressed as a formula of Łukasiewicz logic.[18] The main result of the paper is that given a Łukasiewicz game, it is possible to construct a formula of Łukasiewicz logic that is satisfiable if and only if the game has a pure-strategy equilibrium. In contrast, in the case of Boolean games that formula would have to be in ∃∀QBF.

*For quantified Boolean formulae see, e.g., Section 17.2 of Papadimitriou [76]. For multimodal* S5*, van Ditmarsch, van der Hoek, and Kooi [77].*

Ågotnes, Harrenstein, van der Hoek, *et al.* [78] introduce epistemic Boolean games, where players have goal formulae in the language of $S5_n$. Given the great number of modal logics out there, and the interest modal logicians have shown towards aspect of game theory, it may be surprising that such an extension took as long as it did to appear. The reason is that providing modal logics with a Boolean games treatment can be tricky – a model of modal logic generally involves a Kripke structure, and it is by no means clear how the players' choice of strategies are to represent one in a graceful way. In the case of iterated Boolean games, as we have saw, the problem was dealt with by the fact that LTL has a canonical model: the straight line. In the case of epistemic Boolean games, Ågotnes, Harrenstein, van der Hoek, *et al.* take a different approach. Each player is provided with a "visibility set" $\Theta_i \subseteq \mathcal{L}(\Phi)$. These together define a Kripke structure with worlds labelled by $2^{\Phi}$ and Player $i$'s indistinguishability relation being $a \sim_i b$ if and only if for all $\varphi \in \Theta_i$, $a \vDash \varphi \iff b \vDash \varphi$. In other words the Kripke structure is implicitly fixed in advance, and the players only select which world to evaluate their formulae at.

18 Known as McNaughton's theorem in the logical community (McNaughton [75]). Not to be confused with the better known McNaughton's theorem of automata theory.

Epistemic Boolean games, being representations of finite win-lose games, are no more expressive than Boolean games, but the authors demonstrate that, unless P=PSPACE, they can be exponentially more succinct. En revanche, the complexity seems to be higher: determining whether a (pure) strategy profile is an equilibrium, or even evaluating a player's goal formula at it, is PSPACE-complete.

In a separate paper, Ågotnes, Harrenstein, Van Der Hoek, *et al.* [79] also consider an epistemic strengthening of an equilibrium in a Boolean game extended with visibility sets: a strategy profile that not only is in equilibrium, but every player can verify that it is.

Finally, the epistemic logic literature contains games similar in spirit to Boolean games such as question-answer games (Ågotnes, van Benthem, van Ditmarsch, *et al.* [80]), knowledge games (van Ditmarsch [81]), and public announcement games (Ågotnes and van Ditmarsch [82], [83]). Like in Boolean games, players are trying to satisfy a goal formula, this time of an epistemic language. A crucial difference is that the players do not, by their choice of strategies, specify a model of the logic; rather, they modify an existing model that is supplied with the game description. For instance, a public announcement game starts at a given vertex in a Kripke structure in which every player then makes a public announcement, with the hope that the upgraded model will be one that will satisfy his goal formula.

#### 2.2.3.3 *Games with circuits and machines*

We address a class of representations known as circuit games in some detail, as the work in this area is the most relevant to the technical part of this thesis – more so than the literature on Boolean games proper. There are three papers on the subject: Feigenbaum, Koller, and Shor [84], Fortnow, Kabanets, Impagliazzo, *et al.* [38] and Schoenebeck and Vadhan [20]. In light of the pedantic stance we have hitherto taken, we feel obliged to point out that all three works give a different definition of the games studied, and Feigenbaum, Koller, and Shor do not use the word "circuit" anywhere in their paper. We begin by giving the definition of Schoenebeck and Vadhan, before discussing how they differ.

**Definition 2.2.8.** A *circuit game*[19] is a representation of a finite strategic game consisting of $n$ disjoint sets of input gates, $I_{i\in N}$, and $n$ Boolean circuits, $C_{i\in N}$, with input gates $I = \biguplus_{i\in N} I_i$ and any number of output gates.

A pure strategy for Player $i$ is an initial configuration of his input gates, $\iota_i : I_i \to \{0, 1\}$, and his payoff from profile $\boldsymbol{\iota} = (\iota_1, \dots, \iota_n)$ is $C_i(\boldsymbol{\iota})$. I.e., the rational number encoded by the output gates of $C_i$ with initial configuration $\boldsymbol{\iota}$. ∎

19 One is tempted to call this a *Boolean circuit game*, but Schoenebeck and Vadhan reserve that term for circuit games with $|I_i| = 1$. We shall avoid the term to avoid confusion.

**Fact 2.2.9.** *Circuit games are adequate with respect to the class of rational-valued games where the cardinality of every player's strategy set is a power of two.*

The definition of Fortnow, Kabanets, Impagliazzo, *et al.* [38] deals with two-player, zero-sum games only, which it specifies via a payoff matrix (see Definition 2.1.13). The succinct representation of the payoff matrix $M$ is a circuit $C$ that on input $i, j$ outputs $m_{i,j}$, the $(i, j)$-entry of $M$. Player One's strategy is then to choose the $i$ value, and Player Two the $j$ value. Strictly speaking, this definition differs from a circuit game in that there is only one circuit and there does not seem to be a restriction on the number of strategies players possess (thought Fortnow, Kabanets, Impagliazzo, *et al.* do not specify what happens if the players choose $i, j$ for which $M_{i,j}$ is undefined), but it is morally equivalent; certainly, there exists a polynomial-time biembedding betwixt the fragment of circuit games á la Fortnow, Kabanets, Impagliazzo, *et al.* with strategy set cardinalities being powers of two, and the two-player zero-sum fragment of the circuit games of Schoenebeck and Vadhan.

The games of Feigenbaum, Koller, and Shor [84], however, are sufficiently different to merit a definition of their own:

**Definition 2.2.10.** A *referee game* is a representation of two-player zero-sum games specified by a polynomial-time deterministic machine augmented with two communication channels, known as the referee, and an initial input string $w$.

The referee can use the channels to privately send messages to Player One and Two (and if the referee does not want the message to be private, it can of course send a copy to the other channel). The recipient can then send a response. The players are arbitrarily powerful, and can perform any computation they need before responding, but because of the constraints of the referee the response needs to be polynomial in the size of the query. Upon terminating, the referee outputs the payoff to Player One. ■

Upon seeing this definition one may be tempted to dismiss the difference on grounds of universal computability; what, after all, is the difference between a circuit and a Turing machine? However, there is a crucial reason why we cannot admit this perspective: the strategy sets are different.

A polynomial-time referee can ask at most a polynomial number of questions, however the response to each question can modify its subsequent behaviour, and thus the game as a whole is better thought as a nondeterministic process (with player choice the only source of nondeterminism), and this process generates a computation tree polynomial in depth but with exponentially many nodes. A pure strategy for Player One is a choice for each of his nodes, which means it is exponential in the size of the referee, and hence there is a double-exponential number

*This tree is in fact the extensive form of the game.*

of strategies. Given the framework we have adopted, we cannot equate referee games with circuit games in any meaningful way.

As such, we cannot freely import arbitrary results of Feigenbaum, Koller, and Shor [84] into the setting of circuit games, but that need not vex us as we are only interested in one – the complexity of DValue.

As it happens, the complexity of DValue is no harder for referee games than circuit games, as the reader can observe by noting that in the game constructed in the proof of Theorem 4.6 of Feigenbaum, Koller, and Shor [84], and restated in the prelude to Theorem 6.1.1 in the current work, the referee asks a grand total of two questions; such a tree is of constant size. This is ultimately due to a result we have already mentioned: despite the increased succinctness of extensive form representations, the complexity of Value for the extensive and normal form is the same. The proof of Feigenbaum, Koller, and Shor [84] will serve as the basis of our argument in Chapter 6.

Other results relevant to us from the work on circuit games is that ∃GuaranteeNash is NEXP-complete for two-player games, and IsNash $\mathrm{coNP}^{\#\mathrm{P}}$-hard if the number of players is unbounded (Schoenebeck and Vadhan [20]). We derive the same results for Boolean games in Chapter 5

The Turing machine games of Àlvarez, Gabarró, and Serna are similar in flavour. The authors study concise representations of both extensive (Àlvarez, Gabarró, and Serna [13]) and strategic games (Àlvarez, Gabarró, and Serna [10]).

Àlvarez, Gabarró, and Serna consider three representations of a strategic game. The *explicit form* involves listing the strategies of the players' and the graphs of their utility functions – the normal form in our parlance. The *implicit form* involves providing a polynomial-time Turing machine and a length constraint on the strategy of each player. Thus a player prescribed to choose a strategy of length of $k$ can choose any string in $\{0,1\}^k$ as his strategy. After the players write their choices on the tape, the machine is activated and computes the payoff vector. We could thus identify the implicit form with a circuit game modulo universal computability, or if we would rather tread the cautious party line we have adopted then we would merely say that there is a natural polynomial-time biembedding between them. The *general form* is intermediate between the two: utility functions are represented implicitly via a Turing machine, but strategy sets are listed explicitly. Players choose an appropriate primitive symbol from their strategy set to inscribe on the input tape, and then the game proceeds as before. As could be expected, the complexity of the general form lies between the two – the problem of ∃PNE is NP-complete, as opposed to $\Sigma_p^2$-complete for the implicit form. For the explicit form, being the normal form, the problem is in P (even in $\mathrm{AC}^0$).

The convenience of an infinite tape, however, allows Àlvarez, Gabarró, and Serna to define game representations that have no parallels in the circuit or Boolean game setting: the *uniform* explicit and general forms. Here a single machine defines a family of games rather than a single instance. For example, a uniform general form of rock-paper-scissors is a Turing machine that on input $\texttt{rock}, \texttt{paper}$ will output $(0, 1)$ and on $\texttt{rock}, \texttt{paper}, \texttt{rock}$ the output will be $(0, 1, 0)$.[20]

The general form of Àlvarez, Gabarró, and Serna [10] appears earlier in Gottlob, Greco, and Scarcello [85]; the authors consider games in the normal form, as a graphical game, and in general form, which they define as the setting where the strategy sets of all players are represented explicitly, but the utility and neighbourhood functions are given as a polynomial time Turing machine. The difference from the general form of Àlvarez, Gabarró, and Serna [10] lies in that to evaluate the utility of Player $i$ in a Gottlob, Greco, and Scarcello game one would first evaluate $i$'s neighbourhood, then write down the strategies of all of $i$'s neighbours on the tape of the utility machine, whereas in a game of Àlvarez, Gabarró, and Serna one would write down the strategies of all players on the utility machine's tape immediately. In the worst case, the players' neighbourhoods are the entire set, and the two game representations become equivalent; indeed, Gottlob, Greco, and Scarcello [85] demonstrate that the complexity of ∃PNE is NP-complete. However, restrictions on the dependency structures of players reduces the complexity of some problems.

20 There are, of course, many ways to extend the scoring system of rock-paper-scissors to more than two players. As exciting as such an exercise may be, here we take the simplistic approach of requiring a player to win every match.

# Part II

# COMPLEXITY RESULTS

# 3

# AUXILIARY RESULTS

*In all the parts of human knowledge, whether terminating in science merely speculative, or operating upon life, private or civil, are admitted some fundamental principles, or common axioms, which, being generally received, are little doubted, and, being little doubted, have been rarely proved.*

— Samuel Johnson, Taxation no Tyranny

We have two goals in this chapter. In Section 3.1 we introduce shorthand expressions for formulae of propositional logic that will occur frequently in the proofs to come. In Section 3.4 we introduce a family of gadget games which have a predetermined value; these will be used to simulate finer-grained preferences than the Boolean game model allows. Section 3.3, included for completeness, reproduces some standard results about the relation of game theory to linear programming, and why two-player games have rational equilibria.

## 3.1 ENCODING INTEGERS IN LOGIC

The goal of this section is to introduce shorthand and notation that will allow us to discuss integers and arithmetic in the language of propositional logic. This will ultimately allow us to interpret a player's strategy as a choice of integers, and a goal formula as an arithmetical constraint.

Our constructions will often involve sequences of propositional variables, so we will use the notation $\overline{p_m}$ to mean $p_1, \ldots, p_m$. Sequences differ from sets in that they have an order, which allows us to interpret a truth assignment to a sequence as an integer. We do this in the standard way – a truth assignment that sets $p_i$ to *true* defines a binary integer where the $i$th most significant bit is 1.

**Definition 3.1.1.** Let $\overline{p_m}$ be a sequence of $m$ propositional variables, and $\nu$ a truth assignment to $\overline{p_m}$. We use $[\![\overline{p_m}]\!]_\nu$ to denote the numeric value associated with $\nu$ via its assignment to $\overline{p_m}$. That is:

*$(\nu \vDash p_i)$ is 1 if $\nu \vDash p_i$, 0 otherwise.*

$$[\![\overline{p_m}]\!]_\nu = \sum_{i=1}^{m} 2^{m-i}(\nu \vDash p_i).$$

If $\nu$ is clear from context, we just write $[\![\overline{p_m}]\!]$. ■

This is quite sufficient to interpret any pure strategy in a Boolean game as a choice of integer. To test whether such an integer satisfies an arithmetical constraint we need to introduce formulae that verify operations of (finite) arithmetic. Throughout the following arguments the reader should bear in mind that what we are trying to establish is that these formulae can be constructed in time polynomial in the bit length of the integers under consideration – that such formulae exist at all follows trivially from the expressive completeness of propositional logic.

Arithmetic is a consequence of equality and successor, so it befits us to start with these.

**Lemma 3.1.2.** *Let* $\boldsymbol{Equal}(\overline{p_m};\overline{q_m})$ *denote a term that is true if and only if* $[\![\overline{p_m}]\!] = [\![\overline{q_m}]\!]$*. I.e.,* $\nu \vDash \boldsymbol{Equal}(\overline{p_m};\overline{q_m})$ *if and only if* $[\![\overline{p_m}]\!]_\nu = [\![\overline{q_m}]\!]_\nu$*.*

*We can construct* $\boldsymbol{Equal}(\overline{p_m};\overline{q_m})$ *in time linear in* $m$*.*

*Proof.* Two binary integers are equal if and only if they are bitwise equal. This gives us the following:

$$\boldsymbol{Equal}(\overline{p_m};\overline{q_m}) = \bigwedge_{1\leq i\leq m} (p_i \leftrightarrow q_i).$$ Q.E.D.

**Lemma 3.1.3.** *Let* $\boldsymbol{Succ}(\overline{p_m};\overline{q_m})$ *denote a term that is true if and only if* $[\![\overline{p_m}]\!] + 1 = [\![\overline{q_m}]\!]$*.*

*We can construct* $\boldsymbol{Succ}(\overline{p_m};\overline{q_m})$ *in time quadratic in* $m$*.*

*Proof.* $2^m - 1$ does not have an $m$ bit successor, so we first require that $\overline{p_m}$ is not all 1s:

$$\boldsymbol{Succ}(\overline{p_m};\overline{q_m}) = \neg(\bigwedge_{1\leq i\leq m} p_i) \wedge Succ'.$$

$Succ'$ will do the heavy lifting.

For intuition, recall that adding a 1 to a binary integer $x$ can have one of two outcomes: either $x$ ends in a 0 and this 0 is replaced with a 1, or $x$ ends with a 1 in which case that 1 is replaced with a 0 and a carry operation occurs – which is, of course, simply the act of adding a 1 to a rightshift of $x$.

In other words, we have a recursive operation that replaces the rightmost consecutive block of 1s in $x$ with 0s, the subsequent 0 with a 1, and leaves the rest of the integer unchanged.

This gives us $Succ'$:

$$\begin{aligned}
&\Big(\neg p_m \rightarrow (q_m \wedge \boldsymbol{Equal}(\overline{p_{m-1}};\overline{q_{m-1}})\Big)\\
\wedge&\Big((p_m \wedge \neg p_{m-1}) \rightarrow \big(\neg q_m \wedge q_{m-1} \wedge \boldsymbol{Equal}(\overline{p_{m-2}};\overline{q_{m-2}})\big)\Big)\\
&\vdots\\
\wedge&\Big((\neg p_1 \wedge \bigwedge_{1<i\leq m} p_i) \rightarrow (q_1 \wedge \bigwedge_{1<i\leq m} \neg q_i)\Big).
\end{aligned}$$

For the reader who abhors ellipses, we can restate this in one line:

$$Succ' = \bigwedge_{i \le m} \Big( (\neg p_i \wedge \bigwedge_{j > i} p_j) \to (\boldsymbol{Equal}(\overline{p_{i-1}}; \overline{q_{i-1}}) \wedge q_i \wedge \bigwedge_{j > i} \neg q_j) \Big).$$

This is quadratic in the number of variables, giving us the desired result. Q.E.D.

The reader will note that in the proof above the knowledge that we can represent $\boldsymbol{Equal}(\overline{p_m}; \overline{q_m})$ concisely simplified our argument for $\boldsymbol{Succ}(\overline{p_m}; \overline{q_m})$. This is of course perfectly natural mathematical thinking – the more we can say the more we can define, and the more we define the more we can say. Following this line we can now use $\boldsymbol{Succ}(\overline{p_m}; \overline{q_m})$ to deal with the less-than order.

**Lemma 3.1.4.** *Let* $\boldsymbol{Less}(\overline{p_m}; \overline{q_m})$ *denote a term that is true under* $\nu$ *if and only if* $[\![\overline{p_m}]\!]_\nu < [\![\overline{q_m}]\!]_\nu$.

*We can construct* $\boldsymbol{Less}(\overline{p_m}; \overline{q_m})$ *in time cubic in* $m$.

*Proof.* Let $a[i]$ be the $i$th most significant bit of $a$.

Intuitively, if $[\![\overline{p_m}]\!] < [\![\overline{q_m}]\!]$ for two big-endian binary digits then if we read the bits of the two from left to right we will see a 0 in $[\![\overline{p_m}]\!]$ before we see one in $[\![\overline{q_m}]\!]$. That is, there exists a $k$ such that:

$$[\![\overline{p_k}]\!] = [\![\overline{q_k}]\!],$$
$$\nu \not\models p_{k+1}, \quad \nu \models q_{k+1}.$$

It follows that the first bit where the two integers differ is a 1 for $[\![\overline{q_m}]\!]$ and a 0 for $[\![\overline{p_m}]\!]$. This is clearly both necessary and sufficient.

Since there are only $m$ possible values of $k$, this can be replaced by a cubic size formula that looks as follows:

*Recall that the size of* $\boldsymbol{Succ}(\overline{p_m}; \overline{q_m})$ *is quadratic.*

$$\bigvee_{0 \le k < m} \boldsymbol{Succ}(\overline{p_k}; \overline{q_k}).$$

Q.E.D.

**Corollary 3.1.5.** *Let* $\boldsymbol{LessEq}(\overline{p_m}; \overline{q_m})$ *denote a term that is true under* $\nu$ *if and only if* $[\![\overline{p_m}]\!]_\nu \le [\![\overline{q_m}]\!]_\nu$.

*We can construct* $\boldsymbol{LessEq}(\overline{p_m}; \overline{q_m})$ *in time cubic in* $m$.

**Lemma 3.1.6.** *Let* $\boldsymbol{Add}(\overline{p_m}; \overline{q_m}; \overline{r_m})$ *denote a term that is true if and only if* $[\![\overline{p_m}]\!] + [\![\overline{q_m}]\!] = [\![\overline{r_m}]\!]$.

$\boldsymbol{Add}(\overline{p_m}; \overline{q_m}; \overline{r_m})$ *can be replaced by a formula of propositional logic cubic in* $m$.

*Proof.* Binary addition is easy: 0 is identity, and $1 + 1$ is 0. In other words if the summands are the same the answer is 0, if they are different the answer is 1. Thus if no carry occurred at position $i + 1$, then $r_i \leftrightarrow (\neg(p_i \wedge q_i))$. If a carry has occurred, we add another 1 to this result, which gives us $r_i \leftrightarrow (p_i \wedge q_i)$.

The trick is determining whether a carry has occurred – we have no memory mechanism to store this information. However, our constraint is polynomial time, and this is sufficiently generous to allow us to explicitly verify whether a carry bit has reached $i$ by checking every $j > i$. The necessary subformula is the following:

$$Carry(i) = \bigvee_{j \geq i+1} \Big( (p_j \wedge q_j) \wedge \bigwedge_{i+1 \leq k < j} (p_k \vee q_k) \Big).$$

In words, $Carry(i)$ holds just if there exists a $j > i$ such that the bits at the $j$th position are both 1, and every single position between $j$ and $i$ has at least one of the bits being 1.

Our final formula is:

$$\begin{aligned} Add = \bigwedge_{i \leq m} \Big[ & \Big( \neg Carry(i) \rightarrow (r_i \leftrightarrow (\neg(p_i \leftrightarrow q_i))) \Big) \\ & \wedge \Big( Carry(i) \rightarrow (r_i \leftrightarrow (p_i \leftrightarrow q_i)) \Big) \Big] \\ & \wedge \neg \Big( (p_1 \wedge q_1) \vee (Carry(1) \wedge (p_1 \vee q_1)) \Big). \end{aligned}$$

The last line states that integer overflow has not occurred. This is added because we are interested in exact addition, not addition modulo $2^m$.

This is cubic in $m$, proving the lemma. Q.E.D.

**Lemma 3.1.7.** *Let* $\boldsymbol{Sub}(\overline{p_m}; \overline{q_m}; \overline{r_m})$ *denote a term that is true if and only if* $[\![\overline{p_m}]\!] - [\![\overline{q_m}]\!] = [\![\overline{r_m}]\!]$.

$\boldsymbol{Sub}(\overline{p_m}; \overline{q_m}; \overline{r_m})$ *can be replaced by a formula of propositional logic cubic in* $m$.

*Proof.* We begin by verifying that the difference is not negative. This is rather simple:

$$NoNegative = \boldsymbol{LessEq}(\overline{q_m}; \overline{p_m}).$$

Borrow bits behave a little differently from carry bits because subtraction does not commute. A borrow case takes place at position $i$ just if at some $j > i$ the minuend contains a 0 and the subtrahend a 1, and in every succeeding position the subtrahend bit is at least as big as the minuend.

$$Borrow(i) = \bigvee_{j \geq i+1} \Big( (\neg p_j \wedge q_j) \wedge \bigwedge_{i+1 \leq k < j} (p_k \rightarrow q_k) \Big).$$

In the absence of a borrow bit, we require $r_i \leftrightarrow (\neg(p_i \leftrightarrow q_i))$. In the presence of a borrow bit, $r_i \leftrightarrow (p_i \leftrightarrow q_i)$. This gives us the result:

$$\begin{aligned} Sub' = \bigwedge_{i \leq m} \Big[ & \Big( \neg Borrow(i) \rightarrow (r_i \leftrightarrow (p_i \leftrightarrow q_i)) \Big) \\ & \wedge \Big( Borrow(i) \rightarrow (r_i \leftrightarrow (\neg(p_i \leftrightarrow q_i))) \Big) \Big]. \end{aligned}$$

$\boldsymbol{Sub}(\overline{p_m}; \overline{q_m}; \overline{r_m}) = NoNegative \wedge Sub'$. Q.E.D.

At times in lieu of testing $[\![\overline{p_m}]\!]$ for one of these three relations against $[\![\overline{q_m}]\!]$, we may wish to test, for example, whether $[\![\overline{p_m}]\!] < 3$. In other words, we would like to have access to constants as well as variables, and that is the purpose of the following definition:

**Definition 3.1.8.** Let $j$ be a binary integer of length $m$. We use $\ulcorner \boldsymbol{j} \urcorner$ to denote a sequence of the logical constants $\top$ and $\bot$, the $i$th element of which is $\top$ if and only if the $i$th most significant bit of $j$ is 1.

The intended use of $\ulcorner \boldsymbol{j} \urcorner$ is as an argument to the parametrised formulae defined so far. For example, we interpret $\boldsymbol{Less}(\overline{p_m}; \ulcorner \boldsymbol{j} \urcorner)$ in the sense of Lemma 3.1.4, except every instance of $q_i$ is replaced with $\top$ or $\bot$ depending on whether the $i$th bit of $j$ is a 1 or 0. ■

Finally, we want some formulae to assert that a certain number of their arguments are true.

**Lemma 3.1.9.** *Let $\boldsymbol{OneOf}(\overline{p_m})$ and $\boldsymbol{NoneOf}(\overline{p_m})$ denote terms that are true just if exactly 1 and 0 respectively of the $p_i$ variables are true. These terms can be constructed in time polynomial (quadratic and linear respectively) in $m$.*

## 3.2 GAMES OF ANY VALUE

We have seen that Theorem 2.1.12 endows two-player zero-sum games with something resembling a unique solution. From a purely extensional point of view, this solution could well be the only thing we want from a game; knowing that Player One, playing rationally, can expect to get a utility of $v$ from a game is already useful enough without caring about the details of how that utility is obtained.

We would like to be able to introduce such games into our proofs without explicitly constructing them each time, so it pays to construct them here. But first, we ought to convince ourselves that such games do, in fact, exist.

In a win-lose game, $v \in [0,1]_{\mathbb{Q}}$ is the probability of Player One winning the game. Given $v = {}^a\!/\!_b$ we can uniformly construct a family of games in which Player One chooses an interval of length $a$ over $[0, b-1]_{\mathbb{N}}$, and Player Two chooses a single integer from the same range. If $a$ and $b$ are coprime, i.e. the fraction ${}^a\!/\!_b$ is maximally reduced, we can show that this game has a unique equilibrium – the profile where Player One randomises equally over every interval and Player Two over every integer – the value of which is clearly ${}^a\!/\!_b = v$. The vocabulary we have developed in Section 3.1 will allow us to express this as a Boolean game.

*Such an interval is allowed to loop around the end points $b-1$ and 0.*

**Lemma 3.2.1.** *For $v \in [0,1]_{\mathbb{Q}}$ we can construct, in time polynomial in $|v|$, a two-player, zero-sum Boolean game with value $v$ and a unique equilibrium.*

*We shall refer to this game as $\mathfrak{G}(v)$.*

*Proof.* Let $v = {}^a\!/_b$. We insist that $a, b$ are coprime.

Consider the game where Player One selects two numbers $c_1, c_2 \in [0, b-1]_{\mathbb{N}}$ with the property that $c_2 - c_1 \equiv a - 1 \mod b$ (the start and end points of an interval of length $a$). Player Two selects $d \in [0, b-1]_{\mathbb{N}}$. The game is won by Player One if $c_2 \geq d \geq c_1$ (Player Two's choice is in a non-looping interval), $d \geq c_1 > c_2$ (Player Two's choice is in a looping interval left of $b-1$) or $c_1 > c_2 \geq d$ (Player Two's choice is in a looping interval right of 0).

To give this game a Boolean rendition we need the following variables:

$$\Phi_1 = \{p_1, \ldots, p_m, q_1, \ldots, q_m, s_1, \ldots, s_m, t_1, \ldots, t_m\},$$
$$\Phi_2 = \{r_1, \ldots, r_m\}.$$

The interpretation is that $[\![\overline{p_m}]\!] = c_1$, $[\![\overline{q_m}]\!] = c_2$ and $[\![\overline{r_m}]\!] = d$. The $s$ and $t$ variables come in to play if Player One wishes to play a looping interval, in which case $[\![\overline{s_m}]\!]$ is the distance between 0 and $c_2$, while $[\![\overline{t_m}]\!]$ is the distance between $c_1$ and $b-1$. These variables are added to give us a way to check that if Player One plays a looping interval, its length is still $a$.

Player One's goal formula is the following:

$$\begin{aligned}
\gamma_1 = &\big(\boldsymbol{Sub}(\overline{q_m}, \overline{p_m}, \ulcorner \boldsymbol{a-1} \urcorner) \wedge \boldsymbol{LessEq}(\overline{q_m}, \ulcorner \boldsymbol{b-1} \urcorner) \\
&\quad \wedge \boldsymbol{LessEq}(\overline{r_m}, \overline{q_m}) \wedge \boldsymbol{LessEq}(\overline{p_m}, \overline{r_m}) \wedge \boldsymbol{Equal}(\overline{s_m}, \ulcorner \boldsymbol{0} \urcorner) \\
&\quad \wedge \boldsymbol{Equal}(\overline{t_m}, \ulcorner \boldsymbol{0} \urcorner)\big) \\
&\vee \Big(\boldsymbol{Add}(\overline{s_m}, \overline{t_m}, \ulcorner \boldsymbol{a-1} \urcorner) \wedge \boldsymbol{Sub}(\overline{q_m}, \ulcorner \boldsymbol{0} \urcorner, \overline{s_m}) \\
&\quad \wedge \boldsymbol{Sub}(\ulcorner \boldsymbol{b-1} \urcorner, \overline{p_m}, \overline{t_m}) \wedge \big(\boldsymbol{LessEq}(\overline{r_m}, \overline{q_m}) \vee \boldsymbol{LessEq}(\overline{p_m}, \overline{r_m})\big)\Big) \\
&\vee \boldsymbol{Less}(\ulcorner \boldsymbol{b-1} \urcorner, \overline{r_m}).
\end{aligned}$$

The first disjunct handles the case where the interval is non-looping: $[\![\overline{q_m}]\!] - [\![\overline{p_m}]\!] = a - 1$ (because the interval is closed) and Player Two's choice falls between them. We require that $[\![\overline{s_m}]\!] = [\![\overline{t_m}]\!] = 0$ to ensure the equilibrium is unique, as otherwise Player One would have been able to assign any value to those variables without affecting his chance of winning. The second disjunct handles the looping case; here $[\![\overline{s_m}]\!]$ and $[\![\overline{t_m}]\!]$ store the length of the interval on either side of zero. The last disjunct of serves to award the game to One should Two name a $d$ outside of $[0, b-1]_{\mathbb{N}}$.

*The reader will note that $[\![\overline{s_m}]\!]$ is superfluous. It is included solely for symmetry with $[\![\overline{t_m}]\!]$.*

As the game is zero-sum, $\gamma_2$ is simply $\neg\gamma_1$.

Now let us verify that the equilibrium is unique. We say the *cover weight* of a number $i$ is the sum of the weights Player One attaches to every interval containing $i$. In other words, the cover weight of $i$ is the probability that Player Two will lose if she plays $i$ with probability 1. If $w_j$ is the weight Player One attaches to the interval starting at $j$ then the cover weight of $i$, $c_i$, can be expressed as follows:

$$c_i = \sum_{j \equiv i-a+1 \mod b}^{i} w_j.$$

That is, the first $w_j$ in the sum is the interval with $i$ at its rightmost point, and hence its leftmost point is at $i - (a - 1) \mod b$.

Now observe that a best response for Player Two to any strategy of Player One is to play $\arg\min_i c_i$ with probability 1, giving Player One a utility of $\min_i c_i$; Player One's maxmin strategy is thus to maximise the smallest of the cover weights. Furthermore, we can see that the sum of all the cover weights is invariant and equal to $a$ (as the sum of all $w_i$ is equal to 1 and each $w_i$ appears in $a$ distinct cover weight terms). It follows that the only way to maximise the smallest cover weight is to make all $c_i$ equal.

We will demonstrate that $c_i = c_{i+1}$ implies that $w_{i-a+1 \mod b} = w_{i+1}$. As $a$ is a generating element in the additive group of $b$ elements, this will establish that all $w_i$ must be the same.

*This is why we assume $a$ and $b$ are coprime.*

The argument itself is trivial. Suppose $c_i = c_{i+1}$. If we expand this we have:

$$w_{i-a+1} + w_{i-a+2} + \cdots + w_i = w_{i-a+2} + w_{i-a+3} + \cdots + w_{i+1}.$$

By subtracting $(w_{i-a+2} + \cdots + w_i)$ from both sides we have that $w_{i-a+1} = w_{i+1}$.

We have thus shown that:

a) In every equilibrium Player One must make all the cover weights equal.

b) The only way to make all the cover weights equal is to play every interval with equal weight.

This settles the argument for Player One. We turn to Player Two.

This time we define the *breadth* of the interval starting at $i$, $b_i$, to be the sum of the weights, $w_j$, that Player Two attaches to every number in that interval:

$$b_i = \sum_{j=i}^{i+a-1 \mod b} w_j.$$

By an argument parallel to the one above, we see that Player Two seeks to choose a strategy that will make all $b_i$ equal. Assuming $b_i = b_{i+1}$, this leads us to:

$$w_i + w_{i+1} + \cdots + w_{i+a-1} = w_{i+1} + w_{i+2} + \cdots + w_{i+a}.$$

This allows us to invoke the same argument.

In sum, we have shown that every equilibrium involves all $c_i$ and $b_i$ being equal, and the only way to achieve this is by randomising equally over all intervals and integers. This equilibrium is thus unique. Q.E.D.

$\mathfrak{G}(v)$ will serve us as a gadget game to get around the win-lose restriction of Boolean games; it may well be the case that in any pure-strategy profile Player One either wins or loses, but Player One is an expected-utility maximiser – he is indifferent between a utility of $1/2$ and a probability $1/2$ of getting a utility of 1. This will allow us to simulate an outcome of $1/2$ for Player One by giving him the opportunity to play $\mathfrak{G}(1/2)$ instead.

### 3.2.1 *Subgames*

If $\mathfrak{G}(v)$ is to be used as a gadget in a larger construction, it behoves us to establish a means to refer to the constituents of $\mathfrak{G}(v)$.

**Definition 3.2.2.** Let $G = (\Phi_1, \ldots, \Phi_n, \gamma_1, \ldots, \gamma_n)$ be a Boolean game. We use $\gamma_i(G)$ to refer to $\gamma_i$ and $\mathrm{var}_i(G)$ to refer to $\Phi_i$. This notation will be useful in the presence of multiple games, as it will allow us to distinguish between $\gamma_i(G)$ and $\gamma_i(G')$.

If $G$ is a game or $\varphi$ is a formula, we will also write $\mathrm{var}(G)$ and $\mathrm{var}(\varphi)$ to refer to the set of all variables present in the game or formula respectively. ■

*Note that Player Three in the overarching game controls the variables that would have belonged to Player Two in a gadget game. Our notation allows this.*

Let $\mathfrak{G}(1/2)$ and $\mathfrak{G}'(1/2)$ be two disjoint games with value $1/2$. Consider the following Boolean "interpretation" of Battle of the Sexes:

$$\begin{aligned}
\Phi_1 &= \{\, Box_1 \,\} \cup \mathrm{var}_1(\mathfrak{G}(1/2)),\\
\Phi_2 &= \{\, Box_2 \,\} \cup \mathrm{var}_1(\mathfrak{G}'(1/2)),\\
\Phi_3 &= \mathrm{var}_2(\mathfrak{G}(1/2)) \cup \mathrm{var}_2(\mathfrak{G}'(1/2)),\\
\gamma_1 &= (\neg Box_1 \wedge \neg Box_2) \vee (Box_1 \wedge Box_2 \wedge \gamma_1(\mathfrak{G}(1/2))),\\
\gamma_2 &= (Box_1 \wedge Box_2) \vee (\neg Box_1 \wedge \neg Box_2 \wedge \gamma_1(\mathfrak{G}'(1/2))),\\
\gamma_3 &= \gamma_2(\mathfrak{G}(1/2)) \wedge \gamma_2(\mathfrak{G}'(1/2)).
\end{aligned}$$

*It is possible to eliminate Player Three by absorbing his rôle by the other two players, but that would make the current example less transparent.*

Intuitively, Player One wins if both players go to see ballet, or if both see boxing and Player One beats Player Three in Matching Pennies. Player Two wins if both see boxing, or both see ballet and she wins the Matching Pennies side game. Player Three is interested in winning both games of Matching Pennies. This ought to have the effect of giving us an equilibrium where both players go to boxing, and play the equilibrium play in the side games. Player One would derive a utility of $1/2$. How could we prove this?

One is tempted to argue as follows:

> Let $\boldsymbol{\sigma}$ be a strategy profile that involves both players playing $Box_i$ with probability 1 and playing the equilibrium strategies against Player Three over the gadget games. This profile is in equilibrium: no player has incentive to deviate in the gadget games, as they are in equilibrium. Player One has no incentive to play $Box_1$ with a probability lower than 1 as any profile obtained by switching his assignment to *false* would yield him a utility of 0, but he is currently getting $1/2$ – the probability of winning $\mathfrak{G}(1/2)$, and hence satisfying $(Box_1 \wedge Box_2 \wedge \gamma_1(\mathfrak{G}(1/2)))$. Mutatis mutandis, for Player Two.

Unfortunately, there are a number of problems with this approach.

First, the strategy profile itself is ill-defined; it makes no sense to speak of "playing $Box_i$ with probability 1 and playing the equilibrium strategies against Player Three over the gadget games", as a player's strategy is not divided into components. A pure strategy is an assignment to *every* variable in $\Phi_i$, and a mixed strategy is a distribution over such *complete* assignments. If we are to use such locutions, we need to be clear about what they mean:

**Definition 3.2.3.** Suppose Player $i$ controls $\Phi_i = A \uplus B$. When we say that Player $i$ plays $A$ with distribution $P$ and $B$ with distribution $Q$, what we mean is that Player $i$ plays the mixed strategy that, for any $\nu_A : A \to \mathbb{B}$ and $\nu_B : B \to \mathbb{B}$, realises $\nu_A \nu_B$ with probability $P(\nu_A)Q(\nu_B)$. ■

In other words, Player $i$ is assigning truth values to the variables in $A$ independently of his assignments to the variables in $B$. In our example, Player One's assignment to $Box_1$ is independent from his assignment to $\text{var}_1(\mathfrak{G}(^1/_2))$.

Next, while our game intuitively contains $\mathfrak{G}(^1/_2)$ and $\mathfrak{G}'(^1/_2)$ as subgames, formally this is meaningless – the notion of a subgame is a property of extensive, rather than strategic games. Hence, we need to be clear as to what we mean when we say that the gadget games are in equilibrium.

**Definition 3.2.4.** Let $\boldsymbol{\sigma}$ be a strategy profile of $G$, and $G'$ a game with $\text{var}(G') \subseteq \text{var}(G)$. We use $\boldsymbol{\sigma}|G'$ to denote the probability distribution over $\text{var}(G')$ that is obtained by marginalising over the variables $\text{var}(G) \setminus \text{var}(G')$ from $\boldsymbol{\sigma}$.

In other words, the probability of $\nu' : \text{var}(G') \to \mathbb{B}$ being realised in $\boldsymbol{\sigma}|G'$ is the sum of the probabilities of every $\nu : \text{var}(G) \to \mathbb{B}$ that agrees with $\nu'$ being realised by $\boldsymbol{\sigma}$.

We say that $G'$ is in equilibrium in $G$ just if $\boldsymbol{\sigma}|G'$, interpreted as a profile of $G'$, is an equilibrium of $G'$. ■

Bear in mind that what follows from the definition is that if $G'$ is in equilibrium in $\boldsymbol{\sigma}$ and Player $i$ controls the variables $\text{var}_j(G')$ (and nothing more in $\text{var}(G')$), then Player $i$ has no strategy available to him that will increase the probability of $\gamma_j(G')$ being realised. No more, and no less. As such, in and of itself, the fact that $G'$ is in equilibrium tells us nothing about the state of the larger game – it is entirely possible for the subgames to be in equilibrium but not $G$ itself. We must first consider how the variables of $G'$ are partitioned among the players and how the goal formulae of $G'$ relate to the goal formulae of $G$.

In the case of our example, the fact that the subgames are in equilibrium tells us only that Player One has no strategy available to him that will increase the probability of him satisfying his right disjunct – that would require a profitable deviation in $\mathfrak{G}(^1/_2)$. That happens to be sufficient, as it is clear that he has no deviation to satisfy the left disjunct either.

Finally, it is not at all clear what we mean when we speak of Player One deviating by changing the assignment to $Box_1$ to *true*.

**Definition 3.2.5.** Suppose Player $i$ controls $\Phi_i = A \uplus B$. Consider a strategy profile $\boldsymbol{\sigma}$ where Player $i$ plays $A$ with distribution $P$ and $B$ with distribution $Q$. When we say Player $i$ deviates by changing his assignment to $A$ to $P'$, we mean he deviates to playing $A$ with distribution $P'$ and $B$ with distribution $Q$. ■

Such a definition may seem redundant, but it serves to highlight an important issue – the locution "Player One deviates by changing the assignment to $Box_1$ to *true*" is only meaningful in the case where his play of $Box_1$ is independent with respect to his other variables. This is the case in our example, but such phrases should not be thrown around without verifying that they actually mean something in the situation at hand.

These definitions will allow us to write proofs in terms of subgames and components of strategies, which will simplify the arguments greatly. However, we must not fall into the trap of assuming that if Player $i$, controlling $\Phi_i = A \uplus B$, has no profitable deviation over $A$ and no profitable deviation over $B$ available to him, then he has no profitable deviation at all. All that means is that Player $i$ has no profitable deviation that involves choosing assignments to $A$ and $B$ *independently*; it could well be the case that there does exist a deviation, but within it the choice of assignment to $A$ would be determined by the choice of assignment to $B$. In the example given this did not apply – Player One had to choose one of two mutually exclusive disjuncts to satisfy, and the left disjunct only depends on $\{ Box_1, Box_2 \}$, so we have no need to consider what Player One's strategy with respect to $\mathrm{var}_1(\mathfrak{G}(1/2))$ is. However, this is something that is worth bearing in mind in the proofs that follow.

### 3.2.2 *Scholium: algebra of games*

The original Boolean game framework (Harrenstein, van der Hoek, Meyer, *et al.*, Harrenstein [2], [41]) included the base cases of **1** and **0**, the games where Player One always wins and always loses respectively. Of course, one could envisage an infinite number games that share this property; **1** and **0** refer specifically to the games where neither player has any actions, but the authors were more interested the essence of a situation where every choice of action leads to invariably to victory or defeat rather than how this is actually realised. This lead to the notion of *strategic equivalence*, $G_1 \equiv G_2$, which holds when every strategy profile $\boldsymbol{s}$ results in a win for Player One in $G_1$ if and only if it results in a win in $G_2$ as well.[1]

The authors introduce algebraic operations on the set of equivalence classes of games under $\equiv$. Addition, $G_1 + G_2$, results in a game where the games are played in parallel and Player One must win at least one. Multiplication, $G_1 \cdot G_2$, results in a game where Player One must win both. Complementation, $\overline{G}$, results in a game where the winning and losing plays are inverted or, equivalently, the game where we rename Player One to Player Two and vice versa. A key result of the authors is that Boolean games form a Boolean algebra modulo strategic equivalence, under the operations $+, \cdot$ and $\overline{\phantom{x}}$.

In the setting of Boolean games the algebraic operations can be implemented using logical connectives in the natural way. To wit, $G_1+G_2$ can be implemented as follows:

$$\begin{aligned}\Phi =& \Phi_1 \uplus \Phi_2, \\ \gamma_1 =& \gamma_1^1 \vee \gamma_1^2.\end{aligned}$$

Likewise, $G_1 \cdot G_2$ can be implemented via conjunction and $\overline{G}$ with negation.

Now, if we feed 0 into Lemma 3.2.1 the resulting game, $\mathfrak{G}(0)$, is one where Player One is asked to name an interval of length 0. As this is not possible, he loses no matter what strategies are picked – this game is strategically equivalent to $\mathbf{0}$. As we have unwittingly introduced $\mathbf{0}$ into our framework, it is worth asking whether other parallels exist.

However, the notion of strategic equivalence is not very convenient for us as we have not treated the names of the propositional variables in our games in any canonical fashion, but this would be necessary for us to compare strategy profiles across games. Given our focus on two-player zero-sum games, a more natural notion is $G_1 \cong G_2$, which holds just if $G_1$ and $G_2$ have the same value. We would expect the following behaviour:

$$\begin{aligned}\mathfrak{G}(v) + \mathfrak{G}(w) &\cong \mathfrak{G}(v + w - vw), \\ \mathfrak{G}(v) \cdot \mathfrak{G}(w) &\cong \mathfrak{G}(vw), \\ \overline{\mathfrak{G}(v)} &\cong \mathfrak{G}(1 - v),\end{aligned}$$

and, indeed, combining disjoint $\mathfrak{G}(v), \mathfrak{G}(w)$ with disjunction and conjunction results in games with value $v + w - vw$ and $vw$ respectively. Negation, however, does not in general yield us $\overline{\mathfrak{G}(v)}$; $\cong$ is a weaker notion of equivalence than $\equiv$, and simply negating Player One's goal formula (i.e. giving him Player Two's goal formula) is insufficient. We need to give him Player Two's variable set as well; algebraic complementation needs to be implemented via rôle switch.

---

1 This, of course, presupposes some reasonable manner of dealing with strategies that are not present in both games, which we will not concern ourselves with here.

This suggests an alternative approach to defining $\mathfrak{G}(v)$: we could start with some base cases, and let the rest follow by induction. It turns out the generating set is 0, plus the reciprocals of primes.

**Proposition 3.2.6.** *For every $v \in [0,1]_{\mathbb{Q}}$, the game $\mathfrak{G}(v)$ can be obtained by a finite number of $+, \cdot$ and $\overline{\phantom{x}}$ operations on games of the form $\mathfrak{G}(0)$ and $\mathfrak{G}(1/b)$, where $b$ is a prime.*

*Proof.* Clearly, once we have $\mathfrak{G}(1/b)$ for primal $b$, we also have $\mathfrak{G}(1/b)$ for every $b \in \mathbb{N}$.

Now, suppose for contradiction that $\mathfrak{G}(a/b)$ cannot be represented as a combination of such games. We can assume that $a$ is the largest $i$ for which $\mathfrak{G}(i/b)$ has no such representation.

*The "largest integer principle" is applicable here because we know for a fact that $\mathfrak{G}(1) \cong \overline{\mathfrak{G}(0)}$.*

Consider the following:

$$\begin{aligned}\mathfrak{G}(a/b) &\cong \mathfrak{G}\Big(\frac{a(a+1)}{b(a+1)}\Big),\\ &\cong \mathfrak{G}\Big(\frac{a}{a+1}\Big) \cdot \mathfrak{G}\Big(\frac{a+1}{b}\Big),\\ &\cong \overline{\mathfrak{G}\Big(\frac{1}{a+1}\Big)} \cdot \mathfrak{G}\Big(\frac{a+1}{b}\Big).\end{aligned}$$

This gives us our contradiction, as $a+1 > a$ and if $\mathfrak{G}(a+1/b)$ can be represented, then so can $\mathfrak{G}(a/b)$. Q.E.D.

While neat, the utility of this result is limited. On the one hand we could have avoided defining ***Add*** and ***Sub***, as $\mathfrak{G}(1/b)$ would have only needed ***Less*** and ***Equal*** to construct, but on the other the direct construction guarantees that $\mathfrak{G}(v)$ has a Boolean representation that is polynomial in $|v|$, whereas the iteration of algebraic operations here is in general much longer. As it turns out, if $v$ is a dyadic fraction then the number of necessary operations is indeed polynomial in $|v|$, and we could rewrite all our proofs to only use dyadic values, so this would be sufficient. However, there is little reason to do so, especially since the ***Add*** and ***Sub*** formulae are useful to us irrespective of their rôle in constructing $\mathfrak{G}(v)$.

## 3.3 TWO-PLAYER GAMES AND RATIONAL EQUILIBRIA

It is a well-known fact that rational-valued two-player games admit rational equilibria, the size of which is polynomial in the normal form of the game. As with all well-known facts, figuring out precisely why this is the case can be difficult for newcomers to the field, especially if the space of time between the origin of such facts and the present is such that the mathematical background of the aforementioned newcomers is significantly different from that which was readily assumed in the days when the facts were established. It is for the benefit of such newcomers (and certainly not because the author had any such difficulty himself) that we present the arguments in this section. A reader familiar with the material may find the following proofs unnecessarily meandering; this is because they are intended to be comprehensible to those with no previous knowledge of linear programming other than that the following problem can be solved in polynomial time:

| LinearProgramming | |
|---|---|
| Input: | A matrix of values $A \in \mathbb{Q}^{s \times t}$, and vectors $\boldsymbol{b} \in \mathbb{Q}^s$, $\boldsymbol{c} \in \mathbb{Q}^t$. |
| Output: | A solution, $\boldsymbol{y}$, to the following optimisation problem: $$\begin{aligned} A\boldsymbol{y} &\leq \boldsymbol{b}, \\ \boldsymbol{y} &\geq 0, \\ \max \quad & \boldsymbol{c}^T \boldsymbol{y}. \end{aligned}$$ |

**Theorem 3.3.1** (Khachiyan [24])**.** LinearProgramming *is in* FP.

As we shall see, Value can be reduced to LinearProgramming directly, whereas various problems in the variable-sum case can be solved with a nondeterministic choice of a linear program to solve.

### 3.3.1 *Zero-sum games*

Consider a matrix representation of a two-player zero-sum game. That is, a matrix $M$ with $m_{i,j} = u_1(s_i \in S_1, s_j \in S_2)$. Let $|S_1| = k$ and $|S_2| = q$. If Player One is playing a maxmin strategy, then he seeks to pick strategy weights $x_1, \ldots, x_k$ satisfying:

$$\begin{aligned} \sum_{i \leq k} x_i &= 1, \\ \boldsymbol{x} &\geq 0, \\ \max_{\boldsymbol{x}} \min_{j} \sum_{i \leq k} & x_i m_{i,j}. \end{aligned}$$

This is, of course, not yet a valid instance of LinearProgramming – the objective function is not linear. However, it is not far off from being so: once the min function has made its selection, we are left with a sum of linear terms. The choice made by the min function is also easy to characterise, as we are looking for a $v$ that is smaller than all $q$ possible choices. So what we do is introduce $v$ as a new variable with the requirement that $v \leq \min_j \sum_{i \leq k} x_i m_{i,j}$. The objective function is then simply to maximise $v$, giving us the following linear program:

$$\begin{aligned} v &\leq \sum_{i \leq k} x_i m_{i,1}, \\ &\vdots \\ v &\leq \sum_{i \leq k} x_i m_{i,q}, \end{aligned}$$

$$\sum_{i \le k} x_i = 1,$$
$$\boldsymbol{x} \ge 0,$$
$$\max_{\boldsymbol{x},v} \quad v.$$

We can now construct the LinearProgramming instance:

$$\begin{bmatrix} 1 & -m_{1,1} & \dots & -m_{k,1} \\ \vdots & \vdots & \ddots & \vdots \\ 1 & -m_{1,q} & \dots & -m_{k,q} \\ 0 & 1 & \dots & 1 \\ 0 & -1 & \dots & -1 \end{bmatrix} \begin{bmatrix} v \\ x_1 \\ \vdots \\ x_k \end{bmatrix} \le \begin{bmatrix} 0 \\ \vdots \\ 0 \\ 1 \\ -1 \end{bmatrix},$$
$$\max v.$$

The vector to be optimised is $(v, x_1, \dots, x_k)$. $A$ is a $(q+2) \times (k+1)$ matrix, the first $q$ rows of it deal with the $v \le \min_j \sum_{i \le k} x_i m_{i,j}$ constraint and the last two ensure that $\sum_{i \le k} x_i = 1$. The function to be maximised is just $v$. The fact that $v$ is restricted to be positive is not a restriction as, given Theorem 2.1.22, we can always shift the payoffs into the positive region.

The following is immediate:

**Proposition 3.3.2.** DValue *is in* P *and* Value *is in* FP *for matrix games, and* EXP, FEXP *respectively for Boolean games.*

### 3.3.2 *Variable-sum games*

The argument below is based on the support enumeration algorithm (Kaplan and Dickhaut, von Stengel [87], [88]). The differences are due to the fact that we do not assume that we are dealing with non-degenerate games.

**Lemma 3.3.3.** *Let $(A, B)$ be a bimatrix representation of a two-player zero-sum game $G$ and $X \subseteq S_1, Y \subseteq S_2$. We claim the following:*

1. *$G$ has one, zero or infinity equilibria with support $(X, Y)$.*
2. *We can ascertain whether an equilibrium exists and, if so, produce one in polynomial time. Moreover, we can do this even if we are looking for equilibria satisfying certain payoff constraints.*

*Proof.* Recall from Fact 2.1.10 that a necessary condition for equilibrium is that a player be indifferent between every strategy in his support. For Player One this condition translates to:

$$\sum_{s_j \in Y} a_{i,j} y_j = \alpha, \text{for } s_i \in X,$$
$$\sum_{i \in Y} y_i = 1,$$
$$\boldsymbol{y} \ge 0.$$

$\alpha$ is the (as yet unknown) utility Player One derives from the potential equilibrium, and the $y_i$ variables are the strategy weights of Player Two.

For Player Two the same condition is the following:

$$\begin{aligned}\sum_{s_i \in X} b_{i,j} x_i =& \beta, \text{for } s_j \in Y,\\ \sum_{i \in X} x_i =& 1,\\ \boldsymbol{x} \geq& 0.\end{aligned}$$

The variables $x_i$ being the strategy weights of Player One.

However, this is not sufficient – a player may well be indifferent between the strategies in his support, but that does not mean he does not have a profitable deviation with a strategy outside of it. We need to check that $\alpha$ and $\beta$ are at least as big as what the players can get with any pure strategy:

$$\begin{aligned}\sum_{s_j \in Y} a_{1,j} y_j \leq& \alpha,\\ &\vdots\\ \sum_{s_j \in Y} a_{k,j} y_j \leq& \alpha,\\ \sum_{s_i \in X} b_{i,1} x_i \leq& \beta,\\ &\vdots\\ \sum_{s_i \in X} b_{i,q} x_i \leq& \beta.\end{aligned}$$

Finally, if we want to bound $\alpha$ and $\beta$ to be greater than certain values, we can add this to the list as well.

This gives us a LINEARPROGRAMMING instance (with a trivial objective function, over the variables $\boldsymbol{x}, \boldsymbol{y}, \alpha, \beta$), and we can produce a solution, or verify that none exists, in polynomial time. Q.E.D.

**Proposition 3.3.4.** *$\exists$*GUARANTEENASH *is in* NP *for bimatrix games and in* NEXP *for two-player Boolean games.* $\forall$GUARANTEENASH *and* UNIQUENASH *are in* coNP *for bimatrix games and* coNEXP *for two-player Boolean games.* $\exists$NASHSAT *and* $\forall$NASHSAT *are in* NEXP *and* coNEXP *respectively.*

*Proof.* For $\exists$GUARANTEENASH we need only nondeterministically choose a support. For $\forall$GUARANTEENASH there is a difficulty because the complement involves finding an equilibrium where some player's payoff is *strictly* lower then some threshold, and the basic LINEARPROGRAMMING formulation does not handle strict constraints. Nevertheless, we can address this by asking the instance to minimise $\alpha$ or $\beta$ respectively, and verifying by hand that the values returned by the algorithm are

*That is, maximise $-\alpha$ or $-\beta$.*

indeed strictly lower. For UniqueNash we consider the complement SecondNash. We start by nondeterministically choosing a support and finding an equilibrium. Having done so, we will have obtained the strategy weights used in that equilibrium. Clearly, if there exists a second equilibrium some strategy weight would have to be different; hence, nondeterministically choose another basis, nondeterministically choose a strategy weight, and have the linear programming instance maximise or minimise that weight as the case may be.

For ∃NashSat and ∀NashSat, note that a formula is made true with probability 1 if and only if every truth assignment in the support satisfies that formula – the weights given to such assignments do not matter. Hence, all we need to do is nondeterministically choose a support where either every strategy satisfies the formula, or some strategy does not as the case may be. Q.E.D.

## 3.4 DISCUSSION

Little in this chapter is new. Section 3.3 collates and adapts standard results of the early days of computational game theory into a form that is convenient for our purposes, but nothing we derive there should come as a surprise to anyone well acquainted with the field. I am unaware of the results in Section 3.1 being explicitly stated anywhere in the literature, but the underlying ideas are behind standard constructions throughout computer science. Any of the lemmata within would not be out of place as an exercise in an undergraduate textbook. does present a novel proof technique, by showing how propositional encodings can allow us to build Boolean games of whatever nature we want, but the significance of this will not be clear until we get to the actual proofs in the later chapters.

That said, this is nevertheless one of the most important sections of the present work. It is my desire that this thesis could be read independently; the references to the literatures are intended to satisfy the reader's curiosity, not to clarify any of the arguments contained herein. This chapter was written to serve that purpose.

# 4 KICKING DOWN THE DOOR

*O gatekeeper, open thy gate,*
*Open thy gate so I may enter!*
*If thou openest not the gate so that I cannot enter,*
*I will smash the door, I will shatter the bolt,*
*I will smash the doorpost, I will move the doors,*
*I will raise up the dead eating the living,*
*So that the dead will outnumber the living.*

— Ishtar's descent into the underworld

This chapter contains Theorem 4.2.1, the proof of which demonstrates that ∃GuaranteeNash is NEXP-complete for two-player Boolean games. This is the main technical result of the thesis and all the rest follow from it, either directly by presenting a reduction from some variant of ∃GuaranteeNash, or indirectly by using the proof techniques introduced here.

## 4.1 A CANONICAL PROBLEM

Every mathematical truth is a tautology, but some tautologies are easier to spot than others; half the work in showing the hardness of a problem is in choosing the right problem to reduce from.

Our main focus is on decision problems about Nash equilibria, problems that are typically NP-hard in the normal form. Given the expressivity and succinctness of Boolean games, it would be reasonable to hypothesise that NEXP is the class we are interested in. The trouble, then, is that whereas compendia have been written on the topic of NP-complete problems, the world of NEXP-completeness is a lot more barren – Papadimitriou's textbook (Papadimitriou [76]) spends a mere six pages on exponential time, and aside from a problem of first order logic the only problems considered are succinct versions of NP-complete problems.[1]

The wider literature is hardly better. The fact is that NEXP is of remote interest to most researchers, given the intractability of the problems involved. One is given the choice between concise circuit representations of known NP problems, and increasingly more arcane logical problems. At that point one ought to ask, would it not be simpler to just bite the bullet and encode a Turing machine directly?

This is precisely our approach.

1 Which is, of course, exactly what we are considering here.

| NEXPTM | |
|---|---|
| Input: | A nondeterministic Turing machine $M$, an input word $w$, and a computation bound $K$ in binary. |
| Output: | YES if $M$ accepts $w$ in at most $K$ steps, NO otherwise. |

**Fact 4.1.1.** NEXPTM *is* NEXP*-complete.*

*Proof.* The key here is that $K$ is given in binary, and hence $K$ is exponential in $|K|$. Thus, given an arbitrary nondeterministic exponential-time machine $M$ with clock $f$, we can construct an instance of NEXPTM $(M, w, f(|w|))$. By definition, $(M, w, f(|w|)) \in$ NEXPTM if and only if $M$ accepts $w$ in at most $f(|w|)$ steps, which is exactly to say that the machine $M$ with clock $f$ accepts $w$. Q.E.D.

## 4.2 A PAYOFF-CONSTRAINED EQUILIBRIUM

**Theorem 4.2.1.** $\exists$GUARANTEENASH *for two-player Boolean games is* NEXP*-complete.*

*Proof.* For NEXP membership, the problem is in NP for normal form games so it is in NEXP for Boolean games – we can expand the game into its normal form, which is exponential in the size of the Boolean representation, and guess an equilibrium. For further details, the reader can go back to Section 3.3.

For NEXP-hardness, we will reduce from NEXPTM. That is, we will demonstrate that given a nondeterministic machine $M$, a binary input string $w$, and a computation bound $K$ in binary, we can construct in polynomial time a two-player Boolean game $G$ and a vector of payoffs $\boldsymbol{v}$ such that there exists an equilibrium $\boldsymbol{\sigma}$ of $G$ where each player $i$ attains a utility of at least $\boldsymbol{v}[i]$ if and only if $M$ accepts $w$ in at most $K$ steps.

Without loss of generality, we assume two restrictions on $M$. First, $M$ is augmented with a "do nothing" transition that is only activated from an accepting state. This ensures that every run of $M$ on $w$ is defined for all $K$ computation steps, even if the machine stops computing early. Second, we assume that $M$ has just one accepting state $q_a$, and if $M$ accepts it accepts with the head over the leftmost cell. This is not really a restriction as we can always extend $M$ into an $M'$ that simulates $M$ until $M$ reaches the accepting state, and then $M'$ enters a routine requiring it to move to the left of the tape and then accept.

Let $k = |K|$; and $q = |Q|$, the number of states of the machine. Observe that since $K$ is given in binary, $2^k \geq K$. We can thus envisage a run of $M$ on $w$ as being represented by a $2^k \times 2^k$ table. Every row of the table is a machine configuration (that is, the tape contents, machine state and head location). Thus the $i$th row, $j$th column of the table tells us the contents of the $j$th cell at the $i$th computation step; whether the

$$2^k \left\{ \begin{array}{cccccccc} \multicolumn{8}{c}{\overbrace{\hspace{20em}}^{2^k}} \\ 1^{q_0} & 1^{\leftarrow} & 0^{\leftarrow} & 1^{\leftarrow} & \sqcup^{\leftarrow} & \cdot & \cdot & \cdot \; \sqcup^{\leftarrow} \\ 0^{\rightarrow} & 1^{q_1} & 0^{\leftarrow} & 1^{\leftarrow} & \sqcup^{\leftarrow} & \cdot & \cdot & \cdot \; \sqcup^{\leftarrow} \\ 0^{\rightarrow} & 0^{\rightarrow} & 0^{q_2} & 1^{\leftarrow} & \sqcup^{\leftarrow} & \cdot & \cdot & \cdot \; \sqcup^{\leftarrow} \\ \cdot & \cdot & \cdot & \cdot & \cdot & \cdot & \cdot & \cdot \; \cdot \\ \cdot & \cdot & \cdot & \cdot & \cdot & \cdot & \cdot & \cdot \; \cdot \\ 0^{q_a} & 0^{\leftarrow} & 0^{\leftarrow} & 0^{\leftarrow} & 0^{\leftarrow} & \cdot & \cdot & \cdot \; \sqcup^{\leftarrow} \\ \cdot & \cdot & \cdot & \cdot & \cdot & \cdot & \cdot & \cdot \; \cdot \\ \cdot & \cdot & \cdot & \cdot & \cdot & \cdot & \cdot & \cdot \; \cdot \\ 0^{q_a} & 0^{\leftarrow} & 0^{\leftarrow} & 0^{\leftarrow} & 0^{\leftarrow} & \cdot & \cdot & \cdot \; \sqcup^{\leftarrow} \end{array} \right.$$

Figure 4.1: What a run of $M$ on $w = 1101$ might look like. The row in the middle represents the step in which the machine accepts, after which it enters into a "do nothing" routine right through to step $2^k - 1$.

head is over the cell, to the left of it, or to the right; and, if the head is over the cell, the machine state. This is illustrated in Figure 4.1.

We can say that such a table represents an accepting run if and only if:

1. The 0th row represents the initial configuration. By this we mean that the $(0, j)$-entry is $w[j]$, or blank if $j \geq |w|$; the $(0, 0)$-entry contains the head in state $q_0$; for each $j \geq 1$, the $(0, j)$-entry asserts that the head is to the left.

*We index the steps and cells from 0. Hence the final step is $K - 1$.*

2. The $(K - 1, 0)$-entry contains the head in the accepting state.

*The $(i, j)$-entry is the $i$th row and $j$th column. Think of a matrix rather than a chart.*

3. Every consecutive pair of rows satisfies some transition rule of the machine.

Points 1 and 2 are, of course, a lot simpler to verify than point 3; the requirements are in some sense local – we need only to demonstrate that the $j$th cell at step 0 does not contain $w[j]$ to prove that point 1 is not satisfied. Point 3, however, seemingly requires us to consider a pair of rows in their entirety, which is problematic given that the length of such a row is exponential in $k$. It is for this reason that we label every entry of the table with the direction to the head. This will allow us to reduce the problem of checking two rows for consistency to merely finding a pair of inconsistent arrows.

We expand on this in the following lemma, showing that if the table does not represent an accepting run, then there will exist a $2 \times 2$ square in it witnessing that fact.

**Lemma 4.2.2.** *There exists a finite set of requirements $S$ such that a table of the kind described above represents an accepting run if and only if every $2 \times 2$ square in the table satisfies every requirement in $S$.*

*Perversely, we shall refer to the $(2^k-1, 2^k-1)$-entry as the "upper left" of this square.*

*Proof.* It should first be said that we allow our squares to loop around the edges of the table. Thus, for example, there exists a square consisting of the entries $(2^k - 1, 2^k - 1)$, $(2^k - 1, 0)$, $(0, 2^k - 1)$ and $(0, 0)$.

There are four requirements, although the fifth breaks down into a (finite, polynomially bounded) slew of disjuncts:

1. If the square contains the $(0, 0)$-entry, that entry contains the head in state $q_0$.

2. If the square contains the $(0, j)$-entry, with $j < |w|$, that entry contains $w[j]$ and if $j \geq 1$ asserts the head is to the left.

3. If the square contains the $(0, j)$-entry, with $j \geq |w|$, that entry contains the blank tape symbol and asserts that the head is to the left.

4. If the square contains the $(K - 1, 0)$-entry, that entry contains the head in the accepting state $q_a$.

5. The square is consistent with some transition rule.

Little can be said about the first four requirements, it is clear that they are necessary. To unpack the fifth requirement fully we would need to fix the Turing machine in question and expend a lot of paper. Instead, we will demonstrate what it means by showing how we can verify that a square is consistent with the rule "Upon reading a 0 in state $q_3$ write 1, move left and transition to state $q_5$" by comparing it against a finite number of admissible square schemata. These are illustrated in Figure 4.2.

*Every schema expands to at most 81 concrete squares, as a cell can contain only 1, 0, or ␣.*

It should be clear that if the fifth requirement is not satisfied then the table cannot represent an accepting run. It remains to argue that if the table is not an accepting run one of the requirements is, in fact, violated.

If the table does not contain the initial configuration in the first row or an accepting state on the leftmost cell of the $(K - 1)$th row, one of the first four requirements would be violated. The only remaining possibility for the table to not represent an accepting run is the existence of rows $i$ and $i + 1$ which do not satisfy any transition rule of the machine. We shall see that this is not possible.

Suppose that row $i$ correctly describes a machine configuration (critically, it contains the head in exactly one entry and every other entry correctly asserts the direction to the head), and every $2 \times 2$ square over row $i$ and $i+1$ satisfies the requirements. We will show that there exists a $(a, q_x) \to (b, D, q_y)$ such that:

1. The head is in exactly one position in row $i + 1$. This position is one step in direction $D$ from the location of the head in row $i$ (unless $D = L$ and the head is in cell 0, in which case it remains in cell 0), and the entries in row $i+1$ correctly assert the direction to the head.

$$\begin{array}{ccc} & \boldsymbol{j} & \boldsymbol{j+1} \\ \boldsymbol{i} & x^{\rightarrow} & 0^{q_3} \\ \boldsymbol{i+1} & x^{q_5} & 1^{\leftarrow} \\ \multicolumn{3}{c}{i < 2^k - 1} \\ \multicolumn{3}{c}{j < 2^k - 1} \end{array} \qquad \begin{array}{ccc} & \boldsymbol{j} & \boldsymbol{j+1} \\ \boldsymbol{i} & 0^{q_3} & x^{\leftarrow} \\ \boldsymbol{i+1} & 1^{\leftarrow} & x^{\leftarrow} \\ \multicolumn{3}{c}{i < 2^k - 1} \\ \multicolumn{3}{c}{0 < j < 2^k - 1} \end{array} \qquad \begin{array}{ccc} & \mathbf{0} & \mathbf{1} \\ \boldsymbol{i} & 0^{q_3} & x^{\leftarrow} \\ \boldsymbol{i+1} & 1^{q_5} & x^{\leftarrow} \\ \multicolumn{3}{c}{i < 2^k - 1} \end{array}$$

$$\begin{array}{ccc} & \boldsymbol{j} & \boldsymbol{j+1} \\ \boldsymbol{i} & x^{\leftarrow} & y^{\leftarrow} \\ \boldsymbol{i+1} & x^{\leftarrow} & y^{\leftarrow} \\ \multicolumn{3}{c}{i < 2^k - 1} \\ \multicolumn{3}{c}{1 < j < 2^k - 1} \end{array} \qquad \begin{array}{ccc} & \boldsymbol{j} & \boldsymbol{j+1} \\ \boldsymbol{i} & x^{\rightarrow} & y^{\rightarrow} \\ \boldsymbol{i+1} & x^{\rightarrow} & y^{q_5} \\ \multicolumn{3}{c}{i < 2^k - 1} \\ \multicolumn{3}{c}{j < 2^k - 2} \end{array} \qquad \begin{array}{ccc} & \boldsymbol{j} & \boldsymbol{j+1} \\ \boldsymbol{i} & x^{\rightarrow} & y^{\rightarrow} \\ \boldsymbol{i+1} & x^{\rightarrow} & y^{\rightarrow} \\ \multicolumn{3}{c}{i < 2^k - 1} \\ \multicolumn{3}{c}{j < 2^k - 3} \end{array}$$

$$\begin{array}{ccc} & \mathbf{2^{\boldsymbol{k}} - 1} & \mathbf{0} \\ \boldsymbol{i} & 0^{q_3} & x^{\rightarrow} \\ \boldsymbol{i+1} & 1^{\leftarrow} & x^{\rightarrow} \\ \multicolumn{3}{c}{i < 2^k - 1} \end{array} \qquad \begin{array}{ccc} & \mathbf{2^{\boldsymbol{k}} - 1} & \mathbf{0} \\ \boldsymbol{i} & x^{\leftarrow} & y^{\rightarrow} \\ \boldsymbol{i+1} & x^{\leftarrow} & y^{\rightarrow} \\ \multicolumn{3}{c}{i < 2^k - 1} \end{array} \qquad \begin{array}{ccc} & \mathbf{2^{\boldsymbol{k}} - 1} & \mathbf{0} \\ \boldsymbol{i} & x^{\leftarrow} & y^{\rightarrow} \\ \boldsymbol{i+1} & x^{\leftarrow} & y^{q_5} \\ \multicolumn{3}{c}{i < 2^k - 1} \end{array}$$

$$\begin{array}{ccc} & \mathbf{2^{\boldsymbol{k}} - 1} & \mathbf{0} \\ \boldsymbol{i} & x^{\leftarrow} & 0^{q_3} \\ \boldsymbol{i+1} & x^{\leftarrow} & 1^{q_5} \\ \multicolumn{3}{c}{i < 2^k - 1} \end{array} \qquad \begin{array}{ccc} & \boldsymbol{j} & \boldsymbol{j+1} \\ \mathbf{2^{\boldsymbol{k}} - 1} & x & y \\ \mathbf{0} & z & w \end{array}$$

Figure 4.2: Admissible squares for the rule $(0, q_3) \rightarrow (1, L, q_5)$.

2. The head in row $i$ is in state $q_x$, and in row $i+1$ in $q_y$.

3. The tape in row $i$ contains $a$ under the head. In row $i+1$ the tape contains $b$ in the entry where the head was in row $i$ and elsewhere agrees with row $i$.

It is clear that these three properties will be sufficient to establish that row $i+1$ follows from $i$ via $(a, q_x) \to (b, D, q_y)$.

Let the head in row $i$ be in cell $j$, with $a$ written on the tape. We assume $j > 0$, the boundary case where $j = 0$ differs only that if $D = L$ the head must stand still. Consider two $2 \times 2$ squares, one with upper left corner $(i, j-1)$ and the other with upper left corner $(i, j)$. In order to match an admissible schema, the left square must either contain $a$ in entry $(i, j)$, $b_1$ in entry $(i+1, j)$, the head in entry $(i, j)$ in state $q_x$ and the head in entry $(i+1, j-1)$ in state $q_{y_1}$ in accordance with some $(a, q_x) \to (b_1, L, q_{y_1})$, or both the lower entries of the square must assert the head is to the right (that is, it is an admissible square for some rule that asserts rightward movement). Likewise, the right square must either agree with some $(a, q_x) \to (b_2, R, q_{y_2})$, or assert the head is to the left. We claim that it must be the case that exactly one of these squares agrees with a machine rule in this fashion, and displays the head in both rows; while the other asserts the head is in the correct direction.

Suppose the contrary – first, that the left square claims the head moved to the right and the right square that it moved to the left. In order for the left square to be correct, this means entry $(i+1, j)$ must assert the head is to the right. However, in order for the right square to be correct this entry must assert the head is to the left, and hence there is no way to fill in that entry of the table to satisfy both squares. Likewise, if the left square claims leftward movement and the right square claims rightward movement, then again it will not be possible to fill in $(i+1, j)$ in a manner consistent with both squares.

This is the crux of the argument. We have established property 2, taking $q_y = q_{y_1}$ or $q_{y_2}$ as the case may be, and 3, as the entry under the tape changes from $a$ to $b_1$ or $b_2$ and, as can be seen from Figure 4.2, in order to be correct, every square that does not cover the head's location in row $i$ is obliged to copy the tape contents faithfully. To establish that property 1 holds, we just need to verify that a second head cannot appear in row $i+1$. This can be seen from an inductive argument. Without loss of generality, suppose $D = L$. The square with upper left corner $(i, j)$ contains $\leftarrow$ in its bottom right corner. As such, the square with upper left corner $(i, j+1)$ over must contain $\leftarrow$ in its bottom left corner, and as can be seen from Figure 4.2 this assures that it also has $\leftarrow$ in its bottom right corner, which inductively carries over to all squares to the right.

The square with upper left corner $(i, j-1)$ contains the head in its bottom left corner. Accordingly, the square with upper left corner

$(i, j-2)$ must have the head in its bottom right corner and a $\rightarrow$ in its bottom left corner – which leads to the induction in the leftwards direction.

To recap: we have shown that if every $2 \times 2$ square over rows $i$ and $i+1$ satisfies an admissible square of some transition rule (the number of both being polynomial in the size of the machine), then the properties 1, 2 and 3 hold, which establishes that row $i+1$ follows from row $i$ via some $(a, q_x) \rightarrow (b, D, q_y)$. This establishes that the five requirements given above are in fact sufficient – if every square is an admissible square of some rule, then every row follows by a valid transition rule from the preceding row, and hence the table must represent a computation history. Q.E.D.

This gives us the framework for our construction. Player One, via his choice of strategy, defines such a table and Player Two verifies that every $2 \times 2$ square in it satisfies all the requirements in $S$.

The players control the following variables:

$$\begin{aligned}
\Phi_1 = {} & \{\mathit{Zero}^1, \mathit{One}^1, \mathit{Left}^1, \mathit{Right}^1\} \cup \mathrm{var}_1(\mathfrak{G}({}^3\!/_4)) \\
& \cup \{\overline{\mathit{Time}^1_k}\} \cup \{\overline{\mathit{Tape}^1_k}\} \cup \{\overline{\mathit{State}^1_q}\}. \\
\Phi_2 = {} & \{\mathit{Zero}^2, \mathit{One}^2, \mathit{sZero}^2, \mathit{sOne}^2, \mathit{Left}^2, \mathit{Right}^2, \mathit{sLeft}^2, \mathit{sRight}^2, \\
& \quad \mathit{nZero}^2, \mathit{nOne}^2, \mathit{nsZero}^2, \mathit{nsOne}^2, \mathit{nLeft}^2, \mathit{nRight}^2, \\
& \quad \mathit{nsLeft}^2, \mathit{nsRight}^2\} \cup \mathrm{var}_2(\mathfrak{G}({}^3\!/_4)) \\
& \cup \{\overline{\mathit{Time}^2_k}\} \cup \{\overline{\mathit{Tape}^2_k}\} \cup \{\overline{\mathit{sTime}^2_k}\} \cup \{\overline{\mathit{sTape}^2_k}\} \\
& \cup \{\overline{\mathit{nTime}^2_k}\} \cup \{\overline{\mathit{nTape}^2_k}\} \cup \{\overline{\mathit{nsTime}^2_k}\} \cup \{\overline{\mathit{nsTape}^2_k}\} \\
& \cup \{\overline{\mathit{State}^2_q}\} \cup \{\overline{\mathit{sState}^2_q}\} \cup \{\overline{\mathit{nState}^2_q}\} \cup \{\overline{\mathit{nsState}^2_q}\}.
\end{aligned}$$

A truth assignment to $\Phi_1$ should be interpreted as defining an entry of the table with index $(\llbracket \overline{\mathit{Time}^1_k} \rrbracket, \llbracket \overline{\mathit{Tape}^1_k} \rrbracket)$; its tape contents (as given by $\mathit{One}^1$ and $\mathit{Zero}^1$); and whether the head is to the left or right (as indicated by $\mathit{Left}^1$ and $\mathit{Right}^1$) or directly above the entry, and, if the head directly above, the machine is in the state indicated by $\mathit{State}^1_i$.

Likewise, an assignment to $\Phi_2$ defines a quadruple of entries which constitute a $2 \times 2$ square as in Lemma 4.2.2: the prefix "s" of variable names denotes the successor cell on the tape, while "n" refers to the next computation step, so $\mathit{Zero}^2$ asserts that there is a 0 in the top left corner of the selected square, and $\mathit{nsZero}^2$ that there is a 0 in the bottom right. There is at present no guarantee that these entries are well formed – Player Two's assignment could well assert the presence of both a 1 and a 0, more than one machine state, and other absurdities; these issues will be dealt with in Player Two's goal formula. The $\mathrm{var}(\mathfrak{G}({}^3\!/_4))$ variables do not feature in this interpretation; their purpose is technical, and will not be clear until the proof of Lemma 4.2.3.

We want Player One to describe the entire table, but so far a pure strategy will only result in the description of a single entry. To address this, we will require Player One to randomise over the pure strategies available to him, and in doing so the mixed strategy he chooses will describe a table as a probability distribution. To force randomisation, we could simply give Player One the goal of naming a table entry different from any of the four named by Player Two. However, we shall do something a little more complicated: if Player One names the upper left entry of Player Two's square he loses outright, but should he name any other entry he is given the chance of redeeming himself by winning a side game of $\mathfrak{G}(3/4)$. The explanation for this, too, must await Lemma 4.2.3 in the sequel.

Player One's goal is thus the following:

$$\begin{aligned}
\gamma_1 =& \mathit{AvoidC} \wedge (\mathit{Avoid} \vee \gamma_1(\mathfrak{G}(3/4))),\\
\mathit{AvoidC} =& \neg(\boldsymbol{Equal}(\overline{Tape^1_k}; \overline{Tape^2_k}) \wedge \boldsymbol{Equal}(\overline{Time^1_k}; \overline{Time^2_k})),\\
\mathit{Avoid} =& \neg(\boldsymbol{Equal}(\overline{Tape^1_k}; \overline{sTape^2_k}) \wedge \boldsymbol{Equal}(\overline{Time^1_k}; \overline{sTime^2_k}))\\
&\wedge \neg(\boldsymbol{Equal}(\overline{Tape^1_k}; \overline{nTape^2_k}) \wedge \boldsymbol{Equal}(\overline{Time^1_k}; \overline{nTime^2_k}))\\
&\wedge \neg(\boldsymbol{Equal}(\overline{Tape^1_k}; \overline{nsTape^2_k}) \wedge \boldsymbol{Equal}(\overline{Time^1_k}; \overline{nsTime^2_k})).
\end{aligned}$$

Player Two's goals are more complicated. First, one of the entries she names must be the same as Player One's, and preferably the top left – we shall later see that this ensures that in equilibrium Player One distributes equally over every entry of the table, thus describing the entire computation history. Next, the contents of the matching entry must agree with Player One – this ensures the players are talking about the same table. Finally, the four entries must constitute a $2 \times 2$ square with the upper left entry indexed by $([\![\overline{Time^2_k}]\!], [\![\overline{Tape^2_k}]\!])$, and this square must satisfy all the requirements in $S$ – hence Lemma 4.2.2 will allow us to draw the connection between Player Two's payoff and the output of the machine.

The formula she is trying to satisfy is:

$$\gamma_2 = \Big(\neg \mathit{AvoidC} \vee (\neg \mathit{Avoid} \wedge \gamma_2(\mathfrak{G}(3/4)))\Big) \wedge \mathit{Agree} \wedge \mathit{Require}.$$

The subgoal *Agree* is a conjunction of four implications, each stating that if one of the named entries matches Player One's, the entries must agree on the description of the cell.

The first of these implications is:

$$\begin{aligned}
&(\boldsymbol{Equal}(\overline{Tape^1_k}; \overline{Tape^2_k}) \wedge \boldsymbol{Equal}(\overline{Time^1_k}; \overline{Time^2_k})) \rightarrow\\
&\qquad((\mathit{Zero}^1 \leftrightarrow \mathit{Zero}^2) \wedge (\mathit{One}^1 \leftrightarrow \mathit{One}^2) \wedge (\mathit{Left}^1 \leftrightarrow \mathit{Left}^2)\\
&\qquad\wedge (\mathit{Right}^1 \leftrightarrow \mathit{Right}^2) \wedge \bigwedge_{1 \leq i \leq q} (\mathit{State}^1_i \leftrightarrow \mathit{State}^2_i)).
\end{aligned}$$

The other three follow the same pattern.

*Require* states that the entries specified form a $2 \times 2$ square of the desired type which satisfies the requirements in $S$. To ensure the entries picked are indeed a square the following is sufficient:

$$\begin{aligned}
&\boldsymbol{Succ}([\![\overline{Tape^2_k}]\!]; [\![\overline{sTape^2_k}]\!]) \wedge \boldsymbol{Succ}([\![\overline{Time^2_k}]\!]; [\![\overline{nTime^2_k}]\!]) \\
&\wedge \boldsymbol{Equal}([\![\overline{Tape^2_k}]\!]; [\![\overline{nTape^2_k}]\!]) \wedge \boldsymbol{Equal}([\![\overline{Time^2_k}]\!]; [\![\overline{sTime^2_k}]\!]) \\
&\wedge \boldsymbol{Equal}([\![\overline{sTape^2_k}]\!]; [\![\overline{nsTape^2_k}]\!]) \wedge \boldsymbol{Equal}([\![\overline{nTime^2_k}]\!]; [\![\overline{nsTime^2_k}]\!]) \\
&\wedge \boldsymbol{OneOf}(Left^2, Right^2, \overline{State^2_q}) \wedge \boldsymbol{OneOf}(nsLeft^2, nsRight^2, \overline{nsState^2_q}) \\
&\wedge \boldsymbol{OneOf}(sLeft^2, sRight^2, \overline{sState^2_q}) \wedge \boldsymbol{OneOf}(nLeft^2, nRight^2, \overline{nState^2_q}) \\
&\wedge \neg(One^2 \wedge Zero^2) \wedge \neg(nOne^2 \wedge nZero^2) \wedge \neg(sOne^2 \wedge sZero^2) \\
&\wedge \neg(nsOne^2 \wedge nsZero^2).
\end{aligned}$$

We will not give the final component – the check on requirements – in detail. Recall simply that in Lemma 4.2.2 we have shown that every transition rule corresponds to a constant number of admissible squares.[2] We can thus invoke the expressive completeness of propositional logic to argue that for each rule there is a formula $Rule_i$ that is satisfied if and only if Player Two's strategy matches one of the admissible squares for that rule. Even if the size of this formula is exponential in the number of admissible squares (it isn't), exponential in a constant is still a constant. The final component is thus simply a disjunction of all such $Rule_i$.

We now claim that the existence of an accepting run of $M$ on $w$ in at most $K$ steps is equivalent to the existence of an equilibrium in the game described where Player Two is guaranteed utility:

$$\boldsymbol{v}[2] = \frac{1}{2^{2k}} + \frac{3}{2^{2k} \cdot 4}.$$

Suppose an accepting run exists. Consider the profile where Player One randomises over every entry with equal probability, plays the equilibrium strategy in $\mathfrak{G}(3/4)$, and his description of the table accords with the accepting run. Player Two is randomising over every $2 \times 2$ square with equal probability, playing the equilibrium in $\mathfrak{G}(3/4)$, and her description of the square also accords with the accepting run. Player Two has a $1/2^{2k}$ chance of matching the top left corner of her square to Player One. If she does so, *Agree* is satisfied because both players are describing the same table and *Require* is satisfied because the table they are describing is an accepting run, and hence satisfies every requirement in $S$.

On top of this, she has a $3/2^{2k}$ chance of matching one of the other entries to Player One. In this case, *Agree* and *Require* are satisfied for the same reasons, but $\gamma_2(\mathfrak{G}(3/4))$ has only a $1/4$ chance of being satisfied. This means that Player Two has an expected utility of $\boldsymbol{v}[2]$. Now let us verify that the profile is in equilibrium.

In this particular profile Player One earns $1 - \boldsymbol{v}[2]$ utility as he fails to

*This is not true for an arbitrary profile as the game is not zero-sum.*

2 In Figure 4.2 we have shown 11 schemata, and each schema expands to at most 81 squares. The upper bound is thus 891, regardless of the machine under consideration.

satisfy his formula only if he lands on one of the cells named by Player Two. We need only consider deviations in pure strategies. Whichever cell Player One picks, he will have a $1/2^{2k}$ chance of landing on a top left corner of Player Two's square and a $3/2^{2k}$ chance on one of the other entries, so his payoff would be no better; likewise if Player Two plays $(i, j)$ with probability 1 she will have a $1/2^{2k}$ chance of catching Player One on the top left of her square and a $3/2^{2k}$ chance on the other three entries.

Now, consider a strategy profile where no such run exists. Observe that as Player Two's goal formula is conjunctive, her utility is bounded above by the probability of catching Player One and winning the side game, i.e. of satisfying $(\neg \mathit{AvoidCorner} \vee (\neg \mathit{Avoid} \vee \neg\gamma_2(\mathfrak{G}(3/4))))$. We will show that there is exactly one way to satisfy this with probability at least $\boldsymbol{v}[2]$, which will allow us to reduce the number of equilibria we need to consider – any equilibrium that does not offer a $\boldsymbol{v}[2]$ chance of satisfying $(\neg \mathit{AvoidCorner} \vee (\neg \mathit{Avoid} \vee \gamma_2(\mathfrak{G}(3/4))))$ would yield Player Two less than $\boldsymbol{v}[2]$ utility.

Let $x_{i,j}$ be the weight Player Two attaches to the square with upper left corner $(i, j)$. Note that Player One's payoff for playing $(i, j)$ with probability 1 is $1 - c_{i,j}$, defining $c_{i,j}$ to be:

*The reader may recall the cover weight of Lemma 3.2.1.*

$$c_{i,j} = x_{i,j} + \frac{x_{i-1,j}}{4} + \frac{x_{i,j-1}}{4} + \frac{x_{i-1,j-1}}{4}.$$

As such, Player Two is obliged to play a strategy where $c_{i,j} \geq \boldsymbol{v}[2]$, else Player One could deviate to that square and cause Player Two to earn less than $\boldsymbol{v}[2]$ utility. In fact, $c_{i,j} = \boldsymbol{v}[2]$ is the best she can hope for as can be seen in the sum below:

$$\sum_{i,j<2^k} c_{i,j} = 1 + \frac{3}{4} = \frac{2^{2k}}{2^{2k}} + \frac{2^{2k} \cdot 3}{2^{2k} \cdot 4} = 2^{2k}\boldsymbol{v}[2].$$

As it happens, this system is linearly independent.

**Lemma 4.2.3.** *Consider the following system of $m^2$ linear equations, taking subtraction in the indices to be modulo $m$:*

$$e_{i,j} = 4x_{i,j} + x_{i-1,j} + x_{i,j-1} + x_{i-1,j-1}.$$

*This system is linearly independent.*

*Proof.* Phrased in the right way, the result follows from Gerschgorin's circle theorem (Gerschgorin [90]). Given a complex $m^2 \times m^2$ matrix $A$, we define $m^2$ discs, one for each row. Disc $i$ has centre $a_{i,i}$, the diagonal entry for that row, and radius $\sum_{j \neq i} |a_{i,j}|$, the sum of the absolute values of the non-diagonal entries. The theorem states that every eigenvalue of $A$ lies in at least one such disc.

As a consequence, if $|a_{i,i}| > \sum_{j \neq i} |a_{i,j}|$ for each row $i$, no eigenvalue can be zero, and hence $A$ is non-singular. The reader will note that if we construct a matrix by placing $e_{i,j}$ in the $(i + kj)$th row, we will obtain a matrix where every row has 4 on the diagonal and three 1s elsewhere in the row, hence allowing us to invoke the circle theorem.

As a side note, I believe the lemma statement still holds with a system of the form $e_{i,j} = 2x_{i,j}+x_{i-1,j}+x_{i,j-1}+x_{i-1,j-1}$. However, it is no longer possible to obtain this result via the circle theorem. As a consequence of Nott and Wilson ([91], but see also Section 2.6 of Kozintsev [92]), this would be equivalent to claiming that $e^{ai\frac{2\pi}{m^2}} + e^{bi\frac{2\pi}{m^2}} + e^{abi\frac{2\pi}{m^2}} \neq -2$ for all $a, b \in \mathbb{Z}$. Whilst this is intuitively plausible, a formal argument proved elusive.[3] Q.E.D.

In other words, there is a unique way to set every $c_{i,j}$ to $\boldsymbol{v}[2]$ – equal randomisation.

We have thus established that in order for Player Two to attain at least $\boldsymbol{v}[2]$ utility it is necessary (though not sufficient) that she randomise equally over all possible squares. Let us thus consider any strategy profile where she does so, but an accepting run of $M$ on $w$ does not exist. First, suppose Player One places positive weight on every entry. As what he describes cannot be the table of an accepting run, we can invoke Lemma 4.2.2 to establish that there must be a $2 \times 2$ square which does not satisfy the requirements in $S$, and hence Player Two loses utility.

Next, suppose Player One attaches zero weight to some entries. As Player Two is randomising equally, she must attach the same weight to square with top left corner $(i, j)$, to which Player One attaches zero weight, as to square with top left corner $(i', j')$, which we choose to be the entry to which Player One attaches the highest weight. This profile cannot be in equilibrium: if $x$ is the probability Player One attaches to $(i', j')$ then by transferring the weight Player Two currently attaches to the square with top left corner $(i, j)$ to the square with top left corner $(i', j')$ she will lose at most $x \cdot \frac{3}{2^{2k} \cdot 4}$ utility, and gain at least $x \cdot \frac{1}{2^{2k}}$ utility – which would be a profitable deviation.

This completes the proof. Q.E.D.

With the proof still fresh in our minds, we can observe that we have in fact proved something stronger than the theorem statement. This will prove useful in future proofs:

**Corollary 4.2.4.** *The following problem is* NEXP*-complete:*

| ∃GUARANTEENASH* | |
|---|---|
| Input: | A Boolean game $G$ and some $v \in [0, 1]_{\mathbb{Q}}$. |
| Output: | YES if $G$ has a rational-valued equilibrium $\boldsymbol{\sigma}$ such that $u_1(\boldsymbol{\sigma}) \geq v$. |

3 This would mean we could use $\mathfrak{G}(1/2)$ in the construction instead. This is desirable because $\mathfrak{G}(1/2)$ has a name, which would allow us to use that name to achieve some illusion of effective academic prose. Is it sufficiently desirable to spend more time trying to prove it? No. No it is not.

*Proof.* The equilibrium we used to witness the payoff constraint being met in the proof above was rational-valued; we are not claiming that every equilibrium satisfying that constraint must be, but that there exists at least one. We have also proved that if the machine does not accept the input string in the time provided, then no equilibrium of the game, rational or otherwise, can meet the constraint.

Only one player's utility was necessary in the construction, so we only need to consider one. Q.E.D.

## 4.3 DISCUSSION

In this chapter we have established the first complexity result about mixed equilibria in Boolean games, and introduced the proof ideas on which all subsequent results in this work will rely on.

The reader will recall that we have chosen to study ∃GuaranteeNash rather than MaxPayoff so that our work could be directly comparable with Schoenebeck and Vadhan [9]. However, at this point we can observe that no generality was lost: Player One's utility in the construction dominates Player Two's, so our proof equally establishes the NEXP-completeness of MaxPayoff.

It is worth asking what this result means, if anything? It is tempting to argue that if the complexity of finding an equilibrium is hard, then we cannot expect the players to find it.[4] There are a number of problems with this approach, not the least of which is that ∃GuaranteeNash has nothing to do with actually finding an equilibrium. Second, there is nothing in the theory of games introduced thus far that provides a mechanism for a game arriving at an equilibrium; equilibrium is a place you stay, not a place you go to.[5] A far more relevant question, then, for players actually playing a game is not ∃GuaranteeNash or even FindNash, but IsNash. Finally, worst-case asymptotic complexity is a poor guide to what is and is not actually feasible. In practice, we are not necessarily interested in games that encode computation histories, and typically an approximate solution is just as good as the real thing.

What this result does suggest to us is that Boolean games are, indeed, an appropriate model of succinct games. This is not as obvious as it may not appear; after all, the extensive form allows us to represent an exponential number of strategies via a compact game tree, but the complexity results suggest that for sufficiently complex games the extensive form, too, blooms exponentially. From a more abstract angle, this is also a further hint as to the source of complexity in games. The fact that two-player games are hard tells us that we do not need a large number of players; win-lose games tell us that we do not need complex gradations of payoff, circuit games tell us that we do not need black-

4 "If your laptop can't find it, neither can the market." -Kamal Jain.
5 Even this is not without its caveats. It's easy to construct examples of intuitively unstable equilibria, e.g. the bank run game.

box functions, but something that we can efficiently compute, and now Boolean games tell us that we do not even need universal computation. Propositional logic is enough.

# 5 VARIABLE-SUM GAMES

*This question the Dodo could not answer without a great deal of thought, and it sat for a long time with one finger pressed upon its forehead, (the position in which you usually see Shakespeare, in the pictures of him,) while the rest waited in silence. At last the Dodo said, "Everybody has won, and all must have prizes."*

— Lewis Carroll, Alice's Adventures in Wonderland

In the last chapter we have punctured the piñata, and it is now that the candy starts to spill. In Section 5.1 we demonstrate that six variants of the pseudo-problem ?Nash, non-trivial questions about Nash equilibria, are NEXP-complete for existential properties and coNEXP-complete for universal properties in two-player Boolean games. In Section 5.2 we address rational and irrational equilibria. Finally in Section 5.3 we consider the problem of determining whether a profile is in equilibrium.

The cleaner proofs are by reduction from ∃GuaranteeNash or its complement, while the less elegant involve replicating parts of the earlier construction, and tweaking them to serve the present ends.

## 5.1 PROPERTIES OF EQUILIBRIA IN TWO-PLAYER GAMES

**Theorem 5.1.1.** UniqueNash *is* coNEXP*-complete for two-player Boolean games .*

*Proof.* We will reduce from the complement of ∃GuaranteeNash. The idea is, given an instance $(G, \boldsymbol{v})$ of ∃GuaranteeNash, the players can either play in $G$ or can unilaterally deviate to a game where they get a utility of $\boldsymbol{v}[1]$ or $\boldsymbol{v}[2]$ in a unique equilibrium.[1] The player who does not deviate gets nothing, so there is no equilibrium where one player chooses $G$ and the other to switch. Hence if ∃GuaranteeNash does not have an equilibrium guaranteeing the players at least $\boldsymbol{v}$ utility, there will be a unique equilibrium where both players deviate.

Let $G$ be a Boolean game and $\boldsymbol{v} \in [0,1]^2_{\mathbb{Q}}$. Let $u = \boldsymbol{v}[1]$ and $w = 1 - \boldsymbol{v}[2]$. Construct $\mathfrak{G}(u)$ and $\mathfrak{G}(w)$ over fresh variables. Recall, from Lemma 3.2.1, that $\mathfrak{G}(u)$ and $\mathfrak{G}(w)$ each have a unique equilibrium. Introduce new *Play* and *Dummy* variables for both players.

We claim that $G'$ with the following parameters has a unique Nash equilibrium if and only if $(G, \boldsymbol{v}) \notin$ ∃GuaranteeNash:

1 "Intuition" being the operative word. In the actual construction the payoffs will be $\boldsymbol{v}[1](1 - \boldsymbol{v}[2])$ and $\boldsymbol{v}[2](1 - \boldsymbol{v}[1])$.

$$\begin{aligned}
\Phi_1 =& \operatorname{var}_1(G) \cup \operatorname{var}_1(\mathfrak{G}(u)) \cup \operatorname{var}_1(\mathfrak{G}(w)) \cup \{\mathit{Play}^1, \mathit{Dummy}^1\},\\
\Phi_2 =& \operatorname{var}_2(G) \cup \operatorname{var}_2(\mathfrak{G}(u)) \cup \operatorname{var}_2(\mathfrak{G}(w)) \cup \{\mathit{Play}^2, \mathit{Dummy}^2\},\\
\gamma_1 =& \gamma_1(\mathfrak{G}(w)) \wedge \Big((\neg \mathit{Play}^1 \wedge \neg \mathit{Play}^2 \wedge \gamma_1(G))\\
& \vee (\mathit{Play}^1 \wedge \neg \mathit{Dummy}^1 \wedge (\bigwedge_{p \in \operatorname{var}_1(G)} \neg p) \wedge \gamma_1(\mathfrak{G}(u)))\Big),\\
\gamma_2 =& \gamma_2(\mathfrak{G}(u)) \wedge \Big((\neg \mathit{Play}^1 \wedge \neg \mathit{Play}^2 \wedge \gamma_2(G))\\
& \vee (\mathit{Play}^2 \wedge \neg \mathit{Dummy}^2 \wedge (\bigwedge_{p \in \operatorname{var}_2(G)} \neg p) \wedge \gamma_2(\mathfrak{G}(w)))\Big).
\end{aligned}$$

First, suppose that $(G, \boldsymbol{v}) \in \exists$GuaranteeNash, and $\boldsymbol{\sigma}$ is an equilibrium that satisfies the payoff criterion. Consider the partial profile $\boldsymbol{\sigma'}$ in $G'$ where players play $\boldsymbol{\sigma}$ over the variables in $G$, set $\mathit{Play}^1, \mathit{Play}^2$ to *false* and play the unique equilibria of $\mathfrak{G}(u), \mathfrak{G}(w)$. We claim that any extension of $\boldsymbol{\sigma'}$ to a full profile (that is, with any distribution over $\mathit{Dummy}^1, \mathit{Dummy}^2$) yields an equilibrium of $G'$. As such, $G'$ does not have a unique equilibrium.

First, observe that any extension of $\boldsymbol{\sigma'}$ to a full strategy profile gives Player One $w\boldsymbol{v}[1]$ utility (the probability of satisfying $\gamma_1(\mathfrak{G}(w))$ times the probability of satisfying $\gamma_1$), and Player Two $(1-u)\boldsymbol{v}[2]$. We will show that no deviation can give the players more than that.

Note that given the play of the opposing player, neither player can hope to increase the probability of satisfying $\gamma_1(\mathfrak{G}(w))$ or $\gamma_2(\mathfrak{G}(u))$. If there is to be a deviation, it will have to increase the probability of satisfying one of the disjuncts on the right. Moreover, as given any assignment to $\Phi_1$ Player One will have to choose whether to set $\mathit{Play}^1$ to *true* or *false*, from his point of view the disjuncts are mutually exclusive. We can thus restrict ourselves to considering the probability with which Player One can satisfy the one or the other.

*Recall, from Fact 2.1.10, that we need only consider pure deviations. Any pure strategy involves picking one disjunct.*

Should Player One choose to satisfy $(\neg \mathit{Play}^1 \wedge \neg \mathit{Play}^2 \wedge \gamma_1(G))$, then his payoff will be bounded by his ability to satisfy $\gamma_1(G)$. As we assumed that $\boldsymbol{\sigma'}|G = \boldsymbol{\sigma}$ is an equilibrium of $G$, Player One cannot increase this unilaterally. If Player One chooses to satisfy $(\mathit{Play}^1 \wedge \neg \mathit{Dummy}^1 \wedge (\bigwedge_{p \in \operatorname{var}_1(G)} \neg p) \wedge \gamma_1(\mathfrak{G}(u)))$, then given Player Two's play over $\operatorname{var}_2(\mathfrak{G}(u))$, Player One can hope for no more than $u$.

Player One can thus choose between $w\boldsymbol{v}[1]$ utility or $wu = w\boldsymbol{v}[1]$. He is indifferent between any such deviation and $\boldsymbol{\sigma'}$, and, mutatis mutandis, so is Player Two. This establishes that every extension of $\boldsymbol{\sigma'}$ is an equilibrium.

Now suppose that $(G, \boldsymbol{v}) \notin \exists$GuaranteeNash. We claim that the only equilibrium of $G'$ involves setting the variables in $G$ as well as $\mathit{Dummy}^1, \mathit{Dummy}^2$ to *false*, $\mathit{Play}^1, \mathit{Play}^2$ to *true* and playing the unique equilibria of $\mathfrak{G}(u)$ and $\mathfrak{G}(w)$ over the remaining variables.

First note that this is indeed an equilibrium – any deviation to satisfy the left disjunct yields a utility of 0 to the deviating player, and given the play of the opponent all deviations in the games $\mathfrak{G}(v)$ and $\mathfrak{G}(w)$ yield the same payoff.

Now consider any other profile $\boldsymbol{\sigma'}$. Note that if the profile realises $\neg \mathit{Play}^1 \wedge \neg \mathit{Play}^2$ with non-zero probability, then $\boldsymbol{\sigma'}|G$ must be an equilibrium of $G$ – were a deviation possible for Player One in $G$, he could replicate that deviation on the variables $\mathrm{var}_1(\mathfrak{G}(u))$ in $G'$ without affecting the rest of the game, which would increase the probability of him satisfying his left disjunct. However this leads directly to a contradiction. Suppose, without loss of generality, that Player One obtains less than $\boldsymbol{v}[1]$ utility in $\boldsymbol{\sigma'}|G$. In this case Player One has a deviation available to him: by setting $\mathit{Play}^1$ to *true*, $\mathit{Dummy}^1$ and all variables in $G$ to *false* he can transfer the weight he is currently assigning to $(\neg \mathit{Play}^1 \wedge \neg \mathit{Play}^2 \wedge \gamma_1(G))$, which yields him less than $\boldsymbol{v}[1]$ utility, to $(\mathit{Play}^1 \wedge \neg \mathit{Dummy}^1 \wedge (\bigwedge_{p \in \mathrm{var}_1(G)} \neg p) \wedge \gamma_1(\mathfrak{G}(u)))$, where he can guarantee himself at least $u = \boldsymbol{v}[1]$ by playing the equilibrium play of $\mathfrak{G}(u)$. This does not affect the probability of him satisfying $\gamma_1(\mathfrak{G}(w))$ and hence increases the probability of satisfying $\gamma_1$.

Next suppose that $\neg \mathit{Play}^1 \wedge \neg \mathit{Play}^2$ is realised with probability 0. We can safely assume that the players set $\mathit{Play}^1, \mathit{Play}^2$ to *true* with probability 1 and $\mathrm{var}_1(G), \mathrm{var}_2(G), \mathit{Dummy}^1, \mathit{Dummy}^2$ with probability 0. The remaining variables are $\mathrm{var}_1(\mathfrak{G}(u)), \mathrm{var}_2(\mathfrak{G}(u)), \mathrm{var}_1(\mathfrak{G}(w))$ and $\mathrm{var}_2(\mathfrak{G}(w))$, and there is a unique way to play those in equilibrium. But that would means $\boldsymbol{\sigma'}$ is precisely the profile we started with initially, and hence the equilibrium must be unique. Q.E.D.

**Corollary 5.1.2.** SecondNash *for two-player Boolean games is* NEXP-*complete.*

*Proof.* By Theorem 2.1.11, we know that every Boolean game $G$ has an equilibrium. If $G \notin$ UniqueNash, it follows that it has at least two. Q.E.D.

**Theorem 5.1.3.** ∃NashSat *for two-player Boolean games is* NEXP-*complete.*

*Proof.* Consider the construction of Theorem 4.2.1. The reader will note that in the equilibrium described in the proof, $\varphi = \gamma_1 \vee \gamma_2$ already holds with probability 1, so we are halfway done. Unfortunately, in the case where $M$ does not accept $w$, there nevertheless can exist an equilibrium where $\gamma_1$ is true with probability 1, and hence $\varphi$ – consider the case where Player One plays a nonsense configuration that Player Two cannot hope to match. Player Two faces 0 utility no matter what she plays, so she has no incentive to guess the same entry as Player One, which could allow Player One to win unchallenged.

The idea then is to alter the game so that Player Two always has an incentive to guess the same entry as Player One, even if the rest of her formula is unsatisfiable.

With this in mind, let $G$ be the game from Theorem 4.2.1 and construct $\mathfrak{G}(1/2)$. We construct a new game $G'$ with:

$$\gamma_1 = \gamma_1(G) \wedge \gamma_1(\mathfrak{G}(1/2)),$$
$$\gamma_2 = \gamma_2(G) \vee (\neg\gamma_1(G) \wedge \gamma_2(\mathfrak{G}(1/2))).$$

The introduction of $\mathfrak{G}(1/2)$ is to give Player Two a little utility for guessing the same entry as Player One, even if she cannot satisfy $\gamma_2(G)$ in its entirety. Player One has to satisfy both $\gamma_1(G)$ and $\gamma_1(\mathfrak{G}(1/2))$ to induce him to always play the equilibrium play in $\mathfrak{G}(1/2)$.

First, observe that it is still the case that $\varphi = \gamma_1(G) \vee \gamma_2(G)$ holds with probability 1 in some equilibrium if $M$ accepts $w$ – the profile where both players describe an accepting run over the entire table and play the unique equilibrium in $\mathfrak{G}(1/2)$.

Now suppose that $M$ does not accept $w$ in $K$ steps. Consider a strategy profile $\boldsymbol{\sigma}$ where both players attach non-zero weight to all entries of the table. As the computation history Player One describes cannot be an accepting run, it must contain an inadmissible square and therefore there exists a non-zero probability that both players will choose that square, invalidating $\gamma_1(G)$, yet Player Two will be unable to satisfy $\gamma_2(G)$, and therefore $\varphi$ is not realised with probability 1. Hence if there does exist an equilibrium where $\varphi$ holds with probability 1, it must be the case that one of the players does not attach non-zero weight to every entry.

Next, consider the case where Player Two does not attach positive weight to all entries. We claim that this profile is not an equilibrium. As Lemma 4.2.3 establishes that there is a unique way to make every entry equally dangerous for Player One, it follows that in this strategy profile some entries are less dangerous than others. That is, we can assume there exist entries $(i, j)$ and $(k, l)$, with Player Two assigning non-zero probability to the square with upper left corner $(i, j)$, such that the expected utility of Player One playing $(i, j)$ with probability 1 is less than that of playing $(k, l)$ with probability 1. As all strategies in the support of Player One's play must have equal expected utility, we can assume that $(i, j)$ is played with probability 0 and $(k, l)$ with probability $p$; we can further choose $(k, l)$ in such a way that $p$ is non-zero.

Now, the expected utility of Player Two playing the square with upper left corner $(i, j)$ is $p_1a_1/4 + p_2a_2/4 + p_3a_3/4$, where $p_1, p_2, p_3$ are the probabilities with which Player One plays $(i + 1, j)$, $(i, j + 1)$ and $(i+1, j+1)$, and $a_1, a_2, a_3$ are 1 if the associated squares are admissible and $1/2$ otherwise. If all of $p_1a_1, p_2a_2, p_3a_3$ are zero, then Player Two can transfer to $(k, l)$ the mass she currently assigns to $(i, j)$, which will increase her payoff as the expected utility of $(i, j)$ is 0 and $(k, l)$ at least $p/2$. If $p_1a_1, p_2a_2, p_3a_3$ are not all zero, without loss of generality suppose that $p_1a_1$ is the largest of these. By transferring to $(i+1, j)$ the mass she currently assigns to $(i, j)$ Player Two will be trading $p_1a_1/4+p_2a_2/4+p_3a_3/4$ utility for at least $p_1a_1$, which is clearly a profitable deviation.

Finally, suppose Player Two does attach positive weight to all entries but Player One does not. If any entry Player One describes cannot be matched by an admissible square we are done – it is possible for $\gamma_1(G)$ and $\gamma_2(G)$ to both fail to hold. If it is possible to tile the portion of the table Player One describes with admissible squares, then we are not in equilibrium as Player Two could deviate: if Player One attaches zero weight to $(i, j)$, Player Two yields $p_1a_1/4 + p_2a_2/4 + p_3a_3/4$ utility from that entry, and she can replicate the deviation we described above. Q.E.D.

**Theorem 5.1.4.** $\forall$NASHSAT *for two-player Boolean games is* coNEXP-*complete.*

*Proof.* The idea is similar to Theorem 5.1.1. Again, we reduce from the complement of $\exists$GUARANTEENASH, having the players play $G$, with the option to unilaterally deviate to another game where Player $i$ is guaranteed a utility of at least $\boldsymbol{v}[i]$. Thus if $(G, \boldsymbol{v}) \notin \exists$GUARANTEENASH some player will always deviate. The formula $\varphi$ will track if a deviation has occurred.

Let $(G, \boldsymbol{v})$ be an $\exists$GUARANTEENASH instance. Let $u = \boldsymbol{v}[1]$ and $w = 1 - \boldsymbol{v}[2]$.

Construct $G'$ as follows:

$$\begin{aligned}
\Phi_1 =& \mathrm{var}_1(G) \cup \mathrm{var}_1(\mathfrak{G}(u)) \cup \mathrm{var}_1(\mathfrak{G}(w)) \cup \{\mathit{Play}^1\},\\
\Phi_2 =& \mathrm{var}_2(G) \cup \mathrm{var}_2(\mathfrak{G}(u)) \cup \mathrm{var}_2(\mathfrak{G}(w)) \cup \{\mathit{Play}^2\},\\
\gamma_1 =& (\neg \mathit{Play}^1 \wedge \neg \mathit{Play}^2 \wedge \gamma_1(G)) \vee (\mathit{Play}^1 \wedge \gamma_1(\mathfrak{G}(u)))\\
&\vee (\mathit{Play}^2 \wedge \gamma_1(\mathfrak{G}(w))),\\
\gamma_2 =& (\neg \mathit{Play}^1 \wedge \neg \mathit{Play}^2 \wedge \gamma_2(G)) \vee (\mathit{Play}^1 \wedge \gamma_2(\mathfrak{G}(u)))\\
&\vee (\mathit{Play}^2 \wedge \gamma_2(\mathfrak{G}(w))).
\end{aligned}$$

We claim that $\varphi = \mathit{Play}^1 \vee \mathit{Play}^2$ is true in every equilibrium with probability 1 if and only if $(G, \boldsymbol{v}) \notin \exists$GUARANTEENASH.

First, suppose that $(G, \boldsymbol{v}) \in \exists$GUARANTEENASH, and let $\boldsymbol{\sigma}$ be an equilibrium satisfying the payoff criterion. Extend $\boldsymbol{\sigma}$ to a profile of $G'$ by having both players play the equilibrium strategies in $\mathfrak{G}(u)$ and $\mathfrak{G}(w)$, and set $\mathit{Play}^1, \mathit{Play}^2$ to *false*. Observe that this is an equilibrium: neither player has incentive to set $\mathit{Play}^1$ or $\mathit{Play}^2$ to *true* as that will yield them $u$ or $w$ utility respectively, which is no more than what they are getting already. If both players set $\mathit{Play}^1, \mathit{Play}^2$ to *false* then neither player has incentive to change their strategies in $\mathfrak{G}(u), \mathfrak{G}(w)$ as that has no effect on their payoff. Finally, no player has incentive to change their distribution over $\mathrm{var}_1(G), \mathrm{var}_2(G)$ as $\boldsymbol{\sigma}$ is an equilibrium of $G$. Now note that this equilibrium satisfies $\varphi$ with probability 0, so $(G', \varphi) \notin \forall$NASHSAT.

*As before, any pure strategy can satisfy at most one disjunct.*

Next, suppose $(G, \boldsymbol{v}) \notin \exists$GUARANTEENASH. Assume, for contradiction, that there exists an equilibrium $\boldsymbol{\sigma}$ of $G'$ where $\varphi$ is satisfied with probability less than 1. Note that this means that both players must assign a non-zero weight to a truth assignment that assigns $\mathit{Play}^1$ and *Play*$^2$ to *false*. This implies that $\boldsymbol{\sigma}|G$ is an equilibrium of $G$: if it were not, there would be a deviation for, say, Player One in $G$ that would increase his probability of satisfying $\gamma_1(G)$. Since this deviation only uses the variables in $\mathrm{var}_1(G)$, Player One could also perform such a deviation in $G'$, as his distribution over $\mathrm{var}_1(G)$ only affects the probability of $\gamma_1(G)$ being satisfied, and not any other formula. Now, since the event $\neg Play^1 \wedge \neg Play^2$ has a non-zero chance of occurring, this means increasing the probability of $\gamma_1(G)$ being satisfied increases the probability of $(\neg Play^1 \wedge \neg Play^2)$ being satisfied, and hence the probability of $\gamma_1$.

Since $(G, \boldsymbol{v}) \notin \exists$GUARANTEENASH, and $\boldsymbol{\sigma}|G$ is an equilibrium of $G$, we can without loss of generality assume that in this equilibrium Player One satisfies $\gamma_1(G)$ with probability $v' < \boldsymbol{v}[1]$. Suppose Player One deviates from $\boldsymbol{\sigma}$ by setting *Play*$^1$ to *true* (with probability 1) and playing the equilibrium strategy in $\mathfrak{G}(u)$. Note that this does not affect the probability of satisfying $(Play^2 \wedge \gamma_1(\mathfrak{G}(w)))$ as we are keeping Player One's play in $\mathfrak{G}(w)$ the same. On the other hand, it decreases the probability of satisfying $(\neg Play^1 \wedge \neg Play^2 \wedge \gamma_1)$ by at most $pv'$, where $p$ is the combined weight Player One previously attached to truth assignments which set *Play*$^1$ to *false*, and en revanche he increases the probability of satisfying $(Play^1 \wedge \gamma_1(\mathfrak{G}(u)))$ by $pu$. As $pu > pv'$ this is a profitable deviation, so $\boldsymbol{\sigma}$ could not have been an equilibrium. Q.E.D.

**Theorem 5.1.5.** *$\forall$GUARANTEENASH is coNEXP-complete for two-player Boolean games.*

*Proof.* Before we prove the theorem, we would like to point out just how trivial it is to establish hardness for the multiplayer variant – determining whether $(G, \varphi) \in \forall$NASHSAT is coNEXP-hard, so just add another player with goal formula $\varphi$ and test whether that player attains 1 utility in every equilibrium. The discerning reader will no doubt infer that the presence of this aside suggests that the argument to follow is somewhat less elegant, so we ask you to bear with us.

We reduce from the complement of NEXPTM. The construction mirrors Theorem 4.2.1. This time, instead of supplying a legal square Player Two attempts to supply an illegal square and, as Lemma 4.2.2 still holds, in the absence of an accepting run an illegal square must exist, giving Player Two some guaranteed payoff that she would not have were Player One to describe the history of an accepting run.

The players have the following goal formulae:

$$\gamma_1 = AvoidC \wedge (Avoid \vee \gamma_1(\mathfrak{G}(3/4))),$$
$$\gamma_2 = (\neg AvoidC \vee (\neg Avoid \wedge \gamma_2(\mathfrak{G}(3/4)))) \wedge Agree \wedge Illegal.$$

With the exception of *Illegal*, the subformulae are as before. *Illegal* asserts that Player Two plays a square, but that square violates some requirement of Lemma 4.2.2.

At this point we will digress to motivate what is to follow. Consider a profile (not necessarily an equilibrium) where both players randomise equally over all entries. If every square Player One describes is admissible, then Player Two can achieve a utility of $\frac{1}{2^{2k}} + \frac{2}{2^{2k+2}}$: no matter what square she names, she can match three of those entries to Player One's and only modify the last to satisfy *Illegal*. In the case where an accepting run exists, there exists an equilibrium of this kind, thus it follows that our value for $\boldsymbol{v}[2]$ would have to be strictly greater than $\frac{1}{2^{2k}} + \frac{2}{2^{2k+2}}$. If Player One describes at least one inadmissible square in the table, as he must if no accepting run exists, then Player Two's utility in this case is at least $v^* = \frac{1}{2^{2k}} + \frac{2}{2^{2k+2}} + \frac{1}{2^{2k+2}2^{2k}}$. I.e., $v^*$ is the value we get if Player Two randomises equally over $1/2^{2k} - 1$ admissible squares and one inadmissible square. If the inadmissible square is realised then Player Two is able to match all four entries to Player One's description, which accounts for the additional $\frac{1}{2^{2k+2}2^{2k}}$ term.

A naïve approach would then be to let $\boldsymbol{v}[2] = v^*$. This is problematic as there is no guarantee that Player One will distribute over every square equally in equilibrium. He may choose to attach less weight to the inadmissible square, and thereby deny Player Two the additional utility. However, a rising tide lifts all boats – the weight Player One chooses to place everywhere will increase the probability of some other square being realised, which will increase Player Two's utility to some $v'$. This $v'$ will be smaller than $v^*$, but nevertheless strictly greater than $\frac{1}{2^{2k}} + \frac{2}{2^{2k+2}}$.

Now suppose $(M, w, K) \notin$ NEXPTM. If every $2 \times 2$ square in the described table is inadmissible we are done – Player Two's utility is at least $v^*$, as that is what she can get by randomising equally throughout the table. Suppose then there exists an admissible square, with upper left corner $(i, j)$, and an inadmissible square, with upper left corner $(k, l)$. Let $p_{i,j}$ be the probability with which Player One plays entry $(i, j)$. Note that the utility Player Two would get by playing $(k, l)$ with probability 1 is $p_{k,l} + p_{k+1,l}/4 + p_{k,l+1}/4 + p_{k+1,l+1}/4$, and for playing $(i, j)$ is $p_{i,j} + p_{i+1,j}/4 + p_{i,j+1}/4$ (we are, without loss of generality, assuming that $p_{i+1,j}, p_{i,j+1} \geq p_{i+1,j+1}$. Player Two can pick whichever of the two entries she wants to match, and it stands to reason that she will pick those most likely to be realised). We furthermore assume that we picked an $(i, j)$ such that $p_{i,j} + p_{i+1,j}/4 + p_{i,j+1}/4$ is maximal. These are the lower bounds on Player Two's utility in every equilibrium in which Player One plays with probabilities $p_{0,0}, \ldots, p_{2^k-1,2^k-1}$ (recall, every pure strategy in the support of Player Two's strategy must maximise her utility).

If it is the case that $p_{k,l} + p_{k+1,l}/4 + p_{k,l+1}/4 + p_{k+1,l+1}/4 \geq v^*$, there is nothing more to be said. Suppose that this is not the case. This where the rising tide comes in to play; if Player One does not attach enough mass to the entries $(k, l)$ through $(k + 1, l + 1)$ to give Player Two $v^*$ utility, he must attach that excess mass somewhere else. The question we now answer is, how much mass do we have to remove from this sum (relative to the equal distribution of $1/2^{2k}$ per entry) in order for it to fall below $v^*$?

We assume this transfer is done adversarially, and hence the mass is first removed from $p_{k,l}$ (if this is not the case, all it means is that the $\delta$ we will arrive at will be larger, which will still establish the theorem). This leads us to:

*The 3 is in the numerator on the left because we remove mass from $p_{k,l}$ before any other weight.*

$$p_{k,l} + \frac{3}{2^{2k+2}} < \frac{1}{2^{2k}} + \frac{2}{2^{2k+2}} + \frac{1}{2^{2k+2}2^{2k}},$$
$$p_{k,l} < \frac{1}{2^{2k}} - \frac{1}{2^{2k+2}} + \frac{1}{2^{2k+2}2^{2k}}.$$

So the total mass transfer is at least $\delta = \frac{1}{2^{2k+2}} - \frac{1}{2^{2k+2}2^{2k}}$. Now observe that this $\delta$ must be shared out in some way among the entries that are not $(k,l), (k,l+1), (k+1,l), (k+1,l+1)$. This raises the average weight of those entries to $1/2^{2k} + \delta/2^{2k} - 4$. As we picked $p_{i,j} + p_{i+1,j}/4 + p_{i,j+1}/4$ to be maximal, we can assume this sum to be at least:

$$\frac{1}{2^{2k}} + \frac{\delta}{2^{2k}-4} + \frac{2}{2^{2k+2}} + \frac{2\delta}{2^{2k+2}-16}.$$

We now claim that $(M, w, K) \in$ NEXPTM if and only if $(G, \boldsymbol{v}) \notin$ ∀GuaranteeNash, for the following $\boldsymbol{v}$:

$$\boldsymbol{v}[1] = 0,$$
$$\boldsymbol{v}[2] = \frac{1}{2^{2k}} + \frac{\delta}{2^{2k}-4} + \frac{2}{2^{2k+2}} + \frac{2\delta}{2^{2k+2}-16}.$$

One direction we have already done. It remains to see what happens when $(M, w, K) \in$ NEXPTM.

Consider the strategy profile where Player One randomises over all squares equally, describing the accepting run, and Player Two randomises equally over all entries in a manner where the upper left, upper right and lower left entry of each square agrees with Player One and the lower right disagrees in order to ensure the described square is not admissible. The utilities derived by the players are:

$$u_1 = 1 - \frac{1}{2^{2k}} - \frac{3}{2^{2k+2}},$$
$$u_2 = \frac{1}{2^{2k}} + \frac{2}{2^{2k+2}}.$$

This is also the most they can hope to gain by deviating to any pure strategy, so this is an equilibrium, and as $u_2 < \boldsymbol{v}[2]$ we have $(G, \boldsymbol{v}) \notin$ ∀GuaranteeNash. Q.E.D.

## 5.2 RATIONAL AND IRRATIONAL EQUILIBRIA

The question of whether a game has a rational equilibrium is of a fundamentally different nature to the questions studied so far: the two-player variant can be solved in constant time by a machine that ignores the input and prints "YES". We need at least three players to have any chance of a hardness result. The proof below, in fact, uses four, as that makes the argument more transparent; a three-player variant involves absorbing the rôle of the fourth player by the first three players, and at that point the logic of the construction becomes harder to follow. We relegate that argument to the appendix.

**Theorem 5.2.1.** RationalNash *for four-player Boolean games is* NEXP*-hard.*

*Proof.* Theorem 3 in Bilò and Mavronicolas [36] gives an example of a three-player win-lose game, call it $G'_1$, that has only irrational equilibria. Players One and Two have two strategies each, but Player Three has three, and as such the game as given does not have a Boolean representation. However, the game has the *positive utility property*: for any choice of strategies by two players, the third has a response that will yield him strictly positive utility. We can thus extend $G'_1$ into $G_1$ that has a fourth strategy for Player Three, which operates as follows:

1. Player Three's payoff for choosing the fourth strategy is zero.
2. Player One and Two's payoff from any profile where Player Three chooses the fourth strategy are the same as if Player Three chose his first strategy instead.

*This is an arbitrary choice to ensure that payoffs for One and Two remain defined even in "illegal" play.*

Observe that in no equilibrium of $G_1$ would Player Three attach positive weight to his fourth strategy – the positive utility property of the game ensures that he can always do better than zero utility. Thus $G_1$ does not introduce any new equilibria; they are the same as the equilibria of $G'_1$, and ergo irrational. As every player now has either two or four strategies, $G_1$ has a Boolean representation.

We will reduce from ∃GuaranteeNash*, from Corollary 4.2.4. Let the pair $(G_2, v)$ be an instance of ∃GuaranteeNash* with three players.

The idea of the construction is to let the players choose to play in $G_1$ or $G_2$ in such a way that every equilibrium involves all the players choosing the same game to play in, and that there is an equilibrium in which the players choose to play in $G_2$ if and only if $(G_2, v)$ is a positive instance of ∃GuaranteeNash*. This will prove the theorem – if $(G_2, v) \in$ ∃GuaranteeNash* then there is an equilibrium where the players play in $G_2$, and hence there is a rational-valued equilibrium (namely, the equilibrium of $G_2$ that guarantees Player One a payoff of $v$). If $(G_2, v) \notin$ ∃GuaranteeNash* then every equilibrium of the game must be in $G_1$, and those are irrational.

The fourth player, $O$, exists only to serve as an opponent in the corresponding gadget games.

The game is defined as follows:

$$\Phi_{i\leq 3} = \mathrm{var}_i(G_1) \cup \mathrm{var}_i(G_2) \cup \{\, \mathit{Choice}_i \,\} \cup \mathrm{var}_1(\mathfrak{G}_i),$$
$$\Phi_O = \mathrm{var}_2(\mathfrak{G}_1(v)) \cup \mathrm{var}_2(\mathfrak{G}_2(1/2)) \cup \mathrm{var}_2(\mathfrak{G}_3(1/2)),$$

$$
\begin{aligned}
\gamma_1 &= \Big(\bigwedge_{i \leq 3} \mathit{Choice}_i \wedge \gamma_1(G_2)\Big) \vee \Big(\bigwedge_{i \leq 3} \neg \mathit{Choice}_i \wedge (\gamma_1(G_1) \vee \gamma_1(\mathfrak{G}(v)))\Big) \\
&\quad \vee \big(\neg \mathit{Choice}_1 \wedge (\mathit{Choice}_2 \vee \mathit{Choice}_2) \wedge \gamma_1(\mathfrak{G}_1(v))\big), \\
\gamma_2 &= \Big(\bigwedge_{i \leq 3} \mathit{Choice}_i \wedge (\gamma_2(G_2) \vee \gamma_1(\mathfrak{G}_2(1/2)))\Big) \vee \Big(\bigwedge_{i \leq 3} \neg \mathit{Choice}_i \wedge \gamma_2(G_1)\Big) \\
&\quad \vee \big(\neg \mathit{Choice}_2 \wedge (\mathit{Choice}_1 \vee \mathit{Choice}_3) \wedge \gamma_1(\mathfrak{G}_2(1/2))\big), \\
\gamma_3 &= \Big(\bigwedge_{i \leq 3} \mathit{Choice}_i \wedge (\gamma_3(G_2) \vee \gamma_1(\mathfrak{G}_3(1/2)))\Big) \vee \Big(\bigwedge_{i \leq 3} \neg \mathit{Choice}_i \wedge \gamma_3(G_1)\Big) \\
&\quad \vee \big(\neg \mathit{Choice}_3 \wedge (\mathit{Choice}_1 \vee \mathit{Choice}_2) \wedge \gamma_1(\mathfrak{G}_3(1/2))\big), \\
\gamma_O &= \gamma_2(\mathfrak{G}_1(v)) \wedge \gamma_2(\mathfrak{G}_2(1/2)) \wedge \gamma_2(\mathfrak{G}_3(1/2)).
\end{aligned}
$$

Intuitively, what we are doing is separating the pure[2] profiles into three categories: the players coordinate on $G_1$, $G_2$, or a miscoordination occurs. It should be clear how the $\mathit{Choice}_i$ variables implement this. The purpose of the gadget games is perhaps a bit more opaque at the moment, but observe that since winning $\mathfrak{G}(v)$ is a disjunctive for Player One in the case of $G_1$ it acts as a floor on his utility in that state, whereas being conjunctive in the case of miscoordination it acts as a ceiling. For Players Two and Three, we likewise add a floor of $1/2$ to $G_2$, and a ceiling of $1/2$ to miscoordination.

*Do note that only the players opting to play in $G_1$ get any utility from miscoordination.*

Suppose $(G_2, v) \in \exists$GuaranteeNash*. We claim there is a rational-valued equilibrium where Players One through Three set $\mathit{Choice}_i$ to *true*, play the rational-valued equilibrium of $G_2$ that satisfies the payoff constraint over $\mathrm{var}_i(G_2)$, the equilibrium strategy over $\mathrm{var}_1(\mathfrak{G}_i)$ and every other variable to *false*. Player $O$ plays the equilibrium strategy in all his gadget games.

Player $O$ has no incentive to deviate as he wins if and only if he wins three independent games, and all those are currently in equilibrium. Player One has no incentive to deviate while $\mathit{Choice}_1$ is *true*: he is indifferent about what he does with $\mathrm{var}_1(G_1)$ and $\mathrm{var}_1(\mathfrak{G}_1(v))$ as those variables do not affect his ability to satisfy $\gamma_1(G_2)$, and he has no incentive to deviate over $\mathrm{var}_1(G_2)$ as $G_2$ is in equilibrium. Should he set $\mathit{Choice}_1$ to *false*, then his utility will depend only on his ability to satisfy $\gamma_1(\mathfrak{G}_1(v))$. The probability of that is $v$, the "ceiling" which we have set earlier, and that is at most how much he is getting in the current profile. Player Two (symmetrically, Three) has no incentive to deviate over $\mathrm{var}_1(\mathfrak{G}_2(1/2))$ or $\mathrm{var}_2(G_2)$ as those games are in equilibrium, and the variables in $\mathrm{var}_2(G_1)$ do not affect her current utility. If she deviates by setting $\mathit{Choice}_2$ to *false*, then she will be getting a utility of $1/2$; whereas in the current profile she is getting $1 - 1/2 \cdot x$, where $x$ is her probability of losing $G_2$. As $x$ is at most 1, $1/2$ is indeed the floor on her utility, and hence such a deviation could do no better.

Now suppose $(G_2, v) \notin \exists$GuaranteeNash*. There are two cases to consider. In the first, $G_2$ does have equilibria for which $u_1(\boldsymbol{\sigma}) \geq v$, but all such equilibria are irrational. In the second, $G_2$ has no equilibrium

2 Modulo equilibrium play in the gadget games.

which affords Player One a utility of at least $v$. We will deal with each as it arises, by considering an arbitrary equilibrium of $G$ and demonstrating that no matter what the case, the equilibrium cannot be rational.

Consider an equilibrium of $G$ where $p, q$ and $r$ are the probabilities of players One, Two and Three respectively opting to play in $G_2$, i.e. the marginal probabilities of setting $Choice_i$ to *true*. We argue that in every equilibrium all three must be zero or non-zero – the player opting to play in $G_2$ gets no utility from miscoordination, so if a play in $G_2$ has no hope of realising (because one of $p$, $q$, or $r$ is 0), any pure strategy that sets $Choice_i$ to *true* has a payoff of zero and hence can only be included in an equilibrium if the player receives a payoff of zero from *every* strategy available to him. From the point of view of Two (mutatis mutandis, Three) this is impossible: she could expect $1/2$ from miscoordination and something strictly positive in $G_1$ were she to change her strategy to one setting $Choice_2$ to *false*. From the point of One we could have a problem where $v = 0$, $q = 0$ and $r = 1$, as he gets a utility of 0 no matter what he plays, but that situation cannot arise as if $q = 0$ and $r = 1$ the profile cannot be in equilibrium – Three would deviate by setting $Choice_3$ to *false*. Hence, if $q = 0$ and $r = 0$, then Player One also has a profitable deviation to whatever strictly positive payoff he can get in $G_1$. We need not consider the case where $p = q = r = 0$, as the equilibrium is clearly irrational. Let us then assume that they are not all 0.

This allows us to handle the case where $G_2$ has no equilibrium that affords Player One a utility of at least $v$ – his current utility would be $pqr \cdot y + (1-p)(1-(1-q)(1-r)) \cdot v + (1-p)(1-q)(1-r) \cdot x$ (payoff from $G_2$, from miscoordination, and from $G_1$), with $y$ being strictly less than $v$ and $x$ being at least $v$. If he were to set $p$ to 0, he would trade $pqry$ for a weighted basket of $v$ and $x$, both which are at least as large as $y$. Thus if $(G_2, v) \notin \exists$GUARANTEENASH* and $G_2$ has no equilibria with $u_1(\boldsymbol{\sigma}) \geq v$, then $G$ has no rational equilibria.

Now suppose that $G_2$ does have an irrational $\boldsymbol{\sigma}$ in which $u_1(\boldsymbol{\sigma}) \geq v$. Recall that Player One's utility is $pqr \cdot y + (1-p)(1-(1-q)(1-r)) \cdot v + (1-p)(1-q)(1-r) \cdot x$. If $y < v$, then the same argument applies as before. If $y \geq v$, this can only be the case if the players are playing an irrational equilibrium in $G_2$, which establishes that the equilibrium of $G$ under consideration cannot be rational. This completes the proof. Q.E.D.

**Theorem 5.2.2.** RATIONALNASH *for three-player Boolean games is* NEXP*-hard.*

*Proof.* See Appendix A.1. Q.E.D.

IRRATIONALNASH is thematically similar to RATIONALNASH, but as we shall show it is in fact no different from the problems studied in the previous section. While it is true that non-degenerate[3] two-player games have an odd number of rational equilibria, in which case IrrationalNash would be trivial, restricting our study to non-degenerate games only is hard to justify, given that determining which class a game belongs to appears to be no easy task. Once we admit degenerate games into consideration then we find that a two-player game can have any number of equilibria and, critically, it has an irrational equilibrium if and only if it has infinitely many of them.

**Fact 5.2.3.** *A two-player game has an irrational equilibrium if and only if it has an infinite number of equilibria.*

*Proof.* Recall Lemma 3.3.3. If the support pair $(X, Y)$ has a unique equilibrium, it must be rational. If it has an infinite number of equilibria, it has a continuum of such, and hence an irrational equilibrium. Q.E.D.

In other words: if we want to demonstrate that a game has an irrational equilibria, we need to demonstrate that it has an infinite number of them.

**Theorem 5.2.4.** IrrationalNash *is* NP*-complete for two-player games in normal form, and* NEXP*-complete for two-player Boolean games.*

*Proof.* For membership, guess a support pair $(X, Y)$ and verify that it induces infinitely many equilibria.

For hardness, we reduce from ∃GuaranteeNash*. Let $(G, v)$ be a game and a payoff constraint. Without loss of generality, suppose $G$ has non-negative payoffs. For the normal form case, simply extend $G$ into a $G'$ in the following way:

1. Duplicate each of Player One's strategies. That is, for every $s_i \in S_1$ in $G$, $G'$ has $s_i$ and $s_i'$, with the property that $u_j(\boldsymbol{s}_{-1}(s_i)) = u_j(\boldsymbol{s}_{-1}(s_i'))$ for all $\boldsymbol{s}$ and $j \in \{1, 2\}$.
2. Give Player One a new strategy, $a$, that yields him a utility of $v$ in every profile. Give Player Two a new strategy $b$, with the property that $u_2(a, b) = 1$, $u_2(s \neq a, b) = -1$, and $u_2(a, s \neq b) = -1$.

If $(G, v) \in$ ∃GuaranteeNash* then $G'$ has infinitely many equilibria – any equilibrium $\boldsymbol{\sigma}$ of $G$ satisfying the payoff constraint generates infinitely many equilibria in $G'$, as Player One can freely distribute strategy weights between duplicate strategies. If $(G, v) \notin$ ∃GuaranteeNash* then $G'$ has a unique equilibrium: $(a, b)$. Any profile that yields Player One less than $v$ utility cannot be an equilibrium as Player One can

3 The definition of degeneracy – the existence of more than $k$ best pure responses to a mixed strategy with support size $k$ – is difficult to get an intuitive appreciation for, and does not concern us in the present work. It suffices to say that whilst statistically "almost all" games are non-degenerate, in practice the 0 probability event of encountering a degenerate game happens too often to be dismissed out of hand. On the other hand, if an approximate answer is sufficient, one need only perturb a degenerate game's payoffs by a random $\epsilon$ and one is good to go.

deviate to $a$, and any profile that gives him more cannot be an equilibrium of $G$, and hence Player Two would have a profitable deviation. By Fact 5.2.3, $G'$ has an irrational equilibrium if and only if $G$ has an equilibrium whence Player One derives a utility of at least $v$.

For the case of Boolean games, let $(G, v)$ be a game and payoff constraint as before. Construct $\mathfrak{G}(v)$, and let $G'$ be the following:

$$\begin{aligned}
\Phi'_1 =& \Phi_1 \cup \{\, \mathit{Dummy}, \mathit{Choice}^1 \,\} \cup \mathrm{var}_1(\mathfrak{G}(v)),\\
\Phi'_2 =& \Phi_2 \cup \{\, \mathit{Choice}^2 \,\} \cup \mathrm{var}_2(\mathfrak{G}(v)),\\
\gamma'_1 =& (\gamma_1 \wedge \mathit{Choice}^1 \wedge \mathit{Choice}^2)\\
& \vee (\gamma_1(\mathfrak{G}(v)) \wedge \neg \mathit{Choice}^1 \wedge \neg \mathit{Dummy} \wedge \bigwedge_{p \in \Phi_1} \neg p),\\
\gamma'_2 =& (\gamma_2 \wedge \mathit{Choice}^1 \wedge \mathit{Choice}^2)\\
& \vee (\gamma_2(\mathfrak{G}(v)) \wedge \neg \mathit{Choice}^1 \wedge \neg \mathit{Choice}^2 \wedge \bigwedge_{p \in \Phi_2} \neg p).
\end{aligned}$$

*The asymmetry is intentional: Player One needs only $\neg \mathit{Choice}^1$ in his right disjunct, Player Two needs both.*

If $(G, v) \in \exists$GUARANTEENASH$^*$ then $G'$ has infinitely many equilibria – the *Dummy* variable does not occur in $\gamma_1$, and thus has the same effect as duplicating each of Player One's strategies did. If $(G, v) \notin \exists$GUARANTEENASH$^*$ then the unique equilibrium involves the players setting the variables in $\Phi_1$, $\Phi_2$, *Dummy*, $\mathit{Choice}^1$ and $\mathit{Choice}^2$ to *false* and playing the equilibrium strategies in $\mathfrak{G}(v)$. This is because any equilibrium of $G$ would yield Player One less than $v$ utility, and he can unilaterally deviate to $\mathfrak{G}(v)$ where he can guarantee himself at least $v$. Once in $\mathfrak{G}(v)$ the equilibrium is guaranteed as neither player can unilaterally deviate from $\mathfrak{G}(v)$ to $G$ – both *Choice* variables appear in their left disjuncts.

$G'$ thus has infinitely many equilibria if and only if $G$ has an equilibrium where Player One derives at least $v$ utility. With Fact 5.2.3, this proves the theorem. Q.E.D.

As a side note, the zero-sum version can be solved by a call to LINEARPROGRAMMING.

**Proposition 5.2.5.** IRRATIONALNASH *is in* P *for two-player zero-sum games in normal form, and* EXP *for two-player zero-sum Boolean games.*

*Proof.* See Appendix A.1. Q.E.D.

## 5.3 TESTING FOR EQUILIBRIUM

In this section we will consider the complexity of determining whether a given mixed-strategy profile is an equilibrium. The difficulty of the problem stems from the fact that the input assumes that only the non-zero strategy weights are given (otherwise the input would be large enough for us to expand the game to normal form in polynomial time), and that the number of players could be large (the problem is coNP-complete for $k$-player games. See Corollary B.2.4 in the appendix).

Our hardness proof is based on Theorem 5.5 of Schoenebeck and Vadhan [9], the main difference being that wherever the authors recourse to universal computation, we must instead explicitly construct a formula of propositional logic serving the same end.

The relevant complexity class is $\text{coNP}^{\#\text{P}}$, i.e. the complement of those decision problems solved by a nondeterministic polynomial-time machine with access to an oracle to #P. The relevant canonical problem is the following:

| $\text{NP}^{\#\text{P}}\text{TM}$ | |
|---|---|
| Input: | A nondeterministic, polynomial-time machine $M$ with an oracle to #P, an input word $w$, and a computation bound in unary $k$ (i.e., the string $1^k$). |
| Output: | YES if $M$ accepts $w$ in at most $k$ steps, NO otherwise. |

**Fact 5.3.1.** $\text{NP}^{\#\text{P}}\text{TM}$ *is complete for* $\text{NP}^{\#\text{P}}$.

The proof of Schoenebeck and Vadhan [9] relies on a convenient characterisation of $\text{NP}^{\#\text{P}}\text{TM}$ in Lemma 5.7. We have need of a modified version of the same:

*Caveat lector: this is the only section of the thesis where $n$ does not denote the number of players. The Latin alphabet has only so many letters.*

**Lemma 5.3.2.** *For every $L \in \text{NP}^{\#\text{P}}$, there is a nondeterministic polynomial-time machine $M$ that decides $L$, and that makes just a single #*Sat *oracle query on any accepting computation. Moreover, the formula in this query is of the form:*

$$\boldsymbol{OneOf}(p_1, \ldots, p_n) \wedge \bigvee_{i \leq n} (\varphi_i \wedge p_i),$$

*where each $\varphi_i$ is a formula in* 2CNF*.*

*Proof.* Let $M$ be a nondeterministic machine with a #P oracle. Recall that #2CNFSat is #P-complete (Valiant [94]), so we can assume that $M$ only submits formulae in 2CNF to the oracle.

Construct an $M'$ that guesses an accepting computation history of $M$ including any oracle queries made, which we denote $\varphi_1$ through $\varphi_n$, and the answers to those queries. Having done this, $M'$ constructs a $\varphi$ with the property that given the number of satisfying truth assignments of $\varphi$ it is possible, in polynomial time, to derive the number of satisfying truth assignments of each of the $\varphi_i$. $M'$ will then submit $\varphi$ to the oracle, verify that its previous guesses for the oracle answers were correct, and if so, accept. Clearly such an $M'$ will accept $w$ in a polynomial number of steps if and only if $M$ does. It remains to show how such a $\varphi$ can be constructed.

Let $k_i$ be the number of propositional variables in $\varphi_i$ (we, of course, assume that the variables used by different formulae are disjoint). Expand each $\varphi_i$ to $\varphi'_i$ in the following manner:

$$\varphi'_i = \varphi_i \wedge \bigwedge_{j < k_1 + \cdots + k_{i-1}} (q_j \vee \neg q_j), \text{ for fresh } q_j.$$

Note that all these new variables are completely redundant – any assignment to them will satisfy the clauses in which they appear. As a result, if $\varphi_i$ has $m_i$ satisfying truth assignments, then $\varphi_i'$ has $2^{k_1+\cdots+k_{i-1}}m_i$. This means that the binary integer representing the number of satisfying truth assignments of $\varphi_i'$ will have 0s in the $k_1 + \cdots + k_{i-1}$ least significant positions, potentially non-zero entries in the next $k_i$, and nothing beyond that.

Given fresh variables $p_i$, our desired formula is the following:

$$\varphi = \boldsymbol{OneOf}(p_1, \ldots, p_n) \wedge \bigvee_{i \leq n} (\varphi_i' \wedge p_i).$$

Since one and only one of $p_i$ must be true, these variables do not introduce any new satisfying truth assignments and thus the number of assignments that satisfy $\varphi$ is simply $\sum_{i \leq n} 2^{k_1+\cdots+k_{i-1}}m_i$. To extract $m_i$ we need only to look at the bits between positions $k_1 + \cdots + k_{i-1}$ and $k_1 + \cdots + k_{i-1} + k_i$. Q.E.D.

Lemma 3.2.1 has served us well so far, as the subgames $\mathfrak{G}(u)$ have allowed us to pretend we were working in a setting with finer grained preferences than Boolean games. The shortcoming of the construction, however, is that by definition $\mathfrak{G}(u)$ is completely modular and separate from the rest of the game. This, of course, is working as intended – when we come to evaluate the payoffs we want to know that the value of $\mathfrak{G}(u)$ is indeed $u$, regardless of what else has happened in the game. The trouble is that $\mathfrak{G}(u)$ is completely unresponsive to the passage of play; we need to know, in advance, what utility values we need and code them in explicitly. Should we want to associate a different utility to every possible strategy we would need formulae of the form $(p_1 \wedge \neg p_2 \wedge \cdots \wedge p_k) \rightarrow \gamma_1(\mathfrak{G}(u_i))$, which would not be feasible in polynomial time.

To deal with this we introduce a different type of gadget game, the value of which can be altered externally.

**Definition 5.3.3.** Fix a sequence $\overline{u_n}$. Let $\mathcal{G}(\overline{u_n})$ denote a two-player, zero-sum game with value $[\![\overline{u_n}]\!]/2^n$, with Player One controlling the variables $\overline{u_n}$. That is, the value of the game is determined *after* Player One chooses an assignment to $\overline{u_n}$.

Formally, $\mathcal{G}(\overline{u_n})$ looks as follows:

$$\begin{aligned}
\Phi_1 =& \{p_1, \ldots, p_n, q_1, \ldots, q_n, s_1, \ldots, s_n, t_1, \ldots, t_n, u_1, \ldots, u_n\} \\
\Phi_2 =& \{r_1, \ldots, r_n\} \\
\gamma_1 =& (\boldsymbol{Sub}(\overline{q_n}, \overline{p_n}, \overline{u_n}) \wedge \boldsymbol{LessEq}(\overline{q_n}, \ulcorner \mathbf{2^n - 1} \urcorner) \wedge \boldsymbol{LessEq}(\overline{r_n}, \overline{q_n}) \\
& \wedge \boldsymbol{LessEq}(\overline{p_n}, \overline{r_n})) \vee \Big(\boldsymbol{Add}(\overline{s_n}, \overline{t_n}, \overline{u_n})) \wedge \boldsymbol{Sub}(\overline{q_n}, \ulcorner \mathbf{0} \urcorner, \overline{s_n}) \\
& \wedge \boldsymbol{Sub}(\ulcorner \mathbf{2^n - 1} \urcorner, \overline{p_n}, \overline{t_n}) \wedge (\boldsymbol{LessEq}(\overline{r_n}, \overline{q_n}) \vee \boldsymbol{LessEq}(\overline{p_n}, \overline{r_n}))\Big) \\
& \vee \boldsymbol{Less}(\ulcorner \mathbf{2^n - 1} \urcorner, \overline{r_n}).
\end{aligned}$$

■

**Example 5.3.4.** At first glance Definition 5.3.3 may appear a little odd. We are essentially letting Player One rig the dice before he casts them – what behaviour can we possibly expect except that of choosing $[\![\overline{u_n}]\!] = 2^n - 1$ every single time? And if we do want a game of that form, we have already seen how to construct $\mathfrak{G}(\frac{2^n-1}{2^n})$ earlier.

The merit of this definition is that, unlike in Lemma 3.2.1, we do *not* require that all the variables of $\mathcal{G}(\overline{u_n})$ are fresh. We use parameter notation for $\overline{u_n}$ to indicate that these variables can occur elsewhere in the construction. Suppose, for example, that Player One is trying to satisfy a formula of the form:

$$\gamma_1 = \varphi(\overline{v_n}) \wedge \boldsymbol{Equal}(\overline{v_n}, \overline{u_n}) \wedge \gamma_1(\mathcal{G}(\overline{u_n})).$$

Now the choice of $[\![\overline{u_n}]\!]$ will very much depend on what is going on in $\varphi(\overline{v_n})$. ■

**Theorem 5.3.5.** IsNash coNP$^{\#\mathrm{P}}$*-hard for Boolean games.*

*Proof.* We follow the argument of Schoenebeck and Vadhan [9], but as we work in a more restricted setting the coding will be more involved.

The idea is as follows: given a nondeterministic $M$ with access to a #Sat oracle, a $w$, and $1^k$ we construct a game where Player One will choose between specifying a computation history of the machine (which will of course include the oracle query and answer received, if a query is made), or to abstain from the game and earn some base level of utility $u$. If Player One chooses to specify a computation history, the payoffs of the game will be so structured as to give Player One more than $u$ utility if he specifies an accepting history with a valid oracle response, or at most $u$ if either the specified history is non-accepting or the oracle response specified by Player One is incorrect.

If we manage, in polynomial time, to construct such a game and profile then we will have proved the theorem – we map $(M, w, 1^k)$ to $(G, \boldsymbol{\sigma})$ where $\boldsymbol{\sigma}$ will involve Player One abstaining. Thereby, if $(M, w, 1^k) \notin$ NP$^{\#\mathrm{P}}$TM then Player One cannot deviate to any strategy that will yield him more than $u$ utility, so the profile will be in equilibrium and thus $(G, \boldsymbol{\sigma}) \in$ IsNash. On the other hand, if $(M, w, 1^k) \in$ NP$^{\#\mathrm{P}}$TM then Player One will be able to deviate to an accepting computation history with a valid oracle response, netting more than $u$ utility and establishing that $(G, \boldsymbol{\sigma}) \notin$ IsNash.

In total $G$ will have $4 + 6k$ players, although Player One is the only one whose preferences will matter. Player Two will play a zero-sum game of value $u$ with Player One. The next $2k$ players will be dubbed group $A$, of which $k$ are called *proposition-players* and $k$ *clause-players*. Group $B$ consists of the next $2k$ players with the same division into proposition and clause-players. Groups $A$ and $B$ will serve to punish Player One for playing an incorrect oracle response. The final $2k + 2$ players are group $C$, and their rôle is to enable the constructed profile $\boldsymbol{\sigma}$ to be polynomial in size, as will be made clear at the end of the proof.

To facilitate the reduction we will augment the machine with registers that deal with the query to, and the response from, the oracle. The output registers will be read-only, and the input registers write-once.

The output registers are simply $k$ cells that will take a binary representation of the oracle's answer, with the least significant bit first, concluding with an end-of-file character $\perp$. As $k$ is an upper bound on the size of the formula submitted to the oracle, the formula can have at most $2^k$ satisfying assignments and as such $k$ cells are sufficient to store whatever output the oracle makes.

To understand the shape of the input registers, consider the form of the formula in the statement of Lemma 5.3.2. Such a formula is uniquely determined by the subformulae $\varphi_i$, and each $\varphi_i$ is in 2CNF. As such, each $\varphi_i$ can be specified simply by listing the literals in the order in which they appear. We will thus endow the machine with $k^2 + 1$ input register cells: $k$ sets of registers, each consisting of $k$ cells, and a header register. The interpretation of the $i$th register is $\varphi_i$, and the $j$th entry of the $i$th register specifies the $j$th literal of $\varphi_i$. The header register will contain a special integer symbol corresponding to the amount of $\varphi_i$ formulae that are present. To facilitate the encoding of literals we assume the machine alphabet contains the symbols $\{ q_1, \ldots, q_k, \neg q_1, \ldots, \neg q_k \}$ as well as the integer symbols $\{ 1, \ldots, k \}$.

The machine will write to an input register by placing the appropriate symbol at the beginning of the tape, followed by the index of the register to write to, and entering a special write state.

At this point it behoves us to summarise the assumptions we have made to convince the reader that at no point have we assumed away the issue at hand – namely, none of the restrictions we place on $M$ are substantial enough to prevent it from representing an arbitrary polynomial-time machine.

1. We assume the machine issues at most one query to the oracle, and the query is in a very specific format. This is justified by Lemma 5.3.2.

2. We assume the machine is equipped with registers. As random access memory would at most allow a polynomial speed up of the machine, there is no generality lost in assuming that an arbitrary polynomial-time machine has these.

*Obviously this only applies to classes in* PSPACE*, which is the case here.*

3. We assume the machine has to write the query formula, symbol by symbol, to the input registers, rather than just dumping it on the tape. This is at worst a polynomial slow down, so the machine is still polynomial time.

4. We assume the machine has additional tape symbols. This will allow at most a polynomial speed up over a machine working only in binary.

We will introduce the variables Player One controls gradually throughout the proof and give them together at the end. For now, we provide him with the variable *Abstain* and the goal formula:

$$\gamma_1 = \big(\mathit{Abstain} \wedge \gamma_1(\mathfrak{G}(u))\big) \vee \big(\neg \mathit{Abstain} \wedge \mathit{Machine} \wedge \mathit{Oracle}\big).$$

The subformula *Machine* asserts that Player One specifies a valid computation history of $M$ on $w$ (with the possible exception of an incorrect oracle response) that ends in an accepting state. Encoding a polynomial-time machine is trivial so we relegate the construction to the appendix.

**Lemma 5.3.6.** *There exists a formula, Machine, such that:*

1. *$\nu \vDash \mathit{Machine}$ if and only if $\nu$ represents an accepting run of $M$ on $w$ in $k$ steps that is correct with respect to everything except possibly the oracle responses.*
2. *$|\mathit{Machine}|$ is polynomial in the size of $(M, w, 1^k)$.*
3. *Machine has variables $Q_{i,j,l}$ and $NQ_{i,j,l}$ representing that the $i$th register has the symbol $q_l$ or $\neg q_l$ in cell $j$ respectively. That is, the $j$th literal of $\varphi_i$ is $q_l$ or $\neg q_l$.*
4. *Machine has variables $R_i$ to denote the value of the $i$th most significant bit of the oracle's response.*
5. *Machine has variables $\{F_1, \ldots, F_k\}$ to denote the value of the header register.*

*Proof.* See Appendix A.2. Q.E.D.

While the details of the construction are unimportant, the reader should pay attention to points 3, 4 and 5 as those variables will be used in the sequel.

The subformula *Oracle* will be designed in a manner to ensure that $\mathbb{E}[\mathit{Oracle} \mid \boldsymbol{\sigma'}] > u$ if and only if the formula specified by Player One in $\boldsymbol{\sigma'}$ does in fact have the number of satisfying assignments that Player One asserted. To do this we will implement the $g$ function from Theorem 5.5 of Schoenebeck and Vadhan [9], after transforming it to fit into the $[0, 1]$ range required by a Boolean game.

Of course, the difficulty here is that we do not know precisely how many satisfying assignments $\varphi$ has – Player One can submit whichever formula he wants. This is where groups $A$ and $B$ come in. The strategies chosen by the players in a group will define, uniformly at random, a truth assignment to $\varphi$. Thus while we do not know $m$, the number of satisfying assignments that $\varphi$ has, we do know that the assignment chosen by group $A$ or $B$ has precisely a $m/2^k$ chance of satisfying $\varphi$.[4] Some algebraic manipulation can then lead us to $m$.

To specify a truth assignment, we equip the $i$th proposition-player of $A$ with a single variable $A_i$. The truth value of this is to be interpreted as the value assigned to $q_i$. In $\boldsymbol{\sigma}$, each of these players sets their variable to *true* with probability $1/2$.

*The variable names $p_i$ and $q_i$, as well as the subformula names $\varphi_i$, are used in the same sense as in Lemma 5.3.2.*

The clause-players in $A$ are in charge of the $p_i$ variables. In the case where $\varphi$ has $i$ subformulae, the $i$th clause-player chooses which $p_j$ is true; the other clause-players are ignored. The $i$th clause-player in $A$ controls the variables $CA_1, \ldots, CA_i$, and in $\boldsymbol{\sigma}$ he sets $CA_j$ to *true* and every other variable to *false* with probability $1/i$. Player One's choice of $F_i$ will tell us which clause-player to listen to.

Group $B$ is built identically to group $A$.

*$\overline{R_k}$ as in Lemma 5.3.6*

Recall that $[\![\overline{R_k}]\!]$ is the number of satisfying assignments that Player One claims that $\varphi$ possesses. Let $\varphi(A)$ be the truth value of $\varphi$ if evaluated by the truth assignment specified by players $A$, and likewise for $\varphi(B)$. Consider the following $g([\![\overline{R_k}]\!], \varphi(A), \varphi(B))$:

$$\begin{aligned}
g([\![\overline{R_k}]\!], \mathit{true}, \mathit{true}) &= \frac{2^{2k+1} - ([\![\overline{R_k}]\!]^2 - 2^{k+1}[\![\overline{R_k}]\!] + 2^{2k})}{2^{2k+2}}, \\
g([\![\overline{R_k}]\!], \mathit{true}, \mathit{false}) &= \frac{2^{2k+1} - (-2^k[\![\overline{R_k}]\!] + [\![\overline{R_k}]\!]^2)}{2^{2k+2}}, \\
&= g([\![\overline{R_k}]\!], \mathit{false}, \mathit{true}), \\
g([\![\overline{R_k}]\!], \mathit{false}, \mathit{false}) &= \frac{2^{2k+1} - [\![\overline{R_k}]\!]^2}{2^{2k+2}}.
\end{aligned}$$

Note that this is bounded above by 1 and below by 0, so it is within the range of feasible payoffs for a Boolean game.

The expected value of $g$, in any $\boldsymbol{\sigma'}$ where groups $A$ and $B$ choose an assignment uniformly at random, is:

$$\begin{aligned}
\mathbb{E}[g \mid \boldsymbol{\sigma'}] = &(\frac{m}{2^k})^2 g([\![\overline{R_k}]\!], \mathit{true}, \mathit{true}) + 2(\frac{m}{2^k})(1 - \frac{m}{2^k}) g([\![\overline{R_k}]\!], \mathit{false}, \mathit{true}) \\
&+ (1 - \frac{m}{2^k})^2 g([\![\overline{R_k}]\!], \mathit{false}, \mathit{false}).
\end{aligned}$$

Which simplifies to:

$$\frac{2^{2k+1} - (m - [\![\overline{R_k}]\!])^2}{2^{2k+2}}.$$

Clearly this function is maximised when $[\![\overline{R_k}]\!] = m$. Player One controls $\overline{R_k}$, so it remains to figure out how to incentivise him to maximise $g$.

The first component in implementing $g$ is a formula that tells us whether the assignment played by players $A$ does in fact satisfy $\varphi$. Our

4 We are tacitly assuming that $\varphi$ contains all $k$ possible variables. This is of course impossible given the time constraints of the machine, but the assumption is harmless if we treat the unlisted variables as dummies whose truth value does not matter.

requirement that each $\varphi_i$ is in 2CNF makes it easy to verify whether $\varphi_i$ is satisfied:

$$
\begin{aligned}
&\mathit{ASatPhi}_i = \\
&\bigwedge_{j \le k/2} \bigvee_{l \le k} \bigvee_{l' \le k} \Big( (Q_{i,2j,l} \wedge Q_{i,2j+1,l'} \wedge (A_l \vee A_{l'})) \\
&\quad \vee (Q_{i,2j,l} \wedge NQ_{i,2j+1,l'} \wedge (A_l \vee \neg A_{l'})) \\
&\quad \vee (NQ_{i,2j,l} \wedge Q_{i,2j+1,l'} \wedge (\neg A_l \vee A_{l'})) \\
&\quad \vee (NQ_{i,2j,l} \wedge NQ_{i,2j+1,l'} \wedge (\neg A_l \vee \neg A_{l'})) \\
&\quad \vee (\boldsymbol{NoneOf}(Q_{i,2j,1}, \ldots, Q_{i,2j,k}, NQ_{i,2j,1}, \ldots, NQ_{i,2j,k}) \\
&\qquad \wedge \boldsymbol{NoneOf}(Q_{i,2j+1,1}, \ldots, Q_{i,2j+1,k}, NQ_{i,2j+1,1}, \ldots, NQ_{i,2j+1,k}))\Big).
\end{aligned}
$$

The outer conjunction iterates over the clauses in $\varphi_i$ (a conjunction because every clause in a CNF formula needs to be satisfied). The two following disjunctions iterate over the $l$ and $l'$ that identify which $q_l$ and $q_{l'}$ appear in the clause under consideration (is is the job of *Machine* to ensure Player One places two and only two variables in each clause, so these disjunctions will be satisfied in at most one case).

The first four disjuncts that follow treat with the non-degenerate cases: if Player One said the $j$th clause contains $q_l$ and $\neg q_{l'}$ ($Q_{i,2j,l} \wedge NQ_{i,2j+1,l'}$), then we require that the players in $A$ set $q_l$ to *true* or $q_{l'}$ to *false* ($A_l \vee \neg A_{l'}$). The last disjunct handles the case where $\varphi_i$ does not contain a $j$th clause.

Recall that in verity $\varphi$ is definitely not going to contain all $k$ variables. If it actually has $k'$, then in our present formulation $\varphi(A)$ will have $2^{k-k'}$ times more satisfying assignments than $\varphi$. To deal with this, we will add a formula requiring that all unmentioned variables need to be false.

$$\mathit{NoSpurious} = \bigwedge_{i \le k} \Big( \bigwedge_{j,l \le k} \boldsymbol{NoneOf}(Q_{j,i,l}, NQ_{j,i,l}) \to \neg A_i \Big).$$

Now, suppose for a moment that $\varphi$ consists of $j$ such subformulae, $\varphi_i$. We could combine the formulae $\mathit{ASatPhi}_i$ as follows:

$$\mathit{ASatPhi}[j] = \bigwedge_{i \le j} (CA_i^j \to \mathit{ASatPhi}_i).$$

That is, if the $j$th clause-player picks $p_i$ then we check whether $\varphi_i$ is satisfied.

Of course, we do not know how many terms $\varphi$ is composed of. Which is why it is kind of Player One to supply us with that information. The final formula is:

*$F_i$ as in Lemma 5.3.6*

$$\mathit{ASatPhi} = \mathit{NoSpurious} \wedge \bigwedge_{i \le k} (F_i \to \mathit{ASatPhi}[i]).$$

With the canaries sorted, the high level plan is to have something that looks as follows:

$$\boldsymbol{Equal}(\overline{x_n}; \text{numerator of } g(\overline{R_k}, \mathit{ASatPhi}, \mathit{BSatPhi})) \wedge \gamma_1(\mathcal{G}(\overline{x_n})).$$

That is, Player One is to select a number that must match the numerator of $g$, and then play a game from Definition 5.3.3 with that value. This will ensure that Player One's payoff is precisely $g$. Now we just need a logical formula telling us what the numerator is.

It is a lot easier to verify arithmetic than to do it, so we will ask Player One to do the work for us. He will provide the values of the various terms in $g$ as well as of $g$ itself, and we will pass those values into logical formulae that are true if the numbers add up. To this end, we want to define $\xi_{t,t}(\overline{R_k}, \overline{G^1_{2k+2}})$ to be a formula that is true just if $[\![\overline{G^1_{2k+2}}]\!] = 2^{2k+2} \cdot g([\![\overline{R_k}]\!], true, true)$. That is, $[\![\overline{G^1_{2k+2}}]\!]$ is the numerator of $g([\![\overline{R_k}]\!], true, true)$.

Recall that this numerator looks as follows:

$$2^{2k+1} - ([\![\overline{R_k}]\!]^2 - 2^{k+1}[\![\overline{R_k}]\!] + 2^{2k}).$$

The main operation is subtraction, and we have already seen how to define $\boldsymbol{Sub}$. The minuend, $2^{2k+1}$, is a constant that can be represented by a $\top$ followed by $2k+1$ $\bot$s. The trouble is with the subtrahend, but we shall cross that bridge once we reach it. For now, we will ask Player One to tell us what the subtrahend is via an assignment to $\overline{sG^1_{2k+2}}$, and introduce a new formula, $s\xi_{t,t}$, to verify that Player One is not lying. This gives us $\xi_{t,t}$:

$$\xi_{t,t}(\overline{R_k}, \overline{G^1_{2k+2}}) = \boldsymbol{Sub}(\ulcorner \mathbf{2^{2k+1}} \urcorner; \overline{sG^1_{2k+2}}; \overline{G^1_{2k+2}}) \wedge s\xi_{t,t}(\overline{R_k}, \overline{sG^1_{2k+2}}).$$

Now we have to deal with the subtrahend, $[\![\overline{R_k}]\!]^2 - 2^{k+1}[\![\overline{R_k}]\!] + 2^{2k}$. The main operations are subtraction and addition, which we can do. The rightmost term is a constant which we can deal with, and the middle term is just $\overline{R_k}$ with $k+1$ $\bot$s after it. All we need to do now is to square $[\![\overline{R_k}]\!]$.

Recall the high-school multiplication algorithm: splitting up the multiplication of $k$-digit integers into $k$ sums, of size up to $2k$. Applied to binary, computing $[\![\overline{R_k}]\!] \cdot [\![\overline{R_k}]\!]$ would involve computing these $k$ sums, the $i$th being either $[\![\overline{R_k}]\!] \cdot 2^{i-1}$ or 0, depending on whether the $i$th most significant bit of $[\![\overline{R_k}]\!]$ is 1 or 0. If we ask Player One to provide us the summands (via $\overline{s_{2k}}$), the result of the partial sums $\sum_{j<i}[\![\overline{s^j_{2k}}]\!]$ (via $\overline{S_{2k}}$) and the squared value (via $\overline{R^2_{2k}}$), we can verify that the squared value is correct with $k$ $\boldsymbol{Add}$ statements.

$$\begin{aligned}
&\boldsymbol{Square}(\overline{R_k}; \overline{R^2_{2k}}; \overline{s^1_{2k}}; \ldots; \overline{s^k_{2k}}; \overline{S^1_{2k}}; \ldots; \overline{S^{k+1}_{2k}}) = \\
&\quad \bigwedge_{j \le k} ((\boldsymbol{Equal}(\overline{s^j_{2k}}; \overline{R_k} \cdot \ulcorner \mathbf{2^{j-1}} \urcorner) \wedge R_j) \vee (\boldsymbol{Equal}(\overline{s^j_{2k}}; \ulcorner \mathbf{0} \urcorner) \wedge \neg R_j)) \\
&\quad \wedge \bigwedge_{j \le k} (\boldsymbol{Add}(\overline{S^j_{2k}}; \overline{s^j_{2k}}; \overline{S^{j+1}_{2k}})) \wedge \boldsymbol{Equal}(\overline{S^1_{2k}}; \ulcorner \mathbf{0} \urcorner) \wedge \boldsymbol{Equal}(\overline{S^{k+1}_{2k}}; \overline{R^2_{2k}}).
\end{aligned}$$

$\overline{R_k} \cdot \ulcorner \mathbf{2^{j-1}} \urcorner$ is shorthand for a mixed proposition/constant sequence with $j-1$ $\bot$s on the end of $\overline{R_k}$, and $k-j+1$ $\bot$s at the front to ensure the sequence is of length $2k$. This is necessary as we have only defined ***Equal*** on sequences of the same length.

This gives us what we need:

$$
\begin{aligned}
&s\xi_{t,t}(\overline{R_k}, \overline{sG^1_{2k+2}}) = \\
&\quad \boldsymbol{Sub}(\overline{R^2_{2k}}; \overline{R_k} \cdot \ulcorner \mathbf{2k+1} \urcorner; \overline{aG^1_{2k+2}}) \wedge \boldsymbol{Add}(\overline{aG^1_{2k+2}}; \ulcorner \mathbf{2^{2k}} \urcorner; \overline{sG^1_{2k+2}}) \\
&\quad \wedge \boldsymbol{Square}(\overline{R_k}; \overline{R^2_{2k}}; \overline{s^1_{2k}}; \ldots; \overline{s^k_{2k}}; \overline{S^1_{2k}}; \ldots; \overline{S^{k+1}_{2k}}).
\end{aligned}
$$

Armed with ***Square***, the other cases succumb to us with ease.

$$
\begin{aligned}
\xi_{t,f} = \xi_{f,t} = &\boldsymbol{Sub}(\ulcorner \mathbf{2^{2k+1}} \urcorner; \overline{sG^2_{2k+2}}; \overline{G^2_{2k+2}}) \\
&\wedge \boldsymbol{Sub}(\overline{R^2_{2k}}; \overline{R_k} \cdot \ulcorner \mathbf{2^k} \urcorner; \overline{sG^2_{2k+2}}) \\
&\wedge \boldsymbol{Square}(\overline{R_k}; \overline{R^2_{2k}}; \overline{s^1_{2k}}; \ldots; \overline{s^k_{2k}}; \overline{S^1_{2k}}; \ldots; \overline{S^{k+1}_{2k}}). \\
\xi_{f,f} = &\boldsymbol{Sub}(\ulcorner \mathbf{2^{2k+1}} \urcorner; \overline{R^2_{2k}}; \overline{G^3_{2k+2}}) \\
&\wedge \boldsymbol{Square}(\overline{R_k}; \overline{R^2_{2k}}; \overline{s^1_{2k}}; \ldots; \overline{s^k_{2k}}; \overline{S^1_{2k}}; \ldots; \overline{S^{k+1}_{2k}}).
\end{aligned}
$$

The sequences in the above are assumed to be padded with leading $\bot$s whenever necessary.

Now construct three games from Definition 5.3.3: $\mathcal{G}(\overline{G^1_{2k+2}})$, $\mathcal{G}(\overline{G^2_{2k+2}})$ and $\mathcal{G}(\overline{G^3_{2k+2}})$. Note that if Player One were to attempt to satisfy $\xi_{t,t}(\overline{R_k}, \overline{G^1_{2k+2}}) \wedge \mathcal{G}(\overline{G^1_{2k+2}})$, he would have to set $[\![\overline{G^1_{2k+2}}]\!]$ to the numerator of $g(R, \mathit{true}, \mathit{true})$, and hence his expected utility in $\mathcal{G}(\overline{G^1_{2k+2}})$ would be the numerator of $g(R, \mathit{true}, \mathit{true})$ over $2^{2k+2}$, which is precisely $g(R, \mathit{true}, \mathit{true})$. And we have already seen that to maximise $g$, Player One would want to set $[\![\overline{R_k}]\!] = m$, the number of satisfying truth assignments of $\varphi$.

This gives us a first guess for Player One's goal. However, we warn the reader that this is not our final formulation, which is why we mark it with an apostrophe.

$$
\gamma_1' = (\mathit{Abstain} \wedge \gamma_1(\mathfrak{G}(u))) \vee (\neg \mathit{Abstain} \wedge \mathit{Machine} \wedge \mathit{Oracle}'),
$$

$$
\begin{aligned}
\mathit{Oracle}' = &(\mathit{ASatPhi} \wedge \mathit{BSatPhi} \wedge \xi_{t,t}(\overline{R_k}, \overline{G^1_{2k+2}}) \wedge \gamma_1(\mathcal{G}(\overline{G^1_{2k+2}}))) \\
&\vee (\mathit{ASatPhi} \wedge \neg \mathit{BSatPhi} \wedge \xi_{t,f}(\overline{R_k}, \overline{G^2_{2k+2}}) \wedge \gamma_1(\mathcal{G}(\overline{G^2_{2k+2}}))) \\
&\vee (\neg \mathit{ASatPhi} \wedge \mathit{BSatPhi} \wedge \xi_{f,t}(\overline{R_k}, \overline{G^2_{2k+2}}) \wedge \gamma_1(\mathcal{G}(\overline{G^2_{2k+2}}))) \\
&\vee (\neg \mathit{ASatPhi} \wedge \neg \mathit{BSatPhi} \wedge \xi_{f,f}(\overline{R_k}, \overline{G^3_{2k+2}}) \wedge \gamma_1(\mathcal{G}(\overline{G^3_{2k+2}}))).
\end{aligned}
$$

All that remains is the value of $u$.

Seeing how if $M$ accepts any word whatsoever, then Player One would be able to construe some machine history to satisfy *Machine*, Player One's choice comes down to taking $u$ utility in the left disjunct or $\mathbb{E}[\mathit{Oracle}' \mid \boldsymbol{\sigma}] = \mathbb{E}[g \mid \boldsymbol{\sigma}] = \frac{2^{2k+1} - (m - [\![\overline{R_k}]\!])^2}{2^{2k+2}}$ utility in the right.

Note that if $m = [\![\overline{R_k}]\!]$, where the expectation is maximal, this is just $1/2$. Since we want Player One to have a strict preference for the right disjunct in the situation where such a choice of $[\![\overline{R_k}]\!]$ is possible (else there would be no profitable deviation), $u$ needs to be strictly smaller than $1/2$ but larger than any other possible value of $\mathbb{E}[g \mid \boldsymbol{\sigma}]$. The closest $\mathbb{E}[g \mid \boldsymbol{\sigma}]$ can get to $1/2$ without actually reaching it is when the difference between $m$ and $[\![\overline{R_k}]\!]$ is unity, which gives us $\frac{2^{2k+1}-1}{2^{2k+2}}$.

Suppose we set $u$ to $\frac{2^{2k+1}-0.5}{2^{2k+2}}$. We have then constructed a $G$ and a $\boldsymbol{\sigma}$ satisfying the requirement that $\boldsymbol{\sigma}$ is an equilibrium if and only if $M$ does not have an accepting computation in under $k$ steps, and $|G|$ is polynomial in the size of $(M, w, 1^k)$. Are we done? Unfortunately not. We must not forget the size of $\boldsymbol{\sigma}$, and $\boldsymbol{\sigma}$ involves Player One playing an equilibrium strategy in $\mathfrak{G}(u)$ – this would require him to randomise over $2^{2k+1} - 1$ possible strategies.

To address this, we could try and shift the extra $\frac{1}{2^{2k+2}}$ term to the other disjunct. I.e., Player One can abstain and get $u = 1/2$, and $\mathfrak{G}(1/2)$ would only require him to randomise over two strategies, or he can get $\mathbb{E}[g \mid \boldsymbol{\sigma}] + \frac{1}{2^{2k+2}}$ in trying to satisfy *Oracle*. But that does not solve the problem: Player One may have a small strategy in $\mathfrak{G}(u)$, but Player Two would still have to play an equilibrium strategy in all the $\mathcal{G}$ games in the *Oracle* subformula, which is also exponentially large. This is where group $C$ comes in.

We introduce a variant of game $\mathcal{G}$:

**Definition 5.3.7.** Let $\mathcal{G}(\overline{u_n})$ be as in Definition 5.3.3. Define $\mathcal{G}'(\overline{u_n})$ to be where in lieu of Player Two, we have Players Two through $n+1$, each controlling just a single bit of what was earlier Player Two's choice.

Formally, $\mathcal{G}'(\overline{u_n})$ looks as follows:

$$\begin{aligned}\Phi_1 =& \{p_1, \ldots, p_n, q_1, \ldots, q_n, s_1, \ldots, s_n, t_1, \ldots, t_n, u_1, \ldots, u_n\}\\ \Phi_{i>1} =& \{r_{i+1}\}\end{aligned}$$

$$\begin{aligned}\gamma_1 =& (\boldsymbol{Sub}(\overline{q_i}, \overline{p_i}, \overline{u_i}) \wedge \boldsymbol{LessEq}(\overline{q_i}, \ulcorner \mathbf{2^n - 1} \urcorner)\\ &\wedge \boldsymbol{LessEq}(\overline{r_i}, \overline{q_i}) \wedge \boldsymbol{LessEq}(\overline{p_i}, \overline{r_i}))\\ &\vee \Big(\boldsymbol{Add}(\overline{s_i}, \overline{t_i}, \overline{u_i})) \wedge \boldsymbol{Sub}(\overline{q_i}, \ulcorner \mathbf{0} \urcorner, \overline{s_i}) \wedge \boldsymbol{Sub}(\ulcorner \mathbf{2^n - 1} \urcorner, \overline{p_i}, \overline{t_i})\\ &\wedge (\boldsymbol{LessEq}(\overline{r_i}, \overline{q_i}) \vee \boldsymbol{LessEq}(\overline{p_i}, \overline{r_i}))\Big)\\ &\vee \boldsymbol{Less}(\ulcorner \mathbf{2^n - 1} \urcorner, \overline{r_i}).\end{aligned}$$

We are not defining the goal formulae of the other players as they are of no importance to us at the present. ■

If the $i$th player in $C$ plays $r_i$ with probability $1/2$, then every possible integer is realised with probability $1/2^{2k+2}$, as required.

This allows us to state *Oracle* as follows:

$$\begin{aligned}
\mathit{Oracle} = \big(&\mathit{ASatPhi} \wedge \mathit{BSatPhi} \wedge \xi_{t,t}(\overline{R_k}, \overline{G^1_{2k+2}}) \\
&\wedge \boldsymbol{Add}(\overline{G^1_{2k+2}}; \ulcorner \mathbf{1} \urcorner; \overline{fG^1_{2k+2}}) \wedge \gamma_1(\mathcal{G}'(\overline{fG^1_{2k+2}}))) \\
\vee\big(&\mathit{ASatPhi} \wedge \neg \mathit{BSatPhi} \wedge \xi_{t,f}(\overline{R_k}, \overline{G^2_{2k+2}}) \\
&\wedge \boldsymbol{Add}(\overline{G^2_{2k+2}}; \ulcorner \mathbf{1} \urcorner; \overline{fG^2_{2k+2}}) \wedge \gamma_1(\mathcal{G}'(\overline{fG^2_{2k+2}}))) \\
\vee\big(&\neg \mathit{ASatPhi} \wedge \mathit{BSatPhi} \wedge \xi_{f,t}(\overline{R_k}, \overline{G^2_{2k+2}}) \\
&\wedge \boldsymbol{Add}(\overline{G^2_{2k+2}}; \ulcorner \mathbf{1} \urcorner; \overline{fG^2_{2k+2}}) \wedge \gamma_1(\mathcal{G}'(\overline{fG^2_{2k+2}}))) \\
\vee\big(&\neg \mathit{ASatPhi} \wedge \neg \mathit{BSatPhi} \wedge \xi_{f,f}(\overline{R_k}, \overline{G^3_{2k+2}}) \\
&\wedge \boldsymbol{Add}(\overline{G^3_{2k+2}}; \ulcorner \mathbf{1} \urcorner; \overline{fG^3_{2k+2}}) \wedge \gamma_1(\mathcal{G}'(\overline{fG^3_{2k+2}}))).
\end{aligned}$$

And the final goal formula:

$$\gamma_1 = (\mathit{Abstain} \wedge \gamma_1(\mathfrak{G}(1/2))) \vee (\neg \mathit{Abstain} \wedge \mathit{Machine} \wedge \mathit{Oracle}).$$

This formula is clearly polynomial in size. The goal formula of every other player is simply:

$$\gamma_i = \bot.$$

Their preferences do not matter, and so they can never be the cause of a deviation from $\boldsymbol{\sigma}$.

The final list of variables used is:

$$\begin{aligned}
\Phi_1 =& \mathrm{var}(\mathit{Machine}) \cup \{\overline{Q_k}, \overline{NQ_k}, \overline{F_k}, \overline{R_k}, \overline{G^1_{2k+2}}, \overline{sG^1_{2k+2}}, \overline{R^2_{2k}}, \\
&\overline{s^1_{2k}}, \dots, \overline{s^k_{2k}}, \overline{S^1_{2k}}, \dots, \overline{S^{k+1}_{2k}}, \overline{aG^1_{2k+2}}\} \\
&\cup \{ \overline{G^2_{2k+2}}, \overline{sG^2_{2k+2}}, \overline{G^3_{2k+2}}, \overline{fG^1_{2k+2}}, \overline{fG^2_{2k+2}}, \overline{fG^3_{2k+2}} \} \\
&\cup \mathrm{var}_1(\mathcal{G}(\overline{fG^1_{2k+2}})) \cup \mathrm{var}_1(\mathcal{G}(\overline{fG^2_{2k+2}})) \cup \mathrm{var}_1(\mathcal{G}(\overline{fG^3_{2k+2}})) \\
&\cup \mathrm{var}_1(\mathfrak{G}(1/2)), \\
\Phi_2 =& \mathrm{var}_2(\mathfrak{G}(1/2)), \\
\Phi_a =& \{A_i\} \text{ or } \{CA_i\}, \\
\Phi_b =& \{B_i\} \text{ or } \{CB_i\}, \\
\Phi_c =& \{r_i\}.
\end{aligned}$$

As there is a polynomial number of these, it remains to check the size of $\boldsymbol{\sigma}$.

Player One is playing *Abstain*, the equilibrium strategy in $\mathfrak{G}(1/2)$, and every other variable to *false*.[5] Player Two is playing the equilibrium strategy in $\mathfrak{G}(1/2)$. The proposition-players in $A$ and $B$ randomise over setting $q_i$ to *true* and *false* equally, giving each a total of two strategies in their support. The clause-players have at most $k$ strategies in the support of each, and the players in $C$ have two each. This completes the proof. Q.E.D.

5 Note that Player One does not have to play the equilibrium strategy in the $\mathcal{G}'$ games as he is playing *Abstain*, and hence his assignment to those variables does not matter.

## 5.4 DISCUSSION

In this chapter we demonstrated the flexibility of propositional logic and the proof techniques of the previous chapter. The reader will however note that the proofs presented here can be split into two categories: those that use more-or-less standard game theoretic reasoning to obtain a reduction from a problem we have already shown to be hard, such as UniqueNash or IrrationalNash, and those that require starting from scratch and encoding a Turing machine again such as ∀GuaranteeNash and IsNash. The first are certainly more satisfactory than the second: having done the hard work once, we would like to put it aside for good. But this does demonstrate the difficulty of finding natural reduction between concise problems when one does not have recourse to universal computation. The reader is encouraged to compare Theorem 5.3.5 to Theorem 5.5 of Schoenebeck and Vadhan [9] to see just what a difference having a full-fledged circuit makes.

It is also worth mentioning that Theorem 5.3.5 is a hardness result; the upper bound is still EXP. Theorem 5.5 of Schoenebeck and Vadhan [9] is a completeness result, but that is because the authors consider approximate solutions, where Evaluation can be done in $FP^{\#P}$. In Appendix B we can only show a lower bound on exact Evaluation, so a hardness result is all we have.

In terms of the practical significance of these results, all the caveats we mentioned in the previous chapter still apply. However, it is worth nothing that while the decision problems aside from IsNash have little import to a player, they are very natural questions for a social planner to ask. The high complexity of these problems suggests that the planner may have to take an approximate or indirect approach. At the same time it is worth noting that these results are obtained under the assumption that the players are computationally unbounded. For an applied theory of Boolean games it would be beneficial to consider the case where this is not so, and we briefly discuss this in Chapter 7.

IsNash, as we have said, differs in that it is a very natural question for the players themselves to ask. After all, if a player does not know that he is in equilibrium, who can blame him for trying a different strategy to see if it might do better? However, unlike the other problems studied, the hardness of IsNash depends crucially on the number of players being unbounded. If it is fixed, the complexity drops to coNP (Corollary B.2.4). While we would not normally consider coNP a particularly tractable class, compared to the rest of the problems we have studied it is very much a walk in the park.

# 6 ZERO-SUM GAMES

*Better to wrong the world than have it wrong me!*

— Cao Cao, Romance of the Three Kingdoms

In this chapter we prove that determining whether the value of a two-player zero-sum game exceeds a threshold is EXP-complete. The proof is not original; it is a translation of Theorem 4.6 of Feigenbaum, Koller, and Shor [84] into the setting of Boolean games using the coding machinery we have thus far developed.

## 6.1 THE VALUE OF A GAME

**Theorem 6.1.1.** DVALUE *for Boolean games is* EXP*-complete.*

*Proof.* To see that the problem is in EXP, expand the Boolean game into its normal form and run the linear programming algorithm.

For hardness, consider $(M, K, w)$ where $M$ is a deterministic Turing machine, $K$ a computation bound and $w$ an input word. Feigenbaum, Koller, and Shor [84] demonstrate how to construct, in polynomial time, a game $G(M, K, w)$ and a rational $v$ such that the value of $G(M, K, w)$ is at least $v$ if and only if $M$ accepts $w$ in at most $K$ steps. What we will show is that we can construct a Boolean game with the same value as $G(M, K, w)$, the size of which is polynomial in $|M|$, $|K|$ and $|w|$. This will prove the theorem.

*In the case of Feigenbaum, Koller, and Shor, $v = 0$.*

For the sake of self-containment, we will replicate the construction of Feigenbaum, Koller, and Shor [84] below. The reader well acquainted with the construction can skip straight to Lemma 6.1.2.

We wish to associate with $M$ a set of Horn clauses $S$ over a set of propositional variables $P$ such that $M$ accepts $w$ in at most $K$ steps if and only if there exists an assignment to the variables in $P$ satisfying every clause in $S$.

$P$ contains propositions of the form $p[t, l, a]$ and $p[t, l, (s, a)]$. The intended interpretation of $p[t, l, a]$ is that cell $l$ contains symbol $a$ at computation step $t$. Without loss of generality, we are working with a binary alphabet, so $a$ is 0, 1, or the blank tape symbol ␣. The intended interpretation of $p[t, l, (s, a)]$ is that, in addition to the above, the head is over cell $l$ and in state $s$.

$S$ contains three types of clauses. The first type describe the initial configuration of the machine. These consist of $p[0, 0, (q_1, w[0])]$, $p[0, i, w[i]]$ for $1 \leq i < |w|$, $p[0, i, ␣]$ for $i \geq |w|$, and $\neg p[0, x, y]$ for every $x, y$ not conforming to the preceding types. The second type describe the transition rules of the machine. These take the form:

$$(p[t, l-1, \varsigma_1] \wedge p[t, l, \varsigma_2] \wedge p[t, l+1, \varsigma_3]) \to p[t+1, l, \varsigma],$$
$$(p[t, l-1, \varsigma_1] \wedge p[t, l, \varsigma_2] \wedge p[t, l+1, \varsigma_3]) \to \neg p[t+1, l, \varsigma'],$$

*As in Theorem 4.2.1, we can without loss of generality assume that $M$ will only accept in the first cell with 0 written on the tape.*

choosing appropriate values for $\varsigma_1, \varsigma_2, \varsigma_3$ and $\varsigma' \neq \varsigma$. The last clause is $p[K-1, 0, (q_f, 0)]$, asserting that $M$ accepts at time $K$.

For convenience, we will treat every negative clause as a clause with a consequent of *false*. That is, instead of $\neg p[0, x, y]$ and $(p_1 \wedge p_2 \wedge p_3) \to \neg q$ we will have the clauses $p[0, x, y] \to \mathit{false}$ and $(p_1 \wedge p_2 \wedge p_3 \wedge q) \to \mathit{false}$. This will mean a clause can have anywhere between 0 and 4 proposition in the tail – a true initial condition, a false initial condition, a positive boundary rule, a positive rule, a negative rule.

The game $G(M, K, w)$ proceeds by letting Player One choose $r \in P$ and Player Two an element $C \in S, C = \bigwedge p_i \to q$. Letting $R \subseteq P$ be the set of variables made true in the unique run of $M$ on $w$, the payoff to One is as follows:

$$H(r, C) = \begin{cases} 1 + \alpha, & r = q, \\ -1 + \alpha, & r = p_i, \\ \alpha, & \text{otherwise}. \end{cases}$$

In the above, $\alpha = \frac{j-1}{|R|}$, where $0 \leq j \leq 4$ is the number of literals in the antecedent of $C$. To simplify matters we assume that $|R| = 2^{2k}$, i.e. we consider the first $2^k$ computation steps and $2^k$ tape cells. This can be done by endowing the machine with a "do nothing" transition as before.

**Lemma 6.1.2** (Feigenbaum, Koller, and Shor [84])**.** *$M$ accepts $w$ in at most $K$ steps if and only if the value of $G(M, K, w)$ is at least 0.*

This is where we seek to hijack the rest of their proof. If we can demonstrate that we can build a Boolean game with the same value as $G(M, K, w)$ in polynomial time, we are done. The first hurdle we face is that we clearly cannot – the payoffs of a Boolean game are restricted to $\{0, 1\}$, so we cannot implement $H$. As such we adjust the payoffs as follows:

$$H'(r, C) = \begin{cases} 3/4 + \alpha/4, & r = q, \\ 1/4 + \alpha/4, & r = p_i, \\ 1/2 + \alpha/4, & \text{otherwise}. \end{cases}$$

$H'$ is obtained via the affine transformation $x \mapsto x/4 + 1/2$, so applying Theorem 2.1.22 to Lemma 6.1.2 we can conclude that a game with these payoffs has a value of at least $v = 1/2$ if and only if $M$ accepts $w$ in at most $K$ steps.

Note that now the payoffs are within $[\frac{1}{4} - \frac{1}{4 \cdot 2^{2k}}, \frac{3}{4} + \frac{3}{4 \cdot 2^{2k}}]$, and thus for $k \geq 1$ they are contained in $[0, 1]$, the range of feasible expected utilities in a Boolean game.

Now, suppose we can find sets of variables $\Phi_1'$ and $\Phi_2'$, and fifteen (polynomial size) formulae $\varphi_{r=q}^j$, $\varphi_{r=p_i}^j$, $\varphi_{\neq}^j$, $j \in \{0, 1, 2, 3, 4\}$, with the following properties:

- Every truth assignment to $\Phi_1'$ corresponds to a choice of $r \in P$.
- Every truth assignment to $\Phi_2'$ corresponds to a choice of $C \in S$.
- $\varphi_{r=q}^j$ is true if and only if $C$ has $j$ elements in the tail and $r$ is equal to the head of $C$. Mutatis mutandis, for the other $\varphi$.

We claim that at that point we are done, and the following is the game we desire:

$$\begin{aligned}
\Phi_1 &= \Phi_1' \cup \mathrm{var}_1(\mathfrak{G}(3/4 + \alpha/4)) \cup \mathrm{var}_1(\mathfrak{G}(1/4 + \alpha/4)) \cup \mathrm{var}_1(\mathfrak{G}(1/2 + \alpha/4)),\\
\Phi_2 &= \Phi_2' \cup \mathrm{var}_2(\mathfrak{G}(3/4 + \alpha/4)) \cup \mathrm{var}_2(\mathfrak{G}(1/4 + \alpha/4)) \cup \mathrm{var}_2(\mathfrak{G}(1/2 + \alpha/4)),\\
\gamma_1 &= \bigvee_{j \leq 4} (\varphi_{r=q}^j \wedge \gamma_1(\mathfrak{G}(3/4 + \alpha/4))) \vee \bigvee_{j \leq 4} (\varphi_{r=p_i}^j \wedge \gamma_1(\mathfrak{G}(1/4 + \alpha/4)))\\
&\quad \vee \bigvee_{j \leq 4} (\varphi_{\neq}^j \wedge \gamma_1(\mathfrak{G}(1/2 + \alpha/4))).
\end{aligned}$$

By Lemma 3.2.1, the subgames can be constructed in polynomial time. As such, all that remains is to provide $\Phi_1'$, $\Phi_2'$, and $\varphi_{r=q}^j$, $\varphi_{r=p_i}^j$, $\varphi_{\neq}^j$, for $j \in \{0, 1, 2, 3, 4\}$.

We start with Player One:

*As before, $k$ is the bit length of $K$. We do not use $q$ for $|Q|$ to avoid confusion with the head propositions.*

$$\Phi_1' = \{Zero^1, One^1\} \cup \{\overline{Time_k^1}\} \cup \{\overline{Tape_k^1}\} \cup \{\overline{State_{|Q|}^1}\}.$$

We map a truth assignment to $\Phi_1'$ to $r \in P$ in the following way:

- If both $Zero^1$ and $One^1$ are true, or more than one of $\{\overline{State_{|Q|}^1}\}$ is true, then the assignment is treated as $p[0, 0, 0]$.[1]
- If $State_m^1$ and, without loss of generality, $Zero^1$ is true then the assignment is treated as $p[[\![\overline{Time_k^1}]\!], [\![\overline{Tape_k^1}]\!], (q_m, 0)]$.
- If all the state variables are false and, say, $Zero^1$ is true, then the assignment is treated as $p[[\![\overline{Time_k^1}]\!], [\![\overline{Tape_k^1}]\!], 0]$.

For Player Two we have a larger set of variables:

$$\begin{aligned}
\Phi_2' = &\{pZero^2, pOne^2, Zero^2, One^2, sZero^2, sOne^2, nZero^2, nOne^2\}\\
&\cup \{\overline{Time_k^2}\} \cup \{\overline{Tape_2^k}\} \cup \{\overline{nState_{|Q|}^2}\} \cup \{\overline{pState_{|Q|}^2}\} \cup \{\overline{State_{|Q|}^2}\}\\
&\cup \{\overline{sState_{|Q|}^2}\} \cup \{Negative, Accept\}.
\end{aligned}$$

We map a truth assignment to $\Phi_2'$ to $C \in S$ in the following way:

1 In the previous section we punished a player for making an illegal move by having them lose the game. However, we cannot do this here as if both players play illegally the game would fail to be zero-sum. Instead, we pick an arbitrary legal move for them, in this case $p[0, 0, 0]$.

- An illegal configuration is mapped to $p[K-1, 0, (q_f, 0)]$.

- The $\overline{Time^2_k}$, $\overline{Tape^2_k}$ variables refer to the step/cell specified by the consequent – this uniquely defines the step/cell values of the tail propositions. Thus, if $[\![\overline{Time^2_k}]\!] = 0$, then the clause is treated as an initial configuration clause, $p[0, [\![\overline{Tape^2_k}]\!], w[i]]$. If $[\![\overline{Tape^2_k}]\!]$ is 0 or $2^k - 1$, then the clause is a boundary case and hence has only two propositions in the tail. If *Negative* is set to *true*, then the assignment is mapped to a negative clause. With this in mind, the contents of $\bigwedge p_i \to q$ are derived from the assignment in the natural way: recall, $nOne^2$ refers to the contents of the *n*ext computation step, or the head of the clause. $sOne^2$ and $pOne^2$ are the *s*uccessor and *p*redecessor of the central literal in the tail, and hence refer to the right and left cell.

- *Accept* is a special variable used to mark the fact that Player Two is playing the accepting clause, $p[K-1, 0, (q_f, 0)]$. If *Accept* is set to *true*, and Player Two plays $[\![\overline{Time^2_k}]\!] = K-1$, $[\![\overline{Tape^2_k}]\!] = 0$, $nZero^2$ and $nState^2_{accept}$, the assignment is treated as $p[K-1, 0, (q_f, 0)]$. [2]

At this point the reader should convince themselves that the mapping defined above does, in fact, allow Player One to specify every proposition in $P$ and Player Two every clause in $S$.

Let us now turn to $\varphi^j_{r=q}$. We will deal with $j = 0$ and $j = 3$. The case of $j = 2$ is obtained from $j = 3$ by changing the appropriate cell index and $j = 1, 4$ is simply *false*, as these are negative clauses and Player One is incapable of guessing the consequent in that instance.

For $j = 0$ there are three possibilities to consider. Player Two may have correctly specified a positive initial condition, the accepting clause, or played an illegal configuration. Recall that a negative initial condition clause is treated as $q \to false$, and thus falls under $j = 1$. We also need not consider Player One playing an illegal configuration, as $p[0, 0, 0]$ cannot appear in the head of any clause, and hence cannot satisfy $\varphi^j_{r=q}$.

$$\varphi^0_{r=q} = Init \vee Final \vee Illegal_i.$$

*The initial state is* $q_1$.

*Init* requires that $\overline{Time^2_k}$ encodes 0; the state variables are false unless $\overline{Tape^2_k}$ encodes 0, in which case only $State^2_1$ is true; and if $\overline{Tape^2_k}$ encodes $j$ then the $nZero^2$, $nOne^2$ variables are played in accordance with $w[j]$. *Negative* and *Accept* are both false. Player One plays his time, tape variables such that they encode the same numbers as Player Two's, and likewise the players agree on the state and content variables.

2 This is technically redundant: Player Two could specify the accepting clause by playing any illegal assignment, but for the purposes of transparency we do not wish to make illegal play a necessary aspect of the game.

*Init* can be broken down into a correctness and a matching requirement.

$$\mathit{Init} = \mathit{Init}_c \wedge \mathit{MatchHead}.$$

Line by line, the formula below reads: if the chosen cell is not 0, the head is not over the cell. If the chosen cell is 0, the head is over the cell and in state $q_0$. If the chosen cell is $i < |w|$, then Player Two sets $w[i] \in \{\mathit{nZero}^2 \wedge \neg \mathit{nOne}^2, \mathit{nOne}^2 \wedge \neg \mathit{nZero}^2\}$ to *true*, depending on the bit of $w$. If the chosen cell is $i \geq |w|$, then the cell is blank. As the clause is neither accepting nor negative, both those variables are set to *false*.

$$\begin{aligned}
\mathit{Init}_c = & (\neg \boldsymbol{Equal}(\ulcorner \mathbf{0} \urcorner; \overline{\mathit{Tape}_k^2}) \to \boldsymbol{NoneOf}(\overline{\mathit{nState}_{|Q|}^2})) \\
& \wedge \Big(\boldsymbol{Equal}(\ulcorner \mathbf{0} \urcorner; \overline{\mathit{Tape}_k^2}) \to (\mathit{nState}_1^2 \wedge \boldsymbol{OneOf}(\overline{\mathit{nState}_{|Q|}^2}))\Big) \\
& \wedge \bigwedge_{0 \leq i < |w|} (\boldsymbol{Equal}(\ulcorner \boldsymbol{i} \urcorner; \overline{\mathit{Tape}_k^2}) \to w[i]) \\
& \wedge (\neg \boldsymbol{Less}(\overline{\mathit{Tape}_k^2}; \ulcorner |\boldsymbol{w}| \urcorner) \to (\neg \mathit{nZero}^2 \wedge \neg \mathit{nOne}^2)) \\
& \wedge \neg \mathit{Accept} \wedge \neg \mathit{Negative}.
\end{aligned}$$

Note that the third line expands into $|w|$ conjuncts, so the formula is of polynomial size.

*MatchHead* states that Player One specifies the same proposition as is in the head of Player Two's clause. That is, the cell is the same, the computation step is the same, the tape contents are the same and the machine state is the same. This is a general term that we will reuse in other subformulae.

$$\begin{aligned}
\mathit{MatchHead} = & \boldsymbol{Equal}(\overline{\mathit{Tape}_k^1}; \overline{\mathit{Tape}_k^2}) \wedge \boldsymbol{Equal}(\overline{\mathit{Time}_k^1}; \overline{\mathit{Time}_k^2}) \\
& \wedge (\mathit{Zero}^1 \leftrightarrow \mathit{nZero}^2) \wedge (\mathit{One}^1 \leftrightarrow \mathit{nOne}^2) \\
& \wedge \bigwedge_{1 \leq i \leq |Q|} (\mathit{State}_i^1 \leftrightarrow \mathit{nState}_i^2).
\end{aligned}$$

*Final* likewise has a correctness and a matching requirement. The correctness requirement asks that Player Two set *Accept* to *true* and specify $p[K-1, 0, (q_f, 0)]$. The matching requirement we can reuse from the preceding case.

$$\begin{aligned}
\mathit{Final} = & \mathit{Final}_c \wedge \mathit{MatchHead}, \\
\mathit{Final}_c = & \mathit{Accept} \wedge \neg \mathit{Negative} \wedge \mathit{nState}_{accept}^2 \wedge \boldsymbol{OneOf}(\overline{\mathit{nState}_{|Q|}^2}) \\
& \wedge \boldsymbol{Equal}(\ulcorner \boldsymbol{K-1} \urcorner; \overline{\mathit{Time}_k^2}) \wedge \boldsymbol{Equal}(\ulcorner \mathbf{0} \urcorner; \overline{\mathit{Tape}_k^2}) \\
& \wedge \mathit{nZero}^2 \wedge \neg \mathit{nOne}^2.
\end{aligned}$$

$\mathit{Illegal}_i$ says that Player Two names an illegal configuration and Player One names $p[K-1, 0, (q_f, 0)]$.

$$\mathit{Illegal}_i = \mathit{TwoIllegal} \wedge \mathit{OneFinal}.$$

Let us list everything that could constitute an illegal assignment for Player Two:

1. The presence of both 1 and 0 in any specified cell.
2. The presence of more than one state in any cell.
3. The presence of the head in more than one cell in the tail.
4. Player Two names computation step 0, but supplies an incorrect initial configuration of the machine.
5. Player Two plays *Accept* and does not correctly describe $p[K-1, 0, (q_f, 0)]$.
6. Player Two names computation step $\geq 1$ and supplies a clause inconsistent with the transition rules of the machine.

*The absence of the head is covered by 6.*

We will introduce a formula for each item. *TwoIllegal* will be the disjunction of these formulae.

$$\begin{aligned}
1 =& (pZero^2 \wedge pOne^2) \vee (Zero^2 \wedge One^2) \vee (sZero^2 \wedge sOne^2) \\
& \vee (nZero^2 \wedge nOne^2). \\
2 =& (\neg \boldsymbol{OneOf}(\overline{nState^2_{|Q|}}) \wedge \neg \boldsymbol{NoneOf}(\overline{nState^2_{|Q|}})) \\
& \vee (\neg \boldsymbol{OneOf}(\overline{pState^2_{|Q|}}) \wedge \neg \boldsymbol{NoneOf}(\overline{pState^2_{|Q|}})) \\
& \vee (\neg \boldsymbol{OneOf}(\overline{State^2_{|Q|}}) \wedge \neg \boldsymbol{NoneOf}(\overline{State^2_{|Q|}})) \\
& \vee (\neg \boldsymbol{OneOf}(\overline{pState^2_{|Q|}}) \wedge \neg \boldsymbol{NoneOf}(\overline{sState^2_{|Q|}})). \\
3 =& (\neg \boldsymbol{NoneOf}(\overline{pState^2_{|Q|}}) \wedge \neg \boldsymbol{NoneOf}(\overline{State^2_{|Q|}})) \\
& \vee (\neg \boldsymbol{NoneOf}(\overline{pState^2_{|Q|}}) \wedge \neg \boldsymbol{NoneOf}(\overline{sState^2_{|Q|}})) \\
& \vee (\neg \boldsymbol{NoneOf}(\overline{State^2_{|Q|}}) \wedge \neg \boldsymbol{NoneOf}(\overline{sState^2_{|Q|}})). \\
4 =& \boldsymbol{Equal}(\ulcorner \mathbf{0} \urcorner; \overline{Time^2_k}) \\
& \wedge \Big( \Big( \neg Negative \wedge \big( \bigvee_{0 \leq i < |w|} (\boldsymbol{Equal}(\ulcorner \boldsymbol{i} \urcorner; \overline{Tape^2_k}) \wedge \neg w[i]) \\
& \qquad \vee (\neg \boldsymbol{Less}(\overline{Tape^2_k}; \ulcorner |\boldsymbol{w}| \urcorner) \wedge (nZero^2 \vee nOne^2))\big)\Big) \\
& \quad \vee \Big( Negative \wedge \big( \bigvee_{0 \leq i < |w|} (\boldsymbol{Equal}(\ulcorner \boldsymbol{i} \urcorner; \overline{Tape^2_k}) \wedge w[i]) \\
& \qquad \vee (\neg \boldsymbol{Less}(\overline{Tape^2_k}; \ulcorner |\boldsymbol{w}| \urcorner) \wedge (\neg nZero^2 \wedge \neg nOne^2))\big)\Big)\Big). \\
5 =& \ Accept \\
& \wedge (\neg \boldsymbol{Equal}(\ulcorner \boldsymbol{K-1} \urcorner; \overline{Time^2_k}) \vee \neg \boldsymbol{Equal}(\ulcorner \mathbf{0} \urcorner; \overline{Tape^2_k}) \\
& \vee \neg nState^2_{accept} \vee \neg nZero^2).
\end{aligned}$$

The last formula we will not provide in its entirety. Its general form is a disjunction:

$$6 = \neg \bigvee_{Rule \in M} Rule.$$

That is, we check whether some rule is inconsistent with the clause. A difficulty arises because a rule of the form $(q_i, \varsigma) \to (q_j, D, \varsigma')$ manifests itself in as many as 24 different Horn clauses – boundary cases and locations of the head. Of course 24 is a constant, so as far as our proof goes there is no problem in introducing that many terms into the disjunction for every rule of the machine, but doing so could well double the size of the present document. Instead we will give a concrete example of one specific case: the rule $(q_3, 0) \to (q_4, R, 1)$ where the head is initially in the middle cell and the middle is neither 0 nor $2^k - 1$:

$$\boldsymbol{NoneOf}(\overline{nState^2_{|Q|}}) \wedge State^2_3 \wedge \neg\boldsymbol{Equal}(\ulcorner\mathbf{0}\urcorner; \overline{Tape^2_k})$$
$$\wedge \neg\boldsymbol{Equal}(\ulcorner\mathbf{2^k - 1}\urcorner; \overline{Tape^2_k}) \wedge \neg\boldsymbol{Equal}(\ulcorner\mathbf{0}\urcorner; \overline{Time^2_k}) \wedge Zero^2 \wedge nOne^2.$$

We will also need to introduce "negative rules" to correspond to what the machine does not do. These will be treated in a similar way, all that needs to be mentioned is that for each $(q_i, \varsigma) \to (q_j, D, \varsigma')$ there will be only polynomially (of order $O(|Q| \cdot 2 \cdot |\Sigma|)$) many $(q_i, \varsigma) \to \neg(q'_j, D', \tau)$.

So much for $j = 0$. Let us turn to $j = 3$.

This turns out to be a lot easier as we have already done much of the grunt work. All we need is for Player Two to name a step $\geq 1$, a cell $\geq 1$ and $< 2^k - 1$, a correct configuration, and for Player One to guess the head.

$$\varphi^3_{r=q} = \neg\boldsymbol{Equal}(\ulcorner\mathbf{0}\urcorner; \overline{Time^2_k}) \wedge \neg\boldsymbol{Equal}(\ulcorner\mathbf{0}\urcorner; \overline{Tape^2_k})$$
$$\wedge \neg\boldsymbol{Equal}(\ulcorner\mathbf{2^k - 1}\urcorner; \overline{Tape^2_k}) \wedge \neg\, TwoIllegal \wedge MatchHead.$$

Next up is $\varphi^j_{r=p_i}$. We will deal with $j = 4$. There is no case for $j = 0$, and $j = 1$, $j = 2$, $j = 3$ can be easily obtained from $j = 4$.

Let us start by introducing the formulae checking for Player One guessing the tail. We have already seen *MatchHead*, which is applicable in the case of $j = 4$ because we treat negative clauses as $(\bigwedge p_i \wedge q) \to$ *false*. The others are built similarly.

$$MatchLeft = \boldsymbol{Succ}(\overline{Tape^2_k}; \overline{Tape^1_k}) \wedge \boldsymbol{Succ}(\overline{Time^1_k}; \overline{Time^2_k})$$
$$\wedge (Zero^1 \leftrightarrow pZero^2) \wedge (One^1 \leftrightarrow pOne^2)$$
$$\wedge \bigwedge_{1 \leq i \leq |Q|} (State^1_i \leftrightarrow pState^2_i).$$

$$MatchCentre = \boldsymbol{Equal}(\overline{Tape^1_k}; \overline{Tape^2_k}) \wedge \boldsymbol{Succ}(\overline{Time^1_k}; \overline{Time^2_k})$$
$$\wedge (Zero^1 \leftrightarrow Zero^2) \wedge (One^1 \leftrightarrow One^2)$$
$$\wedge \bigwedge_{1 \leq i \leq |Q|} (State^1_i \leftrightarrow State^2_i).$$

$$MatchRight = \boldsymbol{Succ}(\overline{Tape^1_k}; \overline{Tape^2_k}) \wedge \boldsymbol{Succ}(\overline{Time^1_k}; \overline{Time^2_k})$$
$$\wedge (Zero^1 \leftrightarrow sZero^2) \wedge (One^1 \leftrightarrow sOne^2)$$
$$\wedge \bigwedge_{1 \leq i \leq |Q|} (State^1_i \leftrightarrow sState^2_i).$$

Note that this is all we need to capture the case where Player One plays legally:

$$\begin{aligned}\varphi^4_{r=p_i} =& \mathit{Illegal}_{r=p_i} \vee \Big(\neg \boldsymbol{Equal}(\ulcorner \mathbf{0} \urcorner; \overline{\mathit{Time}^2_k}) \wedge \neg \boldsymbol{Equal}(\ulcorner \mathbf{0} \urcorner; \overline{\mathit{Tape}^2_k})\\ &\wedge \neg \boldsymbol{Equal}(\ulcorner \mathbf{2^k - 1} \urcorner; \overline{\mathit{Tape}^2_k}) \wedge \neg \mathit{TwoIllegal} \wedge \mathit{Negative}\\ &\wedge (\mathit{MatchHead} \vee \mathit{MatchLeft} \vee \mathit{MatchCentre} \vee \mathit{MatchRight})\Big).\end{aligned}$$

$\mathit{Illegal}_{r=p_i}$ is also relatively simple. Player One must make a violation, and Player Two needs to play a legal clause with $p[0,0,0]$ in the tail.

$$\begin{aligned}\mathit{Illegal}_{r=p_i} =& \mathit{OneIllegal} \wedge \neg \mathit{TwoIllegal}\\ &\wedge \Big((\mathit{pZero}^2 \wedge \boldsymbol{NoneOf}(\overline{\mathit{pState}^2_{|Q|}})\\ &\quad \wedge \boldsymbol{Equal}(\ulcorner \mathbf{1} \urcorner; \overline{\mathit{Time}^2_k}) \wedge \boldsymbol{Equal}(\ulcorner \mathbf{1} \urcorner; \overline{\mathit{Tape}^2_k}))\\ &\vee (\mathit{Zero}^2 \wedge \boldsymbol{NoneOf}(\overline{\mathit{State}^2_{|Q|}})\\ &\quad \wedge \boldsymbol{Equal}(\ulcorner \mathbf{1} \urcorner; \overline{\mathit{Time}^2_k}) \wedge \boldsymbol{Equal}(\ulcorner \mathbf{0} \urcorner; \overline{\mathit{Tape}^2_k}))\Big).\end{aligned}$$

Player One does not have a lot of creativity in how to play incorrectly:

$$\begin{aligned}\mathit{OneIllegal} =& (\mathit{One}^1 \wedge \mathit{Zero}^1)\\ &\vee (\neg \boldsymbol{OneOf}(\overline{\mathit{State}^1_k}) \wedge \neg \boldsymbol{NoneOf}(\overline{\mathit{State}^1_k})).\end{aligned}$$

Finally we come to $\varphi^j_{\neq}$, where we look at $j = 3$.

There are four cases: both players play correctly and differ in the step/cell specified. Player One plays incorrectly and Player Two plays a correct clause not covering step/cell $(0,0)$. Player Two plays incorrectly and Player One plays a correct proposition not covering the step/cell $(K-1,0)$, and of course both players could play incorrectly, in which case $p[0,0,0]$ does not cover $p[K-1,0,(q_f,0)]$.

$$\begin{aligned}\varphi^3_{\neq} =& \neg \mathit{Negative} \wedge \neg \mathit{Accept} \wedge \neg \boldsymbol{Equal}(\ulcorner \mathbf{0} \urcorner; \overline{\mathit{Time}^2_k})\\ &\wedge \neg \boldsymbol{Equal}(\ulcorner \mathbf{0} \urcorner; \overline{\mathit{Tape}^2_k}) \wedge \neg \boldsymbol{Equal}(\ulcorner \mathbf{2^k - 1} \urcorner; \overline{\mathit{Tape}^2_k})\\ &\wedge (\mathit{BothCorrect} \vee \mathit{TwoCorrect} \vee \mathit{OneCorrect} \vee \mathit{NoneCorrect}).\end{aligned}$$

We already have all the tools we need.

$$\begin{aligned}\mathit{BothCorrect} =& \neg \mathit{OneIllegal} \wedge \neg \mathit{TwoIllegal}\\ &\wedge \neg(\mathit{MatchHead} \vee \mathit{MatchLeft} \vee \mathit{MatchCentre} \vee \mathit{MatchRight}).\\ \mathit{TwoCorrect} =& \mathit{OneIllegal} \wedge \neg \mathit{TwoIllegal}\\ &\wedge \neg(\boldsymbol{Equal}(\ulcorner \mathbf{1} \urcorner; \overline{\mathit{Time}^2_k}) \wedge \boldsymbol{Equal}(\ulcorner \mathbf{0} \urcorner; \overline{\mathit{Tape}^2_k}) \wedge \mathit{Zero}^2)\\ &\wedge \neg(\boldsymbol{Equal}(\ulcorner \mathbf{1} \urcorner; \overline{\mathit{Time}^2_k}) \wedge \boldsymbol{Equal}(\ulcorner \mathbf{1} \urcorner; \overline{\mathit{Tape}^2_k}) \wedge \mathit{pZero}^2).\end{aligned}$$

$$
\begin{aligned}
\mathit{OneCorrect} &= \neg \mathit{OneIllegal} \wedge \mathit{TwoIllegal} \\
&\quad \wedge \neg(\boldsymbol{Equal}(\ulcorner \boldsymbol{K-1} \urcorner; \overline{\mathit{Time}^1_k}) \wedge \boldsymbol{Equal}(\ulcorner \mathbf{0} \urcorner; \overline{\mathit{Tape}^1_k}) \\
&\qquad \wedge \mathit{Zero}^1 \wedge \mathit{State}^1_{accept}). \\
\mathit{NoneCorrect} &= \mathit{OneIllegal} \wedge \mathit{TwoIllegal}.
\end{aligned}
$$

Let us recap: we have shown that we can interpret truth assignment to $\Phi'_1$ and $\Phi'_2$ as choices of a clause and proposition, and that we can define polynomial-size formulae that are true just if the named proposition correctly matches the head of the clause. Using these formulae we can construct a game where Player One finds himself playing $\mathfrak{G}(v)$ whenever $H'(r, C) = v$. To see that the resulting Boolean game has the same value as the game with payoffs defined by $H'$, one need only consider the equilibrium where Player One and Two play an equilibrium of the Feigenbaum, Koller, and Shor game on the $\Phi'_1 \uplus \Phi'_2$ variables, and the equilibria of the $\mathfrak{G}(v)$ games on the rest. Q.E.D.

## 6.2 DISCUSSION

Apart from yet another demonstration of our coding techniques, a take-home message of this section is the necessity of interpreting complexity results in context. VALUE is often cited as an example of an easy problem in game theory – after all, it is only polynomial. Well, once we remember that this is a function of the input, and the input is the normal form, the picture changes somewhat. After all, chess is a two-player zero-sum game, yet a thousand years down the line we have yet to determine its value.

It is also worth stressing that the fact that we have looked at DVALUE rather than VALUE is significant. Within polynomial space, we tend to switch between a function problem and its decision variant without much thought, but we cannot do so once we go beyond that. In Appendix B.1 we discuss some of the conceptual problems that arise when dealing with this issue, and hence why we deal only with DVALUE in the above.

# 7 CONCLUSION

*A strange game. The only winning move is not to play.*

— WOPR, WarGames

The technical contributions of this thesis are summarised in Table 7.1, alongside with known complexity results for the normal form. The general pattern is that an exponential jump of complexity occurs when we switch between the two representations; as the normal form representation of a game is at most exponential in the size of the smallest Boolean representation of the same game, this can be read as stating that Boolean games are "as hard as possible" – one can always solve a problem concerning a Boolean game by expanding it into its normal form first and, as far as asymptotic worst-case complexity is concerned, one can do no better. A more optimistic reading is that Boolean games offer a high level of succinctness – were it the case that in all non-trivial cases the normal form was not exponential in the size of the Boolean, then the expand-it-into-normal-form algorithm would contradict the hardness results demonstrated here.

Within mathematics the proof is at least as important as the theorem, so the techniques used to derive these results deserve a mention. In Theorem 4.2.1 we have demonstrated how we can use mixed strategies to describe the computation history of a machine. This is a powerful technique as in a concise game a mixed strategy could carry an exponential amount of information and thereby describe exponentially large objects, and variations of this approach could be fruitful in any setting where pure strategies have a significantly smaller representation size than mixed strategies.

The gadget games of Section 3.4, and the variation of Definition 5.3.3, can be used to simplify proofs in Boolean games by allowing us, within limits, to simulate richer payoffs than $\{ 0, 1 \}$. These are unlikely to be of use outside of the Boolean games framework as they benefit only representations that are both succinct and with binary preferences – circuit games have no need of fictional preferences as they can already represent any rational-valued game, while win-lose games in normal form cannot benefit from these gadgets because while $\mathfrak{G}(v)$ has a Boolean form polynomial in $|v|$, the normal form is nevertheless of order $v$, or $2^{|v|}$. This means that even though we have shown that IRRATIONAL-NASH is coNEXP-hard for Boolean games, this is unlikely to help us in proving that the problem is coNP-hard for win-lose normal-form games.

Faced with this pattern of an exponential jump, however, a reader acquainted with complexity theory may ask, is this not obvious? Boolean games, after all, are a succinct representation of games in normal form.

| Problem | Normal form | Boolean games |
|---|---|---|
| ∃GuaranteeNash | NP-complete | NEXP-complete |
| ∀GuaranteeNash | coNP-complete | coNEXP-complete |
| UniqueNash | coNP-complete | coNEXP-complete |
| ∃NashSat | N/A | NEXP-complete |
| ∀NashSat | N/A | coNEXP-complete |
| RationalNash† | NP-hard | NEXP-hard |
| IrrationalNash | NP-complete | NEXP-complete |
| IsNash† | P | $\text{coNP}^{\#\text{P}}$-hard |
| DValue | P-complete | EXP-complete |

Table 7.1: Comparison of the complexity of problems studied between the normal form and Boolean representations. Problems are for the two-player case unless marked by dagger.

Should there not be a general result, in the vein of Papadimitriou and Yannakakis [97], that defines a general class of problems for which NP-hardness in the normal form implies NEXP-hardness for Boolean games? The author had such thoughts before, during, and after obtaining the results of this thesis; the question was also raised at the 3rd International Workshop on Strategic Reasoning by a member of the audience; and a close reading of Mavronicolas, Monien, and Wagner [69][1] suggests that the authors may have had a similar construction in mind. All I can say on this note is that if such a result exists, then I have not been able to find it. The standard arguments tend to rely on universal computation, and hence not applicable in the case of Boolean games as propositional logic is (possibly)[2] less concise than Boolean circuits.

## 7.1 PLAYING BOOLEAN GAMES

The approach taken in this thesis was to study Boolean games strictly as a concise representation of win-lose games, and to address standard problems of algorithmic game-theory in this setting. This was done to firmly establish the place of Boolean games among the many models of strategic behaviour that exist out there, and their relationship to other representations.

1 "It is nevertheless possible to transform a boolean circuit into a polynomial size set of *clauses*; [...] Hence, there is a polynomial time transformation of a circuit game into a boolean formula game". If I understand the authors correctly, then such a construction would run into difficulties in the presence of mixed strategies.

2 In particular, a polynomial translation of an arbitrary Boolean circuit to a formula would imply that the NC hierarchy collapses to the first level. I.e., polylogarithmic time is equal to logarithmic time on a parallel machine.

A shortcoming of this approach is that it fails to capture some of the essence of the framework that makes it so appealing to the multiagent community. People do not view a Boolean game as a representation, but as something of itself, that has more to do with logic than classical game theory. This is demonstrated by the fact that perfectly natural extensions of the framework, such as CP-Boolean games, are very difficult to interpret from the perspective of finite strategic games, and the concept of play in mixed strategies is a lot harder to make sense of.

However, I maintain that mixed strategies are important. Randomisation is a necessary response to uncertainty and games without uncertainty, attractive as they may be to a social planner, are in some sense trivial: children quickly grow bored of tic-tac-toe, but rock-paper-scissors has international tournaments. The only thing saving chess from a similar fate is its colossal size.[3]

All the standard questions of Boolean games, such as implementing desirable equilibria and determining tractable fragments, can and ought to be asked in the setting of mixed strategies. However, if one is to stay relevant to Boolean games as a model of strategic interaction, it pays to consider some computationally friendly restrictions.

To play a pure strategy in a Boolean game, a player need only flick some bits in his set of propositional letters. A mixed strategy involves randomising over exponentially many assignments. Exponential space is a much graver handicap than exponential time. Agents might be willing to wait, or to use approximate reasoning, but if they cannot even store their final plan of action then they cannot possibly be expected to play the game.

In this work we have already seen one way of dealing with: drop the zero weights. One could approach the problem by positing that agents can store a maximum of $k$ strategies in their support and see what comes of it. This is worth considering, but there is also much that is unsatisfactory with this approach. If I have the power to randomise over $k$ plays, surely if I really wanted I could go for $k+1$?

Computational restrictions are perhaps more natural. We could, for instance, assume the agents possess a computational device that on input $s_i$ returns the weight of $s_i$ in their strategy. This would allow us to study a hierarchy of computationally bounded agents, with black-box agents at the top being equivalent to the unbounded players considered in this thesis, down to agents with a very modest circuit in their device, that only allows for limited randomisation.

It has also been brought to my attention that at present there are scholars considering randomised play in Boolean games that operates not by randomising over all possible assignments, but rather setting the probability with which $p_i$ is set to *true*. These are dubbed *behavioural strategies* given their similarity with their namesake in extensive games. I had briefly considered this concept at the start of my doctorate, but

3 Checkers, of course, is no longer worth playing.

in the end found them unsuitable to the goals of this work. The trouble is that behavioural strategies can only capture a subset of all possible mixed strategies, and Nash's theorem does not apply to this subset. Consider, for example, the following game:

$$\begin{aligned}
\Phi_1 &= \{ p_1, p_2 \}, \\
\Phi_2 &= \{ q \}, \\
\gamma_1 &= (p_1 \leftrightarrow q) \wedge (p_2 \leftrightarrow q), \\
\gamma_2 &= \neg\gamma_1.
\end{aligned}$$

This is essentially Matching Pennies, except Player One has two variables in his disposal. He loses if they are assigned different values, so the only equilibrium play would involve him setting *both* $p_1$ and $p_2$ to *true* with probability $^1/_2$, and *both* to *false* otherwise.

Still, in the context of modelling strategic behaviour of computationally bounded agents such behavioural strategies are an interesting approach. It is also worth noting that, even though probability is a very different concept from vagueness, such a strategy has a superficial resemblance to fuzzy logic. It would be interesting to see whether some connection with Łukasiewicz games can be drawn.

## 7.2 OPEN PROBLEMS OF COMPLEXITY

The most obvious lacuna in the current presentation is the complete absence of any function problems, even though VALUE, FINDNASH and EVALUATION are some of the most natural questions one could ask about a game. These problems could prove to be both technically and conceptually difficult, for reasons discussed in Appendix B.1.

The work has also focused on the complexity of exact solutions rather than approximation schemes, in contrast to the focus in the circuit games literature. Given the high complexity of the problems involved, approximate reasoning could well be the only way to deal with Boolean games; it would be interesting to study the complexity of such problems.

A more abstract, though I would argue much more significant issue is the length and complexity of the proofs involved. The problem, of course, pertains not only to the current work but to a great deal of modern mathematics. Man today is much the same as he was two millennia ago, yet a single conference's proceedings could hold more technical content than all the beaches of Attica. Diagrams and explanations to bolster the reader's intuition only go so far, and if the reader must rely on faith alone to carry them through a twenty page argument then the author has failed; the minute a proof stops being self-evident it stops being a proof.

There has been much interest in automatic proof verification in recent years, and perhaps this is the way to go. Though a mixed-strategy space is ostensibly continuous, the proofs throughout this session were essentially combinatorial in nature; it could be the case that they lend themselves well to at least partial automation. However, it is my belief that there is still room for elegance in mathematics, and that writing a readable proof involves nothing more than writing a readable proof. Atrocities like Theorem 5.1.5 could be avoided with the aid of ink, paper, and a great deal of thought.

# Part III

# APPENDIX

# A OMITTED PROOFS

*The beautiful is good,*
*And if a thing's not beautiful it isn't good.*

— Theognis, Elegies

## A.1 FROM SECTION 5.2

**Theorem 5.2.2.** RationalNash *for three-player Boolean games is* NEXP*-hard.*

*Proof.* Let $G_1$ be a game with only irrational equilibria and the positive utility property. Let $(G_2, v)$ be an instance of $\exists$GuaranteeNash$^*$. The desired $G'$ is:

$$\begin{aligned}
&\Phi_1 = \mathrm{var}_1(G_1) \cup \mathrm{var}_1(G_2) \cup \{\, \mathit{Choice}_1 \,\} \cup \mathrm{var}_1(\mathfrak{G}_1(v)),\\
&\Phi_2 = \mathrm{var}_2(G_1) \cup \mathrm{var}_2(G_2) \cup \{\, \mathit{Choice}_2 \,\} \cup \mathrm{var}_1(\mathfrak{G}_2(1/2)) \cup \mathrm{var}_2(\mathfrak{G}_3(1/2)),\\
&\Phi_3 = \mathrm{var}_3(G_1) \cup \mathrm{var}_3(G_2) \cup \{\, \mathit{Choice}_3 \,\} \cup \mathrm{var}_1(\mathfrak{G}_3(1/2)) \cup \mathrm{var}_2(\mathfrak{G}_1(v))\\
&\quad \cup \mathrm{var}_2(\mathfrak{G}_2(1/2)),\\
&\gamma_1' = (\bigwedge_{i\le 3} \mathit{Choice}_i \wedge \gamma_1(G_2)) \vee (\bigwedge_{i\le 3} \neg\mathit{Choice}_i \wedge (\gamma_1(G_1) \vee \gamma_1(\mathfrak{G}(v))))\\
&\quad \vee (\neg\mathit{Choice}_1 \wedge (\mathit{Choice}_2 \vee \mathit{Choice}_2) \wedge \gamma_1(\mathfrak{G}_1(v))),\\
&\gamma_2' = \Big((\bigwedge_{i\le 3} \mathit{Choice}_i \wedge (\gamma_2(G_2) \vee \gamma_1(\mathfrak{G}_2(1/2)))) \vee (\bigwedge_{i\le 3} \neg\mathit{Choice}_i \wedge \gamma_2(G_1))\\
&\quad \vee (\neg\mathit{Choice}_2 \wedge (\mathit{Choice}_1 \vee \mathit{Choice}_3) \wedge \gamma_1(\mathfrak{G}_2(1/2)))\Big) \wedge \gamma_2(\mathfrak{G}_3(1/2)),\\
&\gamma_3' = \Big((\bigwedge_{i\le 3} \mathit{Choice}_i \wedge (\gamma_3(G_2) \vee \gamma_1(\mathfrak{G}_3(1/2)))) \vee (\bigwedge_{i\le 3} \neg\mathit{Choice}_i \wedge \gamma_3(G_1))\\
&\quad \vee (\neg\mathit{Choice}_3 \wedge (\mathit{Choice}_1 \vee \mathit{Choice}_2) \wedge \gamma_1(\mathfrak{G}_3(1/2)))\Big) \wedge \gamma_2(\mathfrak{G}_2(1/2))\\
&\quad \wedge \gamma_2(\mathfrak{G}_1(v)).
\end{aligned}$$

Should the reader still have the proof of Theorem 5.2.1 fresh in their minds, what we have done is the following:

$$\begin{aligned}
\gamma_2' &= \gamma_2 \wedge \gamma_2(\mathfrak{G}_3(1/2)),\\
\gamma_3' &= \gamma_3 \wedge \gamma_2(\mathfrak{G}_2(1/2)) \wedge \gamma_2(\mathfrak{G}_1(v)).
\end{aligned}$$

Suppose $(G_2, v) \in \exists$GuaranteeNash$^*$. Consider the profile where the players play a rational equilibrium satisfying the payoff constraint over the variables in $G$, equilibrium play over the gadget games, and setting the *Choice* variables to *true*. This is clearly an equilibrium, and all the strategy weights are rational.

Now suppose $(G_2, v) \notin \exists$GUARANTEENASH*. As before, let $p$, $q$ and $r$ be the probabilities of the players opting to play in $G_2$. If players Two and Three are playing equilibrium play over the gadget games, then the argument of Theorem 5.2.1 applies. Let us then consider the case where they are not.

Because Player Two's goal is conjunctive, the only reason she would play suboptimally in $\gamma_2(\mathfrak{G}_3(1/2))$ is if her probability of satisfying $\gamma_2$ is 0, and hence it does not matter what she does elsewhere. However, by observing her goal formula one will see that her probability of satisfying $\gamma_2$ is at least $1/2pqr + 1/2(1-q)(1-(1-p)(1-r)) + (1-p)(1-q)(1-r)x$, for a strictly positive $x$. The only way this could be 0 is if $q = 1$ and either $p = 0$ or $r = 0$, but that is not an equilibrium profile as Player Two could deviate to $q = 0$. Mutatis mutandis, for Player Three. Q.E.D.

**Fact A.1.1.** *A two-player zero-sum game has either one or a continuum of equilibria.*

*Arbitrary linear programs may not have solutions, but games do due to Theorem 2.1.12.*

*Proof.* We have seen in Section 3.3 the connection between the equilibria of a zero-sum game and the solutions[1] to a linear program. A solution to a linear program is found where the hyperplane of the objective function is tangent to the convex region determined by the constraints. The hyperplane will either be tangent on a corner, in which case the solution is unique, or on at least one edge, in which case we have a continuum.

For the reader averse to linear programming, a direct argument is possible. Let $\boldsymbol{\sigma} = (\sigma^1, \sigma^2)$ and $\boldsymbol{\tau} = (\tau^1, \tau^2)$ be two equilibria. Choose an $\alpha \in (0,1)$ and consider the profile $\boldsymbol{\rho} = (\alpha\sigma^1 + (1-\alpha)\tau^1, \sigma^2)$. Observe that $u_1(\boldsymbol{\rho}) = \alpha u_1(\boldsymbol{\sigma}) + (1-\alpha)u_1(\tau^1, \sigma^2) = u_1(\boldsymbol{\sigma})$. It can be no higher, as $\boldsymbol{\sigma}$ is an equilibrium, and the only way it could be lower is if $u_1(\tau^1, \sigma^2) < u_1(\boldsymbol{\sigma})$. The latter is impossible as, by Theorem 2.1.12, $u_1(\boldsymbol{\sigma}) = u_1(\boldsymbol{\tau})$, and since $\boldsymbol{\tau}$ is an equilibrium we know that $u_2(\boldsymbol{\tau}) \geq u_2(\tau^1, \sigma^2)$, hence $u_1(\boldsymbol{\tau}) \leq u_1(\tau^1, \sigma^2)$. It follows that Player One has no profitable deviation from $\boldsymbol{\rho}$. Player Two, by deviating to $\delta^2$, would expect a utility of $\alpha u_2(\sigma^1, \delta^2) + (1-\alpha)u_2(\tau^1, \delta^2)$. $u_2(\sigma^1, \delta^2)$ is bounded above by $u_2(\boldsymbol{\sigma})$, and $u_2(\tau^1, \delta^2)$ by $u_2(\boldsymbol{\tau}) = u_2(\boldsymbol{\sigma})$. Thus Player Two derives a utility of at most $u_2(\boldsymbol{\sigma})$, which is what she is already getting. Q.E.D.

**Lemma A.1.2** (Appa [99])**.** *Determining whether a linear program has a unique solution is in* P*.*

**Proposition 5.2.5.** IRRATIONALNASH *is in* P *for two-player zero-sum games in normal form, and* EXP *for two-player zero-sum Boolean games.*

*Proof.* If the equilibrium is unique, then it is rational due to the linear programming characterisation. Hence the only way multiple equilibria can exist if there is a continuum. Q.E.D.

1 In the language of linear programming, we would say the optimal feasible solutions.

## A.2 FROM SECTION 5.3

**Lemma 5.3.6.** *There exists a formula, Machine, such that:*

1. $\nu \vDash$ *Machine if and only if* $\nu$ *represents an accepting run of* $M$ *on* $w$ *in* $k$ *steps that is correct with respect to everything except possibly the oracle responses.*
2. $|\mathit{Machine}|$ *is polynomial in the size of* $(M, w, 1^k)$.
3. *Machine has variables* $Q_{i,j,l}$ *and* $NQ_{i,j,l}$ *representing that the* $i$*th register has the symbol* $q_l$ *or* $\neg q_l$ *in cell* $j$ *respectively. That is, the* $j$*th literal of* $\varphi_i$ *is* $q_l$ *or* $\neg q_l$.
4. *Machine has variables* $R_i$ *to denote the value of the* $i$*th most significant bit of the oracle's response.*
5. *Machine has variables* $\{F_1, \ldots, F_k\}$ *to denote the value of the header register.*

*Proof.* The proof is essentially Cook's Theorem with some register manipulation.

Observe that the computation bound, $k$, is given in unary. As a consequence the $k \times k$ table representing the history of the machine (à la Figure 4.1) is polynomial in the size of the input. As such we have no need to mess around with cell indices, but instead can ask Player One to describe the entire table directly.

We equip Player One with $k^2$ $l[i, j]$ variables for every $l \in \Sigma$, denoting that the $(i, j)$-entry of the computation history is $l$. We also provide $k^2$ state variables for every $q \in Q$, $q[i, j]$, denoting at once the location of the head and the state that it is in. This allows us to formulate the initial configuration of the machine:

$$\mathit{Initial} = q_0[0,0] \wedge \Big( \bigwedge_{j<|w|} w[j][0,j] \Big) \wedge \Big( \bigwedge_{j \geq |w|} \boldsymbol{NoneOf}(\{\, l[0,j] : l \in \Sigma \,\}) \Big).$$

*$w[j][0,j]$ is to be interpreted as $l[0,j]$, $l \in \Sigma$ being the $j$th entry of $w$.*

Now all we need is a *Rules* term which is nothing but a giant or-statement, checking that every square in the history follows from the preceding computation step via the application of an admissible rule, to establish that the machine history is correct. We shall not consider the standard transition rules and instead focus on the ones unique to the machine at hand.

Let $\mathit{Write}[t, i, j, l]$ be the rule that states that the machine at time $t$ writes symbol $q_l$ to the $j$th entry of the $i$th clause of $\varphi$. We can express it as follows:

$$\begin{aligned} \mathit{Write}[t,i,j,l] = &((q_w[t,0] \wedge p_l[t,0] \wedge i[t,1] \wedge j[t,2]) \\ &\wedge (Q_{i,j,l} \wedge q'[t+1,0])) \wedge \mathit{SameTape}[t]. \end{aligned}$$

Recall that we have included the integers $\{1, \ldots, k\}$ (of which there is a polynomial number) as primitives so there is no need to mess

*In Lemma 5.3.2 the proposition variable would be called $q_l$, but we use $p_l$ here to avoid confusion with state*

around with encoding numbers – we simply demand that the symbol $i$ be written on the tape to write to the $i$th clause. The state is $q_w$, the special write state, and $p_l[t, 0]$ is a propositional symbol, which is also included as a primitive. The next state, $q'$, can be chosen arbitrarily: the machine is nondeterministic and we can assume it jumps from $q'$ to wherever it needs to go to resume computation.

*SameTape*$[t]$ asserts that the tape contents are unchanged between steps $t$ and $t+1$, and that the head does not move. This is expressed in the obvious way:

$$\begin{aligned} \mathit{SameTape}[t] = \bigwedge_{i<k, l\in\Sigma} & ((l[t,i] \leftrightarrow l[t+1,i]) \\ & \wedge (\boldsymbol{NoneOf}(\{\, q[t,i] : q \in Q \,\}) \leftrightarrow \boldsymbol{NoneOf}(\{\, q[t,i] : q \in Q \,\}))). \end{aligned}$$

There are no more than $k^4$ such rules, and we analogously include the rules for writing a negative literal.

Next is the oracle query. We assume this is made from state $q_o$, giving us the following formula:

$$\mathit{Oracle}[t,i] = q_o[t,i] \wedge q'[t+1,i] \wedge \mathit{SameTape}.$$

There are $k^2$ possible cases depending on where and when the query is made.

Finally, the machine needs to be able to access the oracle results somehow. There are several ways we could handle this, a simple one is to allow the machine to request that the $i$th bit be written to the start of the tape by writing $i$ at the start of the tape and entering state $q_r$.

$$\begin{aligned} \mathit{Read}[t,i] =& (q_r[t,0] \wedge i[t,0]) \wedge \\ & (q'[t+1,0] \wedge (R_i \to 1[t+1,0]) \wedge (\neg R_i \to 0[t+1,0])) \\ & \wedge \mathit{SameTapeExceptCellZero}[t]. \end{aligned}$$

*SameTapeExceptCellZero*$[t]$ is exactly what it says.

We add the usual consistency constraints such as every cell can only contain one symbol, the machine can only be in one state, and so on. We do not need to worry about what happens if the machine writes an ill-configured formula into the registers, or tries to read the oracle's response before it makes the query – we can assume the machine itself would enter an abort state in such an instance. All that remains to do is verify that Player One behaves sensibly with the variables we use later in the construction.

So far we have guaranteed that if *Write*$[t,i,j,l]$ is issued, then $Q_{i,j,l}$ is true. At present nothing prevents Player One from setting to $Q_{i,j,l}$ to *true* regardless, so we have constraints of the following form for every $j, l$:

$$(\bigwedge_{t,i<k} \neg \mathit{Write}[t,i,j,l]) \to \neg Q_{i,j,l}.$$

Analogously for the negative literals.

We also need Player One to tell us how many clauses $\varphi$ has. As the internal rules of the machine guarantee that there are no gaps in between the clauses, we only need to spot the highest index of an issued write command.

$$\bigvee_{i \leq k} \bigwedge_{k \geq m > i} (\neg(\bigvee_{t,j,l} \mathit{Write}[t,m,j,l]) \wedge (\bigvee_{t,j,l} \mathit{Write}[t,i,j,l]) \wedge F_i)$$
$$\wedge \boldsymbol{OneOf}(\overline{F_i}).$$

The outer or ranges over candidates for the highest $i$ for which $\varphi_i$ is defined. For this to be true it must be the case that for all $m > i$ the write command was never issued, but that it was issued at $i$. In this case, $F_i$ ought to be true.

Putting these together, the lemma is proven. Q.E.D.

# B

# TRIVIAL OBSERVATIONS

*I'm afraid that some times you'll play lonely games too.*
*Games you can't win 'cause you'll play against you.*

— Dr. Seuss, Oh, the Places You'll Go!

## B.1 COMPARING FUNCTIONS

Function, or search problems, have been oft neglected in favour of their decision-version cousins. The extent of this neglect is such that, when the author set down in an (unsuccessful) attempt to prove a hardness result about VALUE, he was faced with the fact that he did not even know what the appropriate reducibility notion is. The purpose of this section is to discuss various functional reducibilities he had come upon in the literature, as well as why their use is problematic for function problems with potentially exponentially long output, such as the natural game theoretic problems applied to concise representations.

The term "function" is potentially misleading, as search problems are more correctly relational in nature; a typical definition of a search problem, in works that bother to define the concept, is by means of a computable predicate, $R$. On input $x$ the problem is to find some $y$ for which $Rxy$ holds, and there could be many, or perhaps none, $y$ satisfying that rôle – consider the task of finding a clique of size $k$ in a graph. We shall, however, assume these problems are functional in the discussion that follows. This will simplify presentation, and restating the arguments in terms of more general search problems would not be a difficult task.

We denote functional complexity classes by appending F to the name of the appropriate class of decision problems, or sets; thus FP and FEXP are the classes of functions computable in deterministic polynomial or exponential time respectively. Less obvious classes, such as $F\Sigma_2^p$, will not interest us in the sequel.

Functions are characterised not only by the computational complexity of producing their result, but also by the representational complexity of displaying that result once it is computed. A natural class of functions is where this result is commensurate with the input.

**Definition B.1.1.** A class of functions $\mathcal{C}$ is said to be *polynomially-balanced* if for every $f \in \mathcal{C}$ there exists a polynomial $p$ such that $p(|x|) \geq |f(x)|$. I.e., the size of the output is polynomial in the size of the input. ■

**Example B.1.2.** Every class in FPSPACE is polynomially-balanced. FEXP is not. ■

We now turn to reducibility between functions. As a starting point, it is reassuring to note that the Turing reduction survives the transition seemingly intact.

**Definition B.1.3.** A *polynomial-time Turing reduction*, or a *Cook reduction*, from function $f$ to $g$ is a polynomial-time oracle machine $M^g$ such that for all $x$:

$$M^g(x) = f(x).$$

We write $f \leq_T^{\mathrm{P}} g$ to indicate that there exists a polynomial-time Turing reduction from $f$ to $g$. ■

Turing reductions are defined identically for both decision and function problems; they are thus the only notion of reducibility that allows us to compare the difficulty of decision to search. Thus, for example, the claim that VALUE is EXP-hard for circuit games (Schoenebeck and Vadhan [20]) can only be interpreted in the context of $\leq_T^{\mathrm{P}}$.

As convenient as this property may be, Turing reductions are not what we want. It would be unsatisfactory to use $\leq_T^{\mathrm{P}}$ for functions when we use the much stricter $\leq_m^{\mathrm{P}}$ for sets, moreover since polynomially-balanced functions trivialise away under $\leq_T^{\mathrm{P}}$ – a polynomially-balanced $f$ is FPSPACE-hard if and only if it is PSPACE-hard, given bireducibility between $f$ and the decision problem "Is the $i$th bit of $f(x)$ 1?".

What we are after is a stricter notion of reducibility. We start with one of Simon [101]:[1]

*$\epsilon$ is chosen as a mnemonic for "encoding". Likewise, $\delta$ will stand for "decoding".*

**Definition B.1.4** (page 83 in Simon [101])**.** A *polynomial-time parsimonious reduction* from function $f$ to $g$ is a function $\epsilon \in \mathrm{FP}$ such that for all $x$:

$$f(x) = g \circ \epsilon(x).$$

We write $f \leq_{par}^{\mathrm{P}} g$ to indicate that there exists a polynomial-time parsimonious reduction from $f$ to $g$. ■

There is, however, a shortcoming to the parsimonious reduction: it is perhaps *too* strong. The following fact trivially follows:

**Fact B.1.5.** *$f \leq_{par}^{\mathrm{P}} g$ only if range$(f) \subseteq$ range$(g)$.*

As a consequence of this $\leq_{par}^{\mathrm{P}}$ is sensitive to the way we choose to encode mathematical objects and present our output – the function that takes as input the index of an exponential time machine $e$ and an input word $w$ and outputs $M_e(w)$ is clearly FEXP-hard, but the function that

1 It is, of course, difficult to determine the origin of concepts that to by now appear self-evident. To my knowledge, Simon was the first to introduce such a reduction in western literature, but in the Eastern Bloc it appears earlier, e.g. in Ershov's theory of numerations (Ершов [102]) on input $(e, w)$ prints "The $e$th machine on input $w$ outputs: ..." is not.

Parsimonious reductions seem more appropriate to "classical" algorithmic problems where the output of the function is meant to be taken at face value – an integer representing the size of the largest clique, for example. But this too is not without its problems. Depending on the biases of our background, we may define a function as $f : \mathbb{N} \to \mathbb{N}$ or $f : \{0,1\}^* \to \{0,1\}^*$, i.e. as a recursive function on integers or as the output of a computational process that manipulates strings. Ideally such a choice should be inconsequential on later theory, but clearly there are problems that would be complete with the definition $f : \mathbb{N} \to \mathbb{N}$ and not $f : \{0,1\}^* \to \{0,1\}^*$, unless we use a non-standard encoding of integers that makes use of leading zeroes.

As our interest is on the complexity of computing a function rather than our particular method of encoding mathematical objects as bit strings, it is natural to simply allow a decoding step to the reduction. This leads to the following definition:

**Definition B.1.6** (Zankó [103]). A *polynomial-time Zankó reduction*[2] from function $f$ to $g$ is a pair of functions $\epsilon, \delta \in \mathrm{FP}$ such that for all $x$:

$$f(x) = \delta \circ g \circ \epsilon(x).$$

We write $f \leq_Z^{\mathrm{P}} g$ to indicate that there exists a polynomial-time metric reduction from $f$ to $g$. ■

This is the notion that seems to be favoured by authors in the algorithmic game theory literature, e.g. Chen and Deng [31].

Levin used a similar, but not identical, notion in his 1973 paper:

**Definition B.1.7** (Левин [106]). A *polynomial-time Levin reduction* from function $f$ to $g$ is three functions $\epsilon, \delta, \tau \in \mathrm{FP}$ such that for all $x$:

1. $g \circ \epsilon(x) = \tau \circ f(x)$.
2. $f(x) = \delta \circ g \circ \epsilon(x)$.

We write $f \leq_L^{\mathrm{P}} g$ to indicate that there exists a polynomial-time Levin reduction from $f$ to $g$. ■

It may be difficult to get an intuition for the concept from the definition alone, and Levin does not offer any motivation himself, but one can view a Levin reduction simply as a strengthening of a Zankó reduction:

2 The author calls this a many-one reduction, on the grounds of similarity to the reducibility notion $\leq_m^{\mathrm{P}}$ between sets. However, Vollmer [104] labelled what we here term a parsimonious reduction as a many-one reduction, on the basis of it being similar to $\leq_m^{\mathrm{P}}$. Likewise, Krentel [105] felt that metric reductions were the "obvious" counterpart of many-one reducibility between sets. In my undergraduate years my lecturer told me that the obvious has no place in mathematics, hence I avoid the usage.

**Fact B.1.8.** *The functions $\epsilon, \delta, \tau \in \mathrm{FP}$ constitute a Levin reduction from $f$ to $g$ if and only if:*

1. *$\epsilon$ and $\delta$ constitute a Zankó reduction from $f$ to $g$.*
2. *$\tau$ is the pseudo-inverse of $\delta$ on the range of $f$.*

*Proof.* Recall the definition of a Levin reduction:

1. $g \circ \epsilon(x) = \tau \circ f(x)$.
2. $f(x) = \delta \circ g \circ \epsilon(x)$.

Condition 2 is precisely that $f \leq^{\mathrm{P}}_{Z} g$ via $\delta$ and $\epsilon$.

Recall that for $\tau$ to be a pseudo-inverse, $\tau(y)$ must be a pre-image of $y$ with respect to $\delta$. That is, $\delta \circ \tau(y) = y$. Observe:

$$\begin{aligned} g \circ \epsilon(x) &= \tau \circ f(x), \\ \delta \circ g \circ \epsilon(x) &= \delta \circ \tau \circ f(x), \\ f(x) &= \delta \circ \tau \circ f(x). \end{aligned}$$

As we are only concerned with the behaviour of $\tau$ on the range of $f$, this proves the claim. Q.E.D.

*Strictly speaking, Levin worked in the more general case of relational rather than purely functional problems.*

Definition B.1.7 is Levin's own, but there is an alternative definition of a "Levin reduction" that comes up perhaps more often the original.[3] This is a weaker notion, so we give it an appropriately creative name:

**Definition B.1.9.** A *polynomial-time weak Levin reduction* from function $f$ to $g$ is three functions $\epsilon, \delta, \tau \in \mathrm{FP}$ such that for all $x$:

1. $g \circ \epsilon(x) = \tau(x, f(x))$.
2. $f(x) = \delta(x, g \circ \epsilon(x))$.

We write $f \leq^{\mathrm{P}}_{wL} g$ to indicate that there exists a polynomial-time weak Levin reduction from $f$ to $g$. ■

Faliszewski and Ogihara [107] gives an identical definition to the one above, but they label it a *polynomial-time metric reversible reduction*. This is in reference to the *metric reduction*, another known reducibility metric. The relationship of a weak Levin reduction to a metric reduction parallels that of a Levin reduction to a Zankó reduction.

**Definition B.1.10** (Krentel [105])**.** A *polynomial-time metric reduction* from function $f$ to $g$ is a pair of functions $\epsilon, \delta \in \mathrm{FP}$ such that for all $x$:

$$f(x) = \delta(x, g \circ \epsilon(x)).$$

We write $f \leq^{\mathrm{P}}_{met} g$ to indicate that there exists a polynomial-time metric reduction from $f$ to $g$. ■

---

3 I have been unable to trace the origins of this notion, but it is rampant in many courses on complexity theory. See, for instance, www.csie.ntu.edu.tw/ lyuu/complexity/2008b/20081028.pdf.

An alternative name for the metric reduction is the *1-Turing reduction*, as $f \leq^{\mathrm{P}}_{met} g$ could be equally defined as the existence of an $M^g(x)$ computing $f(x)$ with at most one call to the oracle to $g$ – $\epsilon$ is the computation of the machine before the oracle query and $\delta$ is the post-computation, which has a dependency on $x$ because the machine would have already read the input by then.

*This equivalence only holds in polynomially-balanced classes*

We present a summary of the relationship between these notions below, which is largely based on the observations of Faliszewski and Ogihara [107]:

**Fact B.1.11** (after Faliszewski and Ogihara [107])**.** *The following diagram depicts the inclusion relations between degree structures of polynomially-balanced classes under various reducibility notions. The arrows point to the stronger*[4] *notion. Solid arrows are strict, dotted arrows weak.*

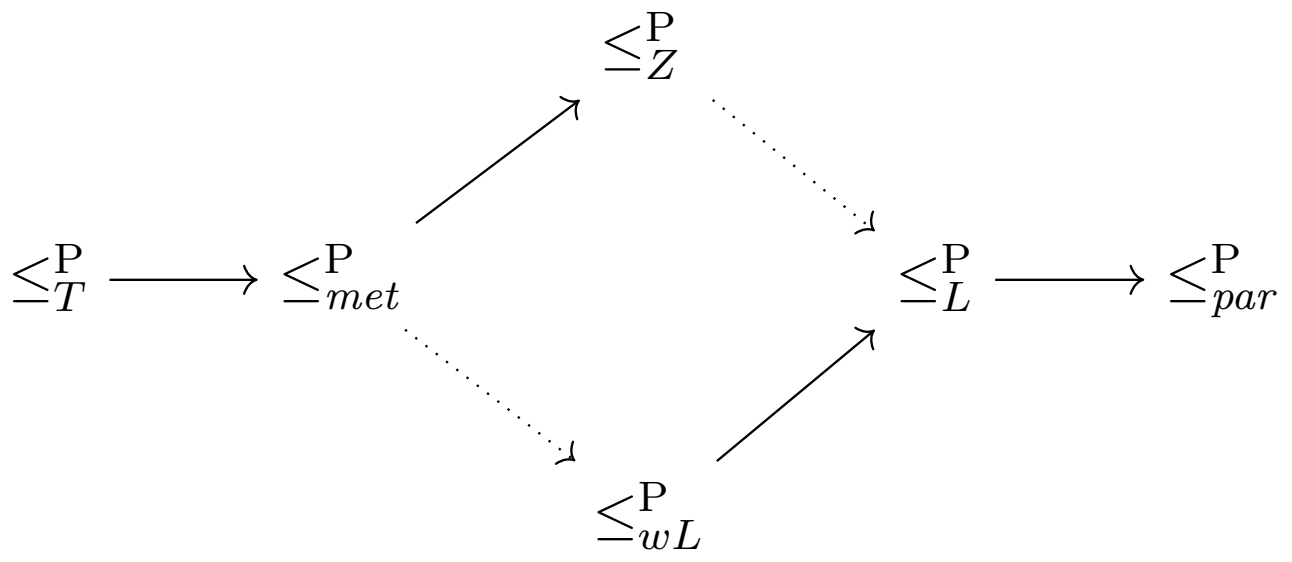

*Proof.* Weak inclusions all follow from definition.

To see that $f \leq^{\mathrm{P}}_{Z} g$ does not imply $f \leq^{\mathrm{P}}_{par} g$, consider $f : x \mapsto x$ and $g : x \mapsto 2x$. We have $f \leq^{\mathrm{P}}_{Z} g$ via $\epsilon : x \mapsto x$ and $\delta : x \mapsto x/2$, but $f \not\leq^{\mathrm{P}}_{par} g$ because the range of $g$ is only the even integers.

To see that $\leq^{\mathrm{P}}_{wL}$ does not imply $\leq^{\mathrm{P}}_{L}$ or $\leq^{\mathrm{P}}_{met}$ does not imply $\leq^{\mathrm{P}}_{Z}$, consider $f : x \mapsto x$ and $g : x \mapsto 0$. We have both $f \leq^{\mathrm{P}}_{wL} g$ and $f \leq^{\mathrm{P}}_{met} g$, with the decoding $\delta : (x, y) \mapsto x$, but $f \not\leq^{\mathrm{P}}_{wL} g$ and $f \not\leq^{\mathrm{P}}_{met} g$ because a univariate decoding function is not much use with a constant $g$.

To see that $f \leq^{\mathrm{P}}_{T} g$ does not imply $f \leq^{\mathrm{P}}_{met} g$, recall that Ladner, Lynch, and Selman [108] demonstrate the existence of sets $A$ and $B$ for which $A \leq^{\mathrm{P}}_{T} B$ but $A \not\leq^{\mathrm{P}}_{tt} B$ (hence, $A \not\leq^{\mathrm{P}}_{1-T} B$). It is clear that this implies that $\chi_A \leq^{\mathrm{P}}_{T} \chi_B$ but $\chi_A \not\leq^{\mathrm{P}}_{met} \chi_B$ (as metric reductions are identical to 1-Turing reductions in a polynomially balanced class).

*$\chi_A$ is the characteristic function of $A$*

Q.E.D.

Furthermore, it turns out that if $\mathrm{P} \neq \mathrm{NP}$, then $\leq^{\mathrm{P}}_{wL}$ and $\leq^{\mathrm{P}}_{Z}$ are in fact incomparable.

---

4 The terminology used in the literature on this point is unfortunately confusing. Some authors call $\leq^{\mathrm{P}}_{par}$ "stronger" because it is a stronger notion – $f \leq^{\mathrm{P}}_{par} g$ implies $f \leq^{\mathrm{P}}_{T} g$. Others call $\leq^{\mathrm{P}}_{T}$ "stronger" because it is a stronger reduction – there exist $f, g$ which $f \leq^{\mathrm{P}}_{T} g$ but about which $\leq^{\mathrm{P}}_{par}$ does not seem to say anything. Others still use "stronger" in both senses in the same paper, because they view their readers with contempt.

**Proposition B.1.12** (Faliszewski and Ogihara [107])**.** *The following relations hold:*

1. $\leq^{\mathrm{P}}_{wL}$ *does not imply* $\leq^{\mathrm{P}}_{Z}$.
2. *If* $\mathrm{P} \neq \mathrm{NP}$, $\leq^{\mathrm{P}}_{Z}$ *does not imply* $\leq^{\mathrm{P}}_{wL}$.

The world looks a little different, however, if the class is not polynomially balanced:

**Fact B.1.13.** *The following diagram depicts the inclusion relations between degree structures of classes of functions that are not polynomially-balanced:*

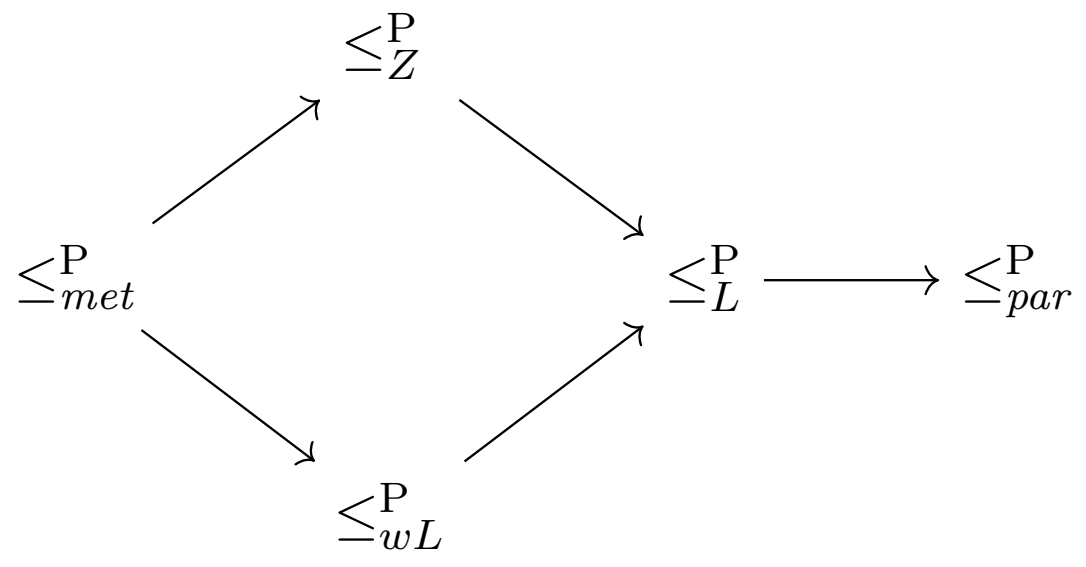

*Proof.* To see that $f \leq^{\mathrm{P}}_{Z} g$ does not imply $f \leq^{\mathrm{P}}_{L} g$ nor $f \leq^{\mathrm{P}}_{met} g$ imply $f \leq^{\mathrm{P}}_{wL} g$, consider $f : x \mapsto x$ and $g : x \mapsto 1^x$. There is both a Zankó and a metric reduction from $f$ to $g$ – $\epsilon$ is identity and $\delta$ takes $1^x$ to $x$. However, while $\delta$ is polynomial-time computable it cannot have a polynomial-time pseudo-inverse as the size of the input is exponential in the size of the output.[5]

Note that this also establishes that $\leq^{\mathrm{P}}_{Z}$ and $\leq^{\mathrm{P}}_{wL}$ are incomparable, even if $\mathrm{P} = \mathrm{NP}$. Q.E.D.

Note that we did not even include $\leq^{\mathrm{P}}_{T}$ on the diagram. In the context of functions whose output could be exponentially long, the very notion of Turing reducibility becomes questionable as it is highly sensitive to the model of computation used. For instance, where does the oracle dump its response? If it is on a special oracle tape, then the machine will simply not have the time to copy it to the working tape; if it is on to the working tape then it is critical to know whether it will overwrite the existing contents or be appended to the end, as the machine will not have the time to shift it. Likewise, it is critical to know whether or not the machine is capable of random access, as otherwise it will only be capable of reading and modifying an initial segment of the oracle response. One would also have to specify whether the tape is one-way or two-way infinite – in the latter case, the machine would be able to append some characters to the left of the oracle response, while in the former it would have to leave it as it is.

5 One could avoid this with a different definition of reducibility, for example where the decoding function is not required to give the entire output but only the $i$th bit. This will do nothing to address the other issues discussed, however.

As such, we argue that Turing reducibility is not particularly meaningful in this context – after all, we use Turing machines not because they are a reasonable model of computation, but because in the context of universal computation it does not matter what model we use. If we find ourselves in a setting where small alterations of the model can lead to significant differences in behaviour, perhaps we ought not to use a Turing machine at all.

The other reducibility notions remain well defined, but the standard intuition behind $\leq_L^{\mathrm{P}}$ and everything below is no longer applicable – the statement "$f$ is reducible to $g$" is typically interpreted as "give me an efficient algorithm for $g$, and I will give you an efficient algorithm for $f$". Not so, if the functions in question have exponential output – I can give you a polynomial-time $\epsilon$ and $\delta$, but if $|g(\epsilon(x))|$ is exponential in the size of the input, then $\delta(g(\epsilon(x)))$ will still run for a number of steps exponential in $|x|$. And if you have to wait that long anyway, why not just compute $f(x)$ directly? Parsimonious reductions, of course, avoid both problems, but they do not become any easier to work with.

Perhaps the trouble is that we have approached the problem from the wrong angle. Rather than asking how one should compare functions that are not polynomially-balanced, we should ask whether we really care about such objects. What profits it a man to study the complexity of $f$ if $|f(x)|$ is astronomical? Perhaps what we are really interested are classes such as the following:

**Definition B.1.14.** The class of *polynomially-balanced exponential-time functions*, $\mathrm{FEXP}_{\mathrm{PB}}$, is the class of functions $f : \{0,1\}^* \to \{0,1\}^*$, such that:

1. $f$ is computable in deterministic exponential time.

2. There exists a polynomial $p$ such that, for all $x$, $p(|x|) \geq |f(x)|$.

■

This is a syntactic class – it is categorised by exponential time machines with bounded output tapes – and as such it has complete problems. It is also a very natural class, as one could imagine many problems that take a long time to compute but do not present any difficulties with the representation of the output. However this is all moot, as the author does not have any $\mathrm{FEXP}_{\mathrm{PB}}$-completeness results to share with the reader, under any notion of reducibility stronger than $\leq_T^{\mathrm{P}}$.

## B.2 BOUNDS AND OBSERVATIONS

**Proposition B.2.1.** EVALUATION *for Boolean games is* #P*-hard (in the number of players) under Levin reductions.*

*Proof.* We will reduce from #SAT.

Let $\varphi$ be a formula with $n$ propositional variables. Consider a Boolean game $G$ with $n$ players, every $\gamma_i = \varphi$ and every $\Phi_i = \{ p_i \}$, i.e. a singleton set. Let $\boldsymbol{\sigma}$ be a profile where every player plays $p_i$ with probability $1/2$. Note that both $G$ and $\boldsymbol{\sigma}$ can be constructed in time polynomial in $|\varphi|$.

Clearly, $\boldsymbol{\sigma}$ realises every truth assignment with probability $1/2^n$, hence $u_i(\boldsymbol{\sigma}) = k/2^n$, where $k$ is the number of satisfying assignments of $\varphi$.

The encoding function of the reduction takes $\varphi$ to $(G, \boldsymbol{\sigma})$, and the decoding function takes $x$ to $x \cdot 2^n$. The decoding function is efficiently invertible by the function that takes $y$ to $y/2^n$, so this is a Levin reduction. Q.E.D.

**Corollary B.2.2.** Evaluation *for Boolean games is* $\mathrm{FP}^{\#\mathrm{P}}$*-hard (in the number of players) under Turing reductions.*

*Proof.* Evaluation is #P-hard, so one can simply trace the execution of $M \in \mathrm{FP}^{\#\mathrm{P}}$, and every time $M$ makes an oracle call replace it with the appropriate call to Evaluation. Q.E.D.

**Fact B.2.3.** Evaluation *for k-player Boolean games is in* FP.

*Proof.* Let $p_i$ denote the number of pure strategies in Player $i$'s support, and $m^i_j$ the representation size of the weight Player $i$ attaches to the $j$th strategy in his support. The representation size of $\boldsymbol{\sigma}$ is thus the sum of all such $m^i_j$. If we use $m^i$ to represent the mean size of the strategy weights chosen by Player $i$, the size of $\boldsymbol{\sigma}$ can be expressed with greater perspicuity as $\sum p_i m^i$. We can thus see that the total number of strategy profiles that could be realised by $\boldsymbol{\sigma}$, $\prod_{i \leq k} p_i$, is $O(p^k)$ for $p = \max_i p_i$, and hence polynomial in the size of $\boldsymbol{\sigma}$.

The naïve algorithm to compute $u_i(\boldsymbol{\sigma})$ will consider each of these $\prod_{i \in N} p_i$ profiles in turn. For every such pure profile, $\boldsymbol{\nu}$, the algorithm will compute the probability of $\boldsymbol{\nu}$ eventuating, $P(\boldsymbol{\nu} \mid \boldsymbol{\sigma})$, and if $\boldsymbol{\nu} \models \gamma_i$ then the algorithm will add $P(\boldsymbol{\nu} \mid \boldsymbol{\sigma})$ to the running total of $u_i(\boldsymbol{\sigma})$.

To compute $P(\boldsymbol{\nu} \mid \boldsymbol{\sigma})$, we need to multiply $k$ rationals. The rationals are given as input so their size is trivially polynomial in the size of the input, and $k$ is constant, so this operation can be done in polynomial time. Deciding whether $\boldsymbol{\nu} \models \gamma_i$ is polynomial in $(|\boldsymbol{\nu}|, |\gamma_i|)$, the latter of the two being part of the input and the former being proportional to $|\Phi|$. Finally, we add a polynomial number (one for each profile) of such weights (which are polynomial in the size of the input, because they are the output of a polynomial-time multiplication algorithm), which is in FP. Q.E.D.

**Corollary B.2.4.** IsNash *for k-player Boolean games is coNP-complete.*

*Proof.* For membership, given $(G, \boldsymbol{\sigma})$ we can guess a pure strategy deviation for Player $i$ and compare $u_i(\boldsymbol{\sigma}_{-i}(s_i))$ in nondeterministic polynomial time. If no such deviation exists, we must be at equilibrium.

For completeness we reduce from TAUTOLOGY: given $\varphi$, construct a game with $\gamma_1 = \neg p \wedge \neg\varphi$ for a fresh $p$, and $\Phi_1$ the entire variable set. Choose any strategy profile that sets $p$ to *true*. If this is at equilibrium, then it must mean that Player One has no deviation to satisfy $\neg\varphi$, which can only be the case if $\varphi$ is a tautology. Q.E.D.

## COLOPHON

This document was typeset using the typographical look-and-feel `classicthesis` developed by André Miede. The style was inspired by Robert Bringhurst's seminal book on typography "*The Elements of Typographic Style*". `classicthesis` is available for both LaTeX and LyX:

http://code.google.com/p/classicthesis/

Happy users of `classicthesis` usually send a real postcard to the author, a collection of postcards received so far is featured here:

http://postcards.miede.de/

*Final Version* as of June 8, 2016 (`classicthesis` version 0.9).